\documentclass[aps,prd,onecolumn,nofootinbib,superscriptaddress,preprintnumbers,notitlepage,11pt]{revtex4-1}
\usepackage{amsmath,amssymb,amsthm,booktabs,graphicx,hyperref}
\usepackage[usenames,dvipsnames]{xcolor}
\usepackage[normalem]{ulem} 
\hypersetup{colorlinks,citecolor=[rgb]{0,0.4,0},linkcolor=[rgb]{0,0.4,0},
            urlcolor=[rgb]{0,0.4,0}}
\allowdisplaybreaks[4]

\def\be{\begin{equation}}
\def\ee{\end{equation}}

\newcommand{\Eq}[1]{Eq.~\eqref{#1}}
\newcommand{\zt}{\zeta}
\newcommand{\zb}{\bar z}
\newcommand{\cC}{\mathcal{C}}
\newcommand{\cF}{\mathcal{F}}
\newcommand{\cL}{\mathcal{L}}
\newcommand{\cM}{\mathcal{M}}
\newcommand{\cN}{\mathcal{N}}
\newcommand{\cO}{\mathcal{O}}

\newcommand{\Fred}{F_{\text{red}_1}}
\newcommand{\FK}{F_{K_1}}
\newcommand{\FRt}{F_{R_2}}

\newcommand{\nn}{\nonumber}

\begin{document}

\preprint{MIT-CTP 6128}

\title{The hidden sunrise in the energy-energy correlator}

\author{Matthew D. Schwartz}
\email{schwartz@g.harvard.edu}
\affiliation{Department of Physics, Harvard University, Cambridge, MA 02138, USA}
\affiliation{The NSF AI Institute for Artificial Intelligence and Fundamental Interactions}

\author{Xiaoyuan Zhang}
\email{xyz2@mit.edu}
\affiliation{The NSF AI Institute for Artificial Intelligence and Fundamental Interactions}
\affiliation{Center for Theoretical Physics -- a Leinweber Institute, Massachusetts Institute of Technology, Cambridge, MA 02139, USA}

\begin{abstract}
The energy-energy correlator (EEC) is one of a handful of collider observables that can be computed analytically to high orders. 
As an energy-weighted cross-section, it exposes features of quantum field theory that scattering amplitudes do not, and it can be compared directly to data.
As with scattering amplitudes, the analytic expressions require special functions beyond polylogarithms. In Ref.~\cite{Henn:2019gkr}, the EEC was computed to next-to-next-to-leading order (NNLO) for $\mathcal{N}=4$ super Yang-Mills (SYM) theory, and the result is expressed in terms of both harmonic polylogarithms (HPLs) and one two-fold integral, which contains elliptic curves.
In this work, we report a complete result including the remaining elliptic sector, which is closely related to the sunrise Feynman integrals. 
We find the $j$-invariants of the EEC and the sunrise
agree identically under a M\"obius map, and thus the elliptic sector in the EEC lives on the $\Gamma_1(6)$ modular curve of the sunrise integral. 
We then express the answer in terms of iterated Eisenstein integrals, which can be evaluated to high precision quickly. 
The analytic form also allows a first study of the EEC Landau bootstrap, and the understanding of its function space in $\mathcal{N}=4$ SYM offers a concrete handle on the elliptic sector of the EEC in QCD.
Many of the technical results in this paper were completed with AI under human supervision.

\end{abstract}

\date{October 7, 2026}
\maketitle
\setcounter{tocdepth}{1}
\newpage
\tableofcontents
\newpage
\section{Introduction}\label{sec:intro}

The energy-energy correlator (EEC) measures the energy deposited in two calorimeter cells at an angular separation $\chi$ and probes correlations in energy flow in a scattering event.
Proposed in the 1970s~\cite{Basham:1978bw,Basham:1978zq,Basham:1979gh,Basham:1977iq}, the EEC has been used as an event-shape observable at colliders and to study different aspects of quantum chromodynamics (QCD). The fact that the EEC is infrared finite, jet-algorithm free, and soft-divergence suppressed makes it one of the simplest observables in perturbative QCD. Such simplicity allows one to study its analytic structure and geometry, and improve our understanding of quantum field theories.

In perturbation theory, QCD and $\cN=4$ super Yang-Mills (SYM) theory share many common features, and the latter is often used as a simpler model for testing techniques. 
The EEC has been computed to next-to-next-to-leading order (NNLO) in $\cN=4$ SYM~\cite{Belitsky:2013ofa,Henn:2019gkr}, and analytically to NLO in both $e^+e^-$ annihilation and hadronic Higgs decays~\cite{Dixon:2018qgp,Luo:2019nig,Gao:2020vyx}. The availability of fixed-order data allows one to study the singular limits of the EEC, namely the collinear limit ($\chi\to 0$) and the back-to-back limit ($\chi\to\pi$), and provides the ingredients for large logarithmic resummation in effective field theories. Specifically, the EEC has been resummed to next-to-next-to-leading logarithmic (NNLL) order in the collinear limit~\cite{Dixon:2019uzg,Kologlu:2019mfz,Korchemsky:2019nzm} and N${}^4$LL in the back-to-back limit~\cite{COLLINS1981381,ELLIS198499,deFlorian:2004mp,Tulipant:2017ybb,Moult:2018jzp,Moult:2019vou,Ebert:2020sfi,Moult:2022xzt,Duhr:2022yyp}.
These results have been applied to QCD studies, for example, the measurement of the strong coupling constant~\cite{Tulipant:2017ybb}. 
Interest in the EEC has grown with the development of multi-point energy correlators. In Refs.~\cite{Chen:2019bpb,Chen:2020vvp}, the EEC is generalized to $N$-point energy correlators, where one weights the differential cross-section by energy deposited in more than two detectors and measures the angles between any pair of them. 
The differential three-point correlator has been calculated at LO in $\cN=4$ SYM~\cite{Yan:2022cye} and QCD~\cite{Yang:2022tgm,Yang:2024gcn}. The four-point correlator has been computed to LO only in $\cN=4$ SYM and in the collinear limit~\cite{Chicherin:2024ifn}, while the mathematical structure for generic angles is explored in Ref.~\cite{Ma:2025qtx}.

Feynman integrals generally give rise to iterated integrals such as polylogarithms.
For the EEC up to NLO and the three-point correlator at LO, which are of order $\mathcal{O}(g_\text{YM}^4)$ in the coupling $g_\text{YM}$, the results can be expressed in terms of generalized polylogarithms up to weight three. 
Starting from the EEC at NNLO or the four-point correlator at LO, polylogarithms run out and elliptic functions appear.
Similar behavior is observed in scattering amplitudes and loop integrals. The first elliptic Feynman integral is the two-loop equal-mass sunrise diagram~\cite{Sabry:1962,Broadhurst:1993mw,Laporta:2004rb,Bloch:2013tra,Adams:2013nia}.
For higher-loop integrals or observables like $N\ge 4$-point energy correlators, richer geometries appear and the analytic results require special functions beyond elliptic ones.
Unlike polylogarithms, whose function space can be well organized using the symbol~\cite{Goncharov:2010jf} and transcendental weights, and which can be evaluated fast using \textsc{GiNaC}~\cite{Bauer:2000cp} or other packages~\cite{Naterop:2019xaf,Wang:2021imw}, elliptic functions and beyond still pose open problems. On one hand, there is not yet a general algorithmic method to calculate Feynman integrals with elliptic curves and systematically express the result in terms of universal basis functions, while for polylogarithmic integrals, one can derive canonical differential equations~\cite{Kotikov:1990kg,Remiddi:1997ny,Gehrmann:1999as,Henn:2013pwa} and write the results in terms of generalized polylogarithms (GPLs)~\cite{Goncharov:1998kja}. 
On the other hand, it is difficult to numerically evaluate elliptic functions and beyond to high precision, though iterated integrals of modular forms and elliptic multiple polylogarithms have been implemented in \textsc{GiNaC}~\cite{Walden:2020odh,Duhr:2026ell}.

Our understanding of Feynman integrals has fostered the development of perturbative bootstrap programs. 
These programs fall into two categories: the amplitude bootstrap and the Landau bootstrap.
The amplitude bootstrap~\cite{Dixon:2016nkn,Caron-Huot:2016owq,Caron-Huot:2019vjl,Dixon:2020bbt,Dixon:2022rse,Basso:2024hlx,Cai:2024znx,Cai:2025atc,Guo:2021bym,Dersy:2022bym} uses the properties of scattering amplitudes like unitarity and supersymmetry to constrain the symbol basis and determine the final expression.
The Landau bootstrap (also called the Feynman integral bootstrap) \cite{Chicherin:2017dob,Caron-Huot:2018dsv,Henn:2018cdp,He:2021non,He:2021eec} focuses on the leading singularities of Feynman integrals and constructs the symbol basis from their analytic properties. 
In both cases, one writes down all possible integrable symbols up to certain weights, lifts them to polylogarithmic functions and uses the constraints to determine the coefficients. Very recently, we also developed an analytic regression method~\cite{Barrera:2025anr}, which can help bypass some consistency constraints. 
These bootstrap methods have also been applied to individual elliptic Feynman integrals~\cite{Morales:2022csr}, but so far not to a collider observable with an elliptic sector. The NNLO EEC, now known in closed form, is a natural testing ground for such a program. 
Many of the steps of such a bootstrap, from building the function space and imposing the constraints to the numerical regression, can be automated and are well suited to agentic AI. Many of the routines have been consolidated in the BootLoops repository~\cite{Schwartz:2026afx}, which was used heavily for the computations in this paper.

In this work, we study the remaining two-fold integral in the NNLO EEC for $\cN=4$ SYM theory. Guided by Galois symmetry~\cite{Brown:2015fyf}, we can perform a partial fraction decomposition on the integrand and separate the elliptic sector from the polylogarithmic sector. The polylogarithmic sector can be integrated directly and the result consists of harmonic polylogarithms (HPLs)~\cite{Remiddi:1999ew}, which are then combined with the existing HPL contribution in Ref.~\cite{Henn:2019gkr}.
For the elliptic sector, we identify a single elliptic curve, which turns out to be the sunrise curve, and write the analytic expression in terms of iterated integrals of $\Gamma_1(6)$ Eisenstein forms~\cite{Adams:2017ejb,Adams:2018yfj,Bogner:2017vim,Broedel:2017kkb}. The iterated Eisenstein integrals are polynomials in $\log q$ whose coefficients are $q$-series with rational coefficients, and can be evaluated to high precision. We implement their evaluation as truncated $q$-series, the method of Refs.~\cite{Adams:2017ejb,Walden:2020odh}, and obtain the NNLO distribution to high precision within seconds. The analytic form also lets us take a first look at a bootstrap of the EEC, in retrospect: with the answer known, we ask how much of it the physical constraints would have fixed and how much must come from numerical data. At LO and NLO~\cite{Belitsky:2013ofa,Henn:2019gkr}, the function space is polylogarithmic and well organized using the EEC alphabet: one writes down all potential symbols with entries from the alphabet and lifts them to polylogarithmic functions using the integrability condition. 
At NNLO, the polylogarithmic sector inherits an increase in transcendental weight, and the same idea generalizes to the elliptic sector, where the elliptic letters can be identified from the underlying curve and the tower of Eisenstein integrals can be generated in the same way. In both cases, we multiply the function basis by unknown rational coefficients. The physical constraints fix a fraction of them; the rest we determine by analytic regression from high-precision samples, generated here from the analytic result itself.

The paper is organized as follows. In Sec.~\ref{sec:eecnnlo}, we review the definition of the EEC and the NNLO result of Ref.~\cite{Henn:2019gkr}, and perform the partial fraction decomposition that separates the remaining two-fold integral into a reducible piece $\Fred$ and two irreducible pieces $\FK$ and $\FRt$. 
In Sec.~\ref{sec:sunrise},  we show that the two irreducible pieces live on the same elliptic curve, which is the curve of the equal-mass sunrise integral, derive the differential system they satisfy, and show that their difference $\Delta$ is polylogarithmic.
In Sec.~\ref{sec:analytic_result}, we provide the analytic expression of the EEC at NNLO. We express the polylogarithmic part in terms of HPLs, and write the elliptic part as iterated Eisenstein integrals. 
In Sec.~\ref{sec:bootstrap}, we analyze the alphabet and function space of the EEC and initiate the perturbative EEC bootstrap program.
We conclude in Sec.~\ref{sec:outlook}. 

\section{EEC at NNLO and partial fractions}\label{sec:eecnnlo}

In this section, we review the definition of the EEC and summarize the NNLO calculation in Ref.~\cite{Henn:2019gkr}. We introduce the energy flow operator in terms of the stress-energy tensor $T_{\mu\nu}$ along the direction $\vec{n}$~\cite{Sterman:1975xv,Sveshnikov:1995vi,Tkachov:1995kk,Korchemsky:1999kt,Hofman:2008ar},
\begin{equation}
\mathcal{E}(\vec n) \;=\; \int_{-\infty}^{\infty}\!d\tau \,\lim_{r\to\infty} r^2\, n^i\, T_{0i}(t=\tau+r,\,r\vec n),
\label{eq:detector}
\end{equation}
and define the EEC as a correlation function~\cite{Belitsky:2013xxa,Belitsky:2013bja},
\begin{equation}
\label{eq:eec_def_operator}
\frac{d\sigma}{d\zt}= \int d\Omega_{\vec n_1} d\Omega_{\vec n_2}\, \delta\left(\zt-\frac{1-\vec n_1\cdot \vec n_2}{2}\right)
\times \frac{\int d^4 x e^{iq \cdot x} \langle 0 |\cO^\dagger (x){\cal E}(\vec n_1){\cal E}(\vec n_2)\cO(0)|0\rangle }{Q^2\int d^4 x e^{iq\cdot x} \langle 0| \cO^\dagger (x) \cO(0)|0\rangle}\,,
\end{equation}
where $\cO(x)$ is the source operator in the underlying theory and we introduce the angular notation $\zt=(1-\cos\chi)/2$. 
In this variable, the collinear limit ($\chi\to 0$) corresponds to $\zt\to 0$ and the back-to-back limit ($\chi\to\pi$) gives $\zt\to 1$.
In perturbation theory, the EEC can also be defined as an energy-weighted cross-section, 
\begin{equation}
\label{eq:eec_def_xs}
\frac{d\sigma}{d\zt}=\frac{1}{\sigma_\text{tot}} \sum_{ij}\int d\text{LIPS}\, |\cM|^2 \,\frac{E_iE_j}{Q^2}\, \delta\left(\zt-\frac{1-\cos\theta_{ij}}{2}\right)\,,
\end{equation}
where $d\text{LIPS}$ is the Lorentz-invariant phase space, $|\cM|^2$ is the squared matrix element and $i,j$ run over all final-state particles.
$\sigma_\text{tot}$ is the total cross-section.
For practical calculations, it is usually more convenient to use the operator definition in Eq.~\eqref{eq:eec_def_operator} for $\cN=4$ SYM theory, where more conformal field theory tools have been developed, and the cross-section definition in Eq.~\eqref{eq:eec_def_xs} for QCD processes.

In Ref.~\cite{Henn:2019gkr}, the EEC in $\cN=4$ SYM is computed to NNLO, and the result is reduced to HPLs (for a review of the conventions, see App.~\ref{app:defs}) of weight up to five plus a single unevaluated piece. Following the notation in Ref.~\cite{Henn:2019gkr}, the normalized EEC is written as
\begin{equation}
    F(\zt)=2\zt^2 (1-\zt)\times\frac{d\sigma}{d\zt}=a F_\text{LO}(\zt)+a^2 F_\text{NLO}(\zt)+a^3 F_\text{NNLO}(\zt)+\cdots,\quad a\equiv \frac{N_c\, g_\text{YM}^2}{4\pi^2}\,,
\end{equation}
with $a$ the coupling in $\cN=4$ SYM.  The NNLO result is then expressed as
\begin{equation}
F_\text{NNLO}(\zt) \;=\; F_\text{HPL}(\zt) \;+\; F_{R}(\zt)\,,
\label{eq:FNNLOsplit}
\end{equation}
where $F_\text{HPL}(\zt)$ is an expression of weight-${\le}5$ HPLs with alphabet $\{\zt,1-\zt,(1-\sqrt{\zt})/(1+\sqrt{\zt})\}$, and $F_{R}(\zt)$ is a finite two-fold integral. Explicitly, the latter term can be written as
\begin{equation}
F_{R}(\zt) = \int_0^1\!\mathrm{d}\zb\int_0^{\zb}\!\mathrm{d}t\;
   \frac{\zt-1}{t(\zt-\zb)+(1-\zt)\zb}\,\bigl[R_1 P_1 + R_2 P_2\bigr]\,,
\label{eq:Izt}
\end{equation}
where $R_{1,2}$ are the rational coefficients,
\begin{equation}
    R_1(z,\bar{z}) = \frac{z\bar{z}}{1 - z - \bar{z}}\,, \qquad
    R_2(z,\bar{z}) = \frac{z^2\bar{z}}{(1-z)^2(1-z\bar{z})}\,,
\end{equation}
and $P_{1,2}$ are also HPL expressions,
\begin{align}
P_1(z,\bar{z}) = & -\frac{1}{8}\,H_0(-z)^2\,H_1(z)
+\frac{1}{8}\,H_0(-z)^2\,H_1(1-\bar z) +\frac{1}{4}\,H_0(-z)\,H_1(z)\,H_1(\bar z)\nn\\
& -\frac{1}{4}\,H_0(-z)\,H_1(1-\bar z)\,H_1(\bar z) +\frac{1}{4}\,H_1(z)\,H_1(1-\bar z)\,H_1(\bar z) +\frac{1}{2}\,H_0(-z)\,H_2(z) \nn\\
&
-\frac{1}{4}\,H_1(\bar z)\,H_2(z) -\frac{1}{2}\,H_0(-z)\,H_2(1-\bar z)
+\frac{1}{4}\,H_1(z)\,H_2(1-\bar z)
+\frac{1}{4}\,H_1(\bar z)\,H_2(1-\bar z) \nn\\
&-\frac{3}{4}\,H_3(z)
+\frac{3}{4}\,H_3(1-\bar z) -\frac{1}{4}\,H_1(1-\bar z)\,H_{1,1}(z)
-\frac{1}{4}\,H_1(\bar z)\,H_{1,1}(z) \nn\\
&+\frac{1}{4}\,H_1(z)\,H_{1,1}(1-\bar z)
-\frac{1}{4}\,H_1(\bar z)\,H_{1,1}(1-\bar z)-\frac{1}{4}\,H_{1,2}(1-\bar z) -\frac{1}{4}\,H_{2,1}(z) \nn\\
&-\frac{1}{4}\,H_{2,1}(1-\bar z)+\frac{1}{4}\,H_{1,1,1}(z)
-\frac{1}{4}\,H_{1,1,1}(1-\bar z)\,,\nn \\
P_2(z,\bar{z}) =  & \frac{1}{6}\pi^2\,H_0(-z)
+\frac{1}{12}\pi^2\,H_1(z)
+\frac{1}{4}\,H_0(-z)^2\,H_1(z)-\frac{1}{12}\pi^2\,H_1(1-\bar z) +\frac{1}{4}\,H_0(-z)^2\,H_1(1-\bar z) \nn\\
& +\frac{1}{2}\,H_0(-z)\,H_1(z)\,H_1(1-\bar z) -\frac{1}{12}\pi^2\,H_1(\bar z)
-\frac{1}{4}\,H_0(-z)^2\,H_1(\bar z) \nn\\
&+\frac{1}{2}\,H_0(-z)\,H_1(1-\bar z)\,H_1(\bar z) +H_1(z)\,H_1(1-\bar z)\,H_1(\bar z)
-H_0(-z)\,H_2(z)
-H_1(\bar z)\,H_2(z) \nn\\
&+H_0(-z)\,H_2(1-\bar z)
+H_1(z)\,H_2(1-\bar z)
+H_1(\bar z)\,H_2(1-\bar z) +2\,H_3(z)+2\,H_3(1-\bar z)\nn\\
&-\frac{1}{2}\,H_1(z)\,H_{1,1}(1-\bar z) +\frac{1}{2}\,H_1(\bar z)\,H_{1,1}(1-\bar z)
+H_{1,2}(z)
-\frac{1}{2}\,H_{1,1,1}(1-\bar z)\,.
\end{align}
Since $P_{1,2}$ carry weight three and each of the two integrations raises the weight by one, $F_{R}(\zt)$ also reaches weight five.
Note that we have traded $z$ for the integration variable $t$,
\begin{equation}
z = \frac{\zeta\, t\, (t - \bar{z})}{t(\zeta - \bar{z}) + (1-\zeta)\bar{z}}\,,
\label{eq:zoft}
\end{equation}
so that $z\le 0$ throughout the integration region $0<t<\bar z<1$.

From now on, we will focus on the remaining two-fold integral $F_{R}(\zt)$. 
To separate the polylogarithmic part from the elliptic sector in $F_{R}(\zt)$, we first perform a partial fraction decomposition of the rational coefficients. For convenience, we further rescale the integration variable to $x=t/\zb\in[0,1]$, which maps the triangle $0<t<\zb<1$ onto the unit square, so that the inner integration boundary does not depend on the outer variable $\zb$. With the new variable, we can define $D \;=\; (\zt-\zb)\,x+(1-\zt)$ and
Eq.~\eqref{eq:zoft} becomes $z=\zt\zb\,x(x-1)/D$. The remaining integral $F_{R}(\zt)$ involves $D$ together with three quadratics in $x$,
\begin{equation}
   D_j(x)=a_j\,x^2+(\zt-\zb+c_j)\,x+e_j\,,\qquad j=0,1,2\,,
\end{equation}
and the coefficients
\begin{equation}
(a_j,c_j,e_j)\;=\;
\begin{cases}
 \big(-\zt\zb,\;\;\zt\zb,\;\;1-\zt\big)\,, & j=0\,,\\[2pt]
 \big(-\zt\zb,\;\;\zb^{2},\;\;(1-\zt)(1-\zb)\big)\,, & j=1\,,\\[2pt]
 \big(-\zt\zb^{2},\;\;\zt\zb^{2},\;\;1-\zt\big)\,, & j=2\,.
\end{cases}
\label{eq:ace}
\end{equation}
Note that these quadratics are related to the original $z$ and $\zb$ variables via $1-z=D_0/D$, $1-z-\zb=D_1/D$ and $1-z\zb=D_2/D$. The discriminant of $D_0$ is a perfect square, so $D_0$ can be factorized as $D_0=\big(1-\zb x\big)\big(\zt x+1-\zt\big)$. The discriminants of $D_1$ and $D_2$ are not squares and they are the source of the elliptic sector. 

With this notation, the $F_{R}(\zt)$ integral can be rewritten as
\begin{equation}
   F_{R}(\zt)=\int_0^1 d\zb\int_0^1 dx\;\big[\ker_1 P_1 + \ker_2 P_2\big]\,,
\end{equation}
where the kernels $\ker_{1,2}$ are rational functions of $x$ and $\zb$,
\begin{equation}
\mathrm{ker}_1=-\frac{\zt(1-\zt)\,\zb^{2}\;x(x-1)}{D\,D_1}\,,\qquad
\mathrm{ker}_2=-\frac{\zt^{2}(1-\zt)\,\zb^{3}\;x^{2}(x-1)^{2}}{D_0^{\,2}\,D_2}\,.
\end{equation}
We then perform the partial fractioning in $x$ for each channel, which separates the pole that carries the square root from the poles that do not. Explicitly we have $\ker_{1,2}=K_{1,2}+\mathrm{red}_{1,2}$, with $x_D=(\zt-1)/(\zt-\zb)$, and 
\begin{align}
   K_1&=\frac{(\zt-1)\,\zb(1-\zb)}{D_1},\quad \mathrm{red}_1=-\frac{\zb\,x_D}{x-x_D}\,,\nn\\
   K_2&=\frac{(\zt-1)\,\zb}{(1-\zb)^{2}\,D_2}\,,\quad \mathrm{red}_2=\frac{\zb\,(\zt-1)\,\big[\,\zt\zb(2-\zb)\,x^{2}
  +\big(\zt\zb^{2}-2\zt\zb+\zb-\zt\big)\,x+\zt-1\,\big]}
  {(1-\zb)^{2}\,\big(1-\zb x\big)^{2}\big(\zt x+1-\zt\big)^{2}}\,.
\end{align}
Here $K_1$ and $K_2$ are irreducible kernels and they were already introduced in Ref.~\cite{Henn:2019gkr}, up to an additional factor $\zb$ due to the change of variable from $t$ to $x$. The remainders $\mathrm{red}_1$ and $\mathrm{red}_2$ are rational in $x$ with rational letters, so they will only give rise to polylogarithms. 

Although the partial fractioning is exact, it does not make both channels separately integrable, and the difference between them dictates how the result must be organized. The obstruction sits at a corner of the integration region.  As the outer variable approaches $\zb=1$, the root $x=1/\zb$ of $D_0$ moves onto the endpoint $x=1$ of the inner integration, and there the two remaining quadratics vanish to different orders,
\begin{equation}
    D_1\big|_{x=1/\zb} = -\zt\,(1-\zb)\,,\qquad
    D_2\big|_{x=1/\zb} = \zt\,(1-\zb)^2/\zb\,,
\label{eq:corner}
\end{equation}
a simple pole and a double pole. 
In the first channel $K_1$ carries the factor $\zb(1-\zb)$, which vanishes as $\zb\to1$, and $\mathrm{red}_1$ has only a simple pole at $x_D$, which reaches the endpoint $x=1$ as $\zb\rightarrow 1$, but produce only logarithms of $1-\zb$.
So both pieces are integrable over $\zb$ on their own.  
In the second channel $K_2$ carries $\zb/(1-\zb)^2$ and $\mathrm{red}_2$ carries the same $(1-\zb)^{-2}$ in its denominator, so each diverges
like $(1-\zb)^{-2}\log^{2}(1-\zb)$ as the endpoint is approached. The divergence will cancel in the sum of the two pieces, but neither piece is integrable by itself. There are two possible ways to resolve this problem. We can introduce a regulator for the $\zb\to 1$ limit and then take the regulator to zero after integration. Alternatively, we can introduce a subtraction term that cancels the divergence in $\mathrm{red}_2$ and add it to $K_2$. However, either approach will make the calculation more complicated. Thus, we will keep $\ker_2$ as a whole and write
\begin{equation}
    F_{R}(\zt)=\Fred(\zt)+\FK(\zt)+\FRt(\zt)\,,
\label{eq:three}
\end{equation}
with three finite pieces defined as
\begin{align}
    \Fred(\zt)&=\int_0^1 d\zb\int_0^1 dx\; \mathrm{red}_1\,P_1(z,\zb)\,,\qquad
    \FK(\zt)  =\int_0^1 d\zb\int_0^1 dx\; K_1\,P_1(z,\zb)\,,\nn\\
    \FRt(\zt) & =\int_0^1 d\zb\int_0^1 dx\; \big[\mathrm{red}_2+K_2\big]\,P_2(z,\zb)\,.
\label{eq:threedef}
\end{align}
The decomposition also respects the Galois symmetry. $\FK$ and $\FRt$ each contain one square root, of the discriminant of $D_1$ and of
$D_2$ respectively, and $\Fred$ contains neither. Flipping the sign of either square root leaves the decomposition unchanged and leaves the integral $F_{R}(\zt)$ invariant, as it must be for a single-valued function on the physical region.
In the next section, we will show that the two square roots in $\FK$ and $\FRt$ are not independent, and the difference
\begin{equation}
    \Delta(\zt)\;\equiv\;\FRt(\zt)-\FK(\zt)
\label{eq:Deltadef}
\end{equation}
is polylogarithmic. Eventually this will allow us to write the final result as
\begin{equation}
      F_{R}(\zt)\;=\;2\,\FK(\zt)\;+\;P(\zt)\,,\qquad P(\zt)\;\equiv\;\Fred(\zt)+\Delta(\zt)\,,
\end{equation}
and our goal in this paper is to obtain the closed analytic form for both the elliptic piece $\FK(\zt)$ and the polylogarithmic piece $P(\zt)$.

\section{The elliptic curve and the sunrise}\label{sec:sunrise}

In this section, we will study the two elliptic curves described by the two irreducible kernels $K_1$ and $K_2$. In particular, we will show that they are the same curve and that it is the sunrise curve. Then we will be able to further extract the polylogarithmic contribution from $\FRt(\zt)$.

\subsection{Elliptic curve in EEC}

After the partial fraction decomposition in the inner variable $x$, the two irreducible kernels carry the square roots of the discriminants of $D_1$ and $D_2$ in that variable. These discriminants are the quartic polynomials
\begin{align}
    Q_1(\zb) &= \big(\zb^2-\zb+2\zt^2-\zt\big)^2+4\zt^3(1-\zt)\nn\\
             &= \zb^4-2\zb^3+(1-2\zt+4\zt^2)\,\zb^2+2\zt(1-2\zt)\,\zb+\zt^2\,,\nn\\
    Q_2(\zb) &= \zt^2\zb^4-2\zt\,\zb^3+(1+4\zt-2\zt^2)\,\zb^2-2\zt\,\zb+\zt^2\,.
\label{eq:Qexpl}
\end{align}
Both quartics are positive for $0<\zt<1$ along the real $\zb$ axis. For $Q_1$ this follows directly from the first line of Eq.~\eqref{eq:Qexpl}, a square plus a positive number. For $Q_2$, one has $Q_2(0)=\zt^2>0$, while for $\zb\ne0$ the variable $v=\zb+1/\zb$ gives the palindromic form $Q_2/\zb^2=(\zt v-1)^2+4\zt(1-\zt)$. The outer integration therefore never approaches a branch point in the physical region, and the integrands of Eq.~\eqref{eq:threedef} are real on the whole domain, with all their singularities on the boundary of the unit square. Moreover, the two quartics have the same discriminant, $\operatorname{disc}_{\zb}Q_1=\operatorname{disc}_{\zb}Q_2=256\,\zt^6(1-\zt)^2(1+8\zt)$, and hence the same singular fibers at $\zt\in\{0,1,-1/8,\infty\}$.

It turns out that the two quartics are related by a M\"obius transformation $w=\zb/(\zb-1)$, which is an involution of the $\zb$ line that fixes $\zb=0$ and sends the endpoint $\zb=1$ to $w=\infty$. Explicitly we find
\begin{equation}
   Q_2(\zb)=(\zb-1)^4\,Q_1(w)\,,\qquad
   \frac{d\zb}{\sqrt{Q_2(\zb)}}=-\frac{dw}{\sqrt{Q_1(w)}}\,.
\label{eq:mobius}
\end{equation}
Then the measure in $\FRt$, $\zb\,d\zb/[(1-\zb)^2\sqrt{Q_2(\zb)}]$, becomes $w(1-w)\,dw/\sqrt{Q_1(w)}$ in the variable $w$, which is the same as the measure of $\FK$. This means both $\FRt$ and $\FK$ are integrals over the same elliptic curve $\cC_\zt: y^2=Q_1(\zb)$, with the same weight, but over different contours. The interval $\zb\in[0,1]$ maps to $w\in(-\infty,0]$. 
The equivalence of the two curves is also manifest from their $j$-invariants, which agree identically in $\zt$,
\begin{equation}
   j(\zt)\;=\;\frac{(1+2\zt)^3\big(8\zt^3-12\zt^2+6\zt+1\big)^3}{\zt^6\,(1-\zt)^2\,(1+8\zt)}\,.
\label{eq:jinv}
\end{equation}

As observed in Ref.~\cite{Henn:2019gkr}, the equal-mass two-loop sunrise integral is also described by the same elliptic curve $\cC_\zt$. Here we make the identification more precise. The sunrise is the first Feynman integral known to require elliptic functions for its evaluation, and it has been studied extensively in the literature. Explicitly, the sunrise integral in $d=2$ dimensions is defined as~\cite{Laporta:2004rb,Adams:2013nia,Bloch:2013tra}
\begin{equation}
   \raisebox{-0.42\height}{\includegraphics[width=1.2in]{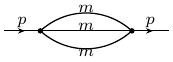}}
   \;\;=\;\;\int\!\frac{d^2\ell_1\,d^2\ell_2}{\pi^{2}}\;
   \frac{1}{(\ell_1^2-m^2)\,(\ell_2^2-m^2)\,\big((p-\ell_1-\ell_2)^2-m^2\big)}\,,
\label{eq:sungraph}
\end{equation}
and we write $p^2=S$ and $m^2\to m^2-i0$ for the $i\epsilon$ prescription. In $d=2$ dimensions, the integral is finite and depends only on the ratio $S/m^2$. With the $\mathcal{U}\mathcal{F}$ polynomials $\mathcal{U}=x_1x_2+x_2x_3+x_3x_1$ and $\cF=m^2\,(x_1{+}x_2{+}x_3)\,\mathcal{U}-S\,x_1x_2x_3$, the Feynman-parametric representation of the sunrise integral takes the form
\begin{equation}
   \raisebox{-0.42\height}{\includegraphics[width=1.2in]{figs/sunrise.pdf}}
   \;=\;\int_0^\infty\! d^3x\;\;\frac{\delta\big(1-x_1-x_2-x_3\big)}{\cF}\,.
\label{eq:sunpar}
\end{equation}
The sunrise curve is read off from $\cF$ alone.  With the Cheng-Wu theorem~\cite{Cheng:1987ga}, we can set $x_3=1$. We can also set $m=1$ for convenience and recover the mass dependence by dimensional analysis in the end. Then we find that $\cF$ is a quadratic in $x_1$,
\begin{equation}
   \cF\big|_{x_3=1,\,m=1}=(x_2{+}1)\,x_1^{2}
   +\big(x_2^{2}+(3-S)\,x_2+1\big)\,x_1+x_2\,(x_2{+}1)\,,
\label{eq:Fquad}
\end{equation}
and completing the square in $x_1$ brings in the square root of its discriminant,
\begin{align}
   \mathrm{disc}_{x_1}\cF&=\big(x_2^{2}+(3-S)\,x_2+1\big)^{2}-4\,x_2\,(x_2{+}1)^{2}
   \;=\;Q_{\rm sun}(x_2;S)\,,\nn\\[2pt]
   \Rightarrow Q_{\rm sun}(u;S)&=u^{4}-2(S-1)\,u^{3}+(S^{2}-6S+3)\,u^{2}-2(S-1)\,u+1\,.
\label{eq:Qsundef}
\end{align}
The sunrise curve is the double cover $y^{2}=Q_{\rm sun}(u;S)$ of the $u$-line branched over the four roots of this quartic. Its $j$-invariant is
\begin{equation}
   j_{\rm sun}(S)=\frac{(S-3)^{3}\big(S^{3}-9S^{2}+3S-3\big)^{3}}{S^{2}\,(S-1)^{3}\,(S-9)}\,,
\label{eq:jsun}
\end{equation}
whose poles sit at the four singular fibers of the family, $S/m^2=0,1,9,\infty$: the normal threshold $S=9m^2$, the pseudo-threshold $S=m^2$, and the two ends. The three finite ones are exactly the zeros of $\mathrm{disc}_u\,Q_{\rm sun}=256\,S^{2}(S-1)^{3}(S-9)$.

To compare the sunrise curve with the EEC curve, we go back to projective coordinates and choose the following parameterization of the Feynman parameters and kinematic variable $S$,
\begin{equation}
   (x_1:x_2:x_3)=\big(\zb\,:\;\zb x-1\,:\;1-\zb\big)\,,\qquad
   S=\frac{\zt-1}{\zt}\,,
\label{eq:sunmap}
\end{equation}
under which one finds the exact identity
\begin{equation}
   \zb\,D_1(x)\;=\;-\zt\;\cF(x_1,x_2,x_3)\,.
\label{eq:symanzik}
\end{equation}
The prefactor $\zb$ on the left of Eq.~\eqref{eq:symanzik} is invisible to the geometry, since $\mathrm{disc}_x(\zb D_1)=\zb^{2}Q_1$ differs from $Q_1$ by a square.
At the level of the quartics themselves the same identification reads
\begin{equation}
   Q_1(\zb)=\zt^{2}(1-\zb)^{4}\,Q_{\rm sun}\!\Big(\frac{\zb}{1-\zb};S\Big)\,,\qquad
   Q_2(\zb)=\zt^{2}\,Q_{\rm sun}(-\zb;S)\,.
\label{eq:Qsun}
\end{equation}
In both relations the prefactor is $\zt^2$ times a perfect square, so both double covers are isomorphic to the sunrise curve: the second channel reaches it by the sign flip $u=-\zb$ alone and needs no M\"obius map, while the first needs $u=\zb/(1-\zb)$, and the two substitutions differ by the involution of Eq.~\eqref{eq:mobius}. Substituting $S=(\zt-1)/\zt$ into Eq.~\eqref{eq:jsun} returns Eq.~\eqref{eq:jinv} identically in $\zt$, and the singular fibers line up one to one: $S=\infty,0,9,1$ correspond to $\zt=0,1,-1/8,\infty$. In particular the singular point $\zt=-1/8$ of the family, which looks accidental in the EEC variable, is the normal threshold $S=9m^2$ of the sunrise.
Under Eq.~\eqref{eq:sunmap}, the physical region $0<\zt<1$ is mapped to $S=(\zt-1)/\zt<0$, namely the Euclidean region of the sunrise integral, below every threshold. This is why the EEC integral is real throughout and why nothing singular happens between the endpoints. The collinear and back-to-back limits correspond to $S\to-\infty$ and $S\to0$, respectively.

The periods of $\cC_\zt$ are annihilated by the sunrise Picard-Fuchs operator~\cite{Laporta:2004rb,MullerStach:2011ru,Adams:2013nia,Bloch:2013tra}
\begin{equation}
   L_{\rm PF}\;=\;\zt(\zt-1)(8\zt+1)\,\partial^2
   \;+\;(24\zt^2-14\zt-1)\,\partial\;+\;2(4\zt-1)\,,
\label{eq:PF}
\end{equation}
with $\partial$ the derivative with respect to $\zt$.
Its four regular singular points $\zt=0,1,-1/8,\infty$ are maximally unipotent: they are the four cusps of $\Gamma_1(6)$, the modular group of the equal-mass sunrise family~\cite{Bloch:2013tra,Adams:2013nia,Broedel:2017kkb}. The same operator annihilates the periods of both $y^2=Q_1$ and $y^2=Q_2$. Explicitly, we have
\begin{equation}
   L_{\rm PF}\big[Q_j^{-1/2}\big]=\frac{d}{d\zb}\Big[U_j\,Q_j^{-3/2}\Big]\,,\qquad j=1,2\,,
   \label{eq:GD2}
\end{equation}
with the following factors
\begin{align}
   U_1&=-\zb(\zb-1)(2\zb-1)\big[(4\zt-1)\,\zb(\zb-1)+\zt(2\zt+1)\big]\,,\nn\\
   U_2&=\zb(\zb-1)(\zb+1)\big[\zt(2\zt+1)\zb^2-(4\zt^2-2\zt+1)\zb+\zt(2\zt+1)\big]\,.
\end{align}
Since $U_j(0)=U_j(1)=0$ identically in $\zt$, the total derivative in Eq.~\eqref{eq:GD2} gives no boundary contribution on the physical contour. 
This is consistent with the two elliptic curves in the NNLO EEC being isomorphic to the sunrise curve.

\subsection{Further decomposition of $\FRt$}
\label{sec:delta}

The equivalence of the two curves does not guarantee that the two integrals $\FK(\zt)$ and $\FRt(\zt)$ will give the same elliptic functions, given that both the contours and the transcendental integrands are different. In this subsection, we will show that the difference $\Delta(\zt)=\FRt(\zt)-\FK(\zt)$ is polylogarithmic, and we will give it in closed form.

To study the transcendentality of $\Delta(\zt)$, we need to derive the differential equations satisfied by $\FK(\zt)$ and $\FRt(\zt)$. It turns out that $\FK(\zt)$ does not satisfy a second-order differential equation by itself. Differentiating it twice in $\zt$ produces one further two-fold integral, which takes the following form 
\begin{equation}
   S_0(\zt)\;=\;\int_0^1 d\zb\int_0^1 dx\;\frac{P_1(z,\zb)}{D_1(x)}\,,
\label{eq:S0def}
\end{equation}
with $D_1$ the quadratic of Eq.~\eqref{eq:ace}, $z=\zt\zb\,x(x-1)/D$, and $P_1$ as in Eq.~\eqref{eq:Izt}. Since $K_1=(\zt-1)\zb(1-\zb)/D_1$, $\FK$ and $S_0$ differ only by the rational factor $(\zt-1)\zb(1-\zb)$ in the integrand. Because $D_1>0$ on the unit square except at the two corners $(x,\zb)=(0,1)$ and $(1,1)$, where it vanishes linearly ($D_1|_{\zb=1}=\zt\,x(1-x)$), $S_0$ is also finite. 
$\FK$ and $S_0$ then satisfy an inhomogeneous second-order system with rational coefficients,
\begin{equation}
\begin{aligned}
  \cL\big[\FK\big] &\;=\; E(\zt)\;+\;\alpha\,S_0\;+\;\beta\,\FK\,,\\[2pt]
  \cL\big[S_0\big] &\;=\; \tilde{E}(\zt)\;+\;\tilde\alpha\,S_0\;+\;\tilde\beta\,\FK\,,
\end{aligned}\;
\label{eq:sys}
\end{equation}
in which the operator $\cL$ is
\begin{equation}
  \cL\;=\;\zt(\zt-1)(8\zt+1)\,\partial^2\;+\;(8\zt^2+1)\,\partial\;-\;\frac{2\zt+1}{\zt}\,,
\label{eq:cLdef}
\end{equation}
and
\begin{equation}
   \alpha=1+\zt-8\zt^2\,,\quad
   \beta=\frac{32\zt^3-18\zt^2+1}{\zt(\zt-1)}\,,\quad
   \tilde\alpha=\frac{8\zt^2-10\zt-1}{\zt}\,,\quad
   \tilde\beta=\frac{6}{\zt(1-\zt)}\,.
\label{eq:abcoef}
\end{equation}
$\cL$ is connected to the Picard-Fuchs operator in Eq.~\eqref{eq:PF} by $\cL=\zt\circ L_{\rm PF}\circ\zt^{-1}$. The fact that $\cL$ contains derivatives of at most second order and that the two equations are coupled only through rational functions indicates that the result is at most elliptic. No geometry beyond the elliptic curve, such as a K3 or Calabi-Yau period, can appear in the final result.

The second elliptic integral in $F_{R}(\zt)$ is treated in the same way, with a source and a companion of its own.  The companion is again a double integral of the type of Eq.~\eqref{eq:S0def}, with
$P_2/D_2$ in place of $P_1/D_1$, so that
\begin{equation}
   \cL\big[\FRt\big]\;=\;E_{R_2}\;-\;\alpha\int_0^1\! d\zb\int_0^1\! dx\;\frac{P_2(z,\zb)}{D_2(x)}\;+\;\beta\,\FRt\,,
\label{eq:rowR}
\end{equation}
with the same coefficients $\alpha,\beta$ as in Eq.~\eqref{eq:sys}.
Both sources $E$ and $E_{R_2}$ are harmonic polylogarithms of $\zt$ over the alphabet $\{0,1\}$ with rational
coefficients in $\zt$: $E$ is of weight at most four, and $E_{R_2}$, which carries a $\zeta_5$ and
a $\pi^2\zeta_3$ channel that $E$ does not, of weight at most five.  They are given in Eqs.~\eqref{eq:Efull} and~\eqref{eq:ER2full} of App.~\ref{app:EEp}.

It turns out that the additional integrals that show up on the RHS after differentiation are the same as in $F_{K_1}$, namely
\begin{equation}
   \int_0^1 d\zb\int_0^1 dx\;\frac{P_2(z,\zb)}{D_2(x)}\;=\;-\,S_0(\zt)\,.
\label{eq:S0eq}
\end{equation}
Then we have the entire differential system for all integrals,
\begin{equation}
\begin{aligned}
  \cL\big[\FK\big]  &\;=\; E\;+\;\alpha\,S_0\;+\;\beta\,\FK\,,\\
  \cL\big[\FRt\big] &\;=\; E_{R_2}\;+\;\alpha\,S_0\;+\;\beta\,\FRt\,,\\
  \cL\big[S_0\big] &\;=\; \tilde{E}\;+\;\tilde\alpha S_0\;+\;\tilde\beta\FK\,, \\
  \cL\big[S_0\big] &\;= \;\tilde{E}_{R_2}\;+\;\tilde\alpha S_0\;+\;\tilde\beta\FRt\,.
\end{aligned}\;
\label{eq:sysboth}
\end{equation}
In particular, the difference of the last two equations gives,
\begin{equation}
\label{eq:form_Delta}
   \tilde\beta \Delta=\tilde{E}-\tilde{E}_{R_2}\quad \Rightarrow\quad  \Delta\;=\;\frac{\zt(1-\zt)}{6}\,\big(\tilde{E}-\tilde{E}_{R_2}\big)\,.
\end{equation} 
Since both $\tilde{E}$ and $\tilde{E}_{R_2}$, given in Eqs.~\eqref{eq:Epfull} and~\eqref{eq:EpR2full} of App.~\ref{app:EEp}, are polylogarithmic, $\Delta$ is also polylogarithmic. Therefore, we arrive at the final form
\begin{equation}
      F_{R}(\zt)\;=\;2\,\FK(\zt)\;+\;P(\zt)\;=\;2\,\FK(\zt)\;+\;\Fred(\zt)+\Delta(\zt)\,.
\label{eq:central}
\end{equation}
We will present their analytic forms in the following sections.

\section{Analytic results}
\label{sec:analytic_result}

In this section, we provide the analytic expression of the EEC at NNLO. As discussed in the previous section, the remaining integral is further written as a single elliptic piece $\FK(\zt)$ and a sum of polylogarithmic pieces $P(\zt)$. We calculate the exact form of $P(\zt)$ and combine it with the existing HPL contribution in Ref.~\cite{Henn:2019gkr}. 
For the elliptic sector, we derive the differential operator and solve the system analytically. The result is then expressed in terms of iterated Eisenstein integrals. 

\subsection{The polylogarithmic sector}\label{sec:poly}

For the polylogarithmic sector, $P(\zt)=\Fred(\zt)+\Delta(\zt)$, we directly integrate $\Fred(\zt)$ using \textsc{HyperInt}~\cite{Panzer:2014caa} in Maple, and add $\Delta(\zt)$ from Eq.~\eqref{eq:form_Delta}.
The result can be expressed in terms of HPLs up to weight five. Explicitly, we find the following analytic form,
\begin{align}
   P(\zt)= &\Big[\,4 \left(\zt - 1\right) + \frac{6 \zt^{3} - 4 \zt^{2} - 5 \zt + 4}{6 \left(\zt - 1\right)}\,\pi^2 + \frac{\left(2 \zt - 3\right) \left(3 \zt^{2} - 11 \zt + 7\right)}{2 \left(\zt - 1\right)}\,\zeta_3 \nn\\
   &\qquad - \frac{\zt}{15 \left(\zt - 1\right)}\,\pi^4 - \frac{\zt \left(\zt^{2} - 2 \zt + 3\right)}{24 \left(\zt - 1\right)}\,\pi^2\zeta_3 - \frac{\zt \left(14 \zt^{2} - 28 \zt - 19\right)}{4 \left(\zt - 1\right)}\,\zeta_5\,\Big]
   \nn\\
   &+\Big[\,\frac{\left(\zt - 2\right) \left(2 \zt - 1\right)}{6 \left(\zt - 1\right)}\,\pi^2 + \frac{6 \zt^{3} - 14 \zt^{2} + 11 \zt - 2}{\zt - 1}\,\zeta_3 + \frac{\zt \left(\zt^{2} - 2 \zt - 1\right)}{60 \left(\zt - 1\right)}\,\pi^4\,\Big]\,H_{0}(\zt)
   \nn\\
   &+\Big[\,-4 \left(\zt - 1\right) - \frac{13 \zt^{2} - 21 \zt + 7}{12 \left(\zt - 1\right)}\,\pi^2 + \frac{\zt \left(12 \zt^{2} - 25 \zt + 15\right)}{2 \left(\zt - 1\right)}\,\zeta_3 \nn\\
   &\qquad + \frac{\zt \left(4 \zt^{2} - 8 \zt - 11\right)}{240 \left(\zt - 1\right)}\,\pi^4\,\Big]\,H_{1}(\zt)
   \nn\\
   &+\Big[\,\left(2 \zt^{2} - 2 \zt + 1\right) - \frac{\left(\zt - 1\right) \left(3 \zt - 1\right)}{12}\,\pi^2 - \frac{\zt \left(\zt^{2} - 2 \zt - 1\right)}{4 \left(\zt - 1\right)}\,\zeta_3\,\Big]\,H_{0,0}(\zt)
   \nn\\
   &+\Big[\,\frac{2 \zt^{3} - 10 \zt^{2} + 9 \zt - 2}{\zt - 1} + \frac{6 \zt^{3} - 14 \zt^{2} + 7 \zt - 2}{6 \left(\zt - 1\right)}\,\pi^2 + \frac{\zt \left(\zt^{2} - 2 \zt + 5\right)}{4 \left(\zt - 1\right)}\,\zeta_3\,\Big]\,H_{2}(\zt)
   \nn\\
   &+\Big[\,\left(\zt - 2\right) \left(2 \zt - 1\right) - \frac{3 \zt^{3} - 6 \zt^{2} + 3 \zt - 2}{12 \left(\zt - 1\right)}\,\pi^2 - \frac{\zt \left(\zt^{2} - 2 \zt - 5\right)}{4 \left(\zt - 1\right)}\,\zeta_3\,\Big]\,H_{1,0}(\zt)
   \nn\\
   &+\Big[\,\left(\zt - 1\right) \left(2 \zt - 3\right) + \frac{12 \zt^{3} - 27 \zt^{2} + 13 \zt - 2}{12 \left(\zt - 1\right)}\,\pi^2 + \frac{\zt \left(\zt^{2} - 2 \zt + 3\right)}{4 \left(\zt - 1\right)}\,\zeta_3\,\Big]\,H_{1,1}(\zt)
   \nn\\
   &+\Big[\,\frac{6 \zt^{3} - 8 \zt^{2} + 2 \zt + 1}{2 \left(\zt - 1\right)} - \frac{\zt \left(\zt^{2} - 2 \zt + 7\right)}{24 \left(\zt - 1\right)}\,\pi^2\,\Big]\,H_{3}(\zt)
   \nn\\
   &+\Big[\,-\frac{5 \zt^{2} - 9 \zt + 2}{2 \left(\zt - 1\right)} + \frac{\zt}{4 \left(\zt - 1\right)}\,\pi^2\,\Big]\,H_{2,0}(\zt)+\Big[\,\frac{\zt}{2} - \frac{\zt}{12 \left(\zt - 1\right)}\,\pi^2\,\Big]\,H_{1,0,0}(\zt)
   \nn\\
   &+\Big[\,\frac{6 \zt^{2} - 11 \zt + 6}{2} - \frac{\zt \left(\zt^{2} - 2 \zt + 3\right)}{24 \left(\zt - 1\right)}\,\pi^2\,\Big]\,H_{2,1}(\zt)+\frac{\zt}{6 \left(\zt - 1\right)}\,\pi^2\,H_{1,1,0}(\zt)
   \nn\\
   &+\Big[\,\frac{6 \zt^{3} - 7 \zt^{2} - \zt + 1}{2 \left(\zt - 1\right)} - \frac{\zt \left(\zt^{2} - 2 \zt + 13\right)}{24 \left(\zt - 1\right)}\,\pi^2\,\Big]\,H_{1,2}(\zt)
   \nn\\
   &+\Big[\,\left(\zt - 1\right) \left(3 \zt - 4\right) - \frac{\zt \left(\zt^{2} - 2 \zt + 7\right)}{24 \left(\zt - 1\right)}\,\pi^2\,\Big]\,H_{1,1,1}(\zt)
   \nn\\
   &-\frac{\zt \left(5 \zt^{2} - 11 \zt + 4\right)}{2 \left(\zt - 1\right)}\,H_{4}(\zt)+\frac{3 \zt^{3} - 7 \zt^{2} + 2 \zt - 1}{\zt - 1}\,H_{3,0}(\zt) +\frac{5 \zt^{2} - 7 \zt + 4}{2}\,H_{3,1}(\zt)
   \nn\\
   &-\frac{\zt^{2} - 2 \zt + 2}{2}\,H_{2,0,0}(\zt)-\frac{2 \left(2 \zt^{3} - 4 \zt^{2} - \zt + 1\right)}{\zt - 1}\,H_{2,2}(\zt)
   +\frac{\zt \left(\zt - 1\right)}{2}\,H_{2,1,0}(\zt)
   \nn\\
   &+\left(3 \zt^{2} - 5 \zt + 4\right)\,H_{2,1,1}(\zt)
   -\frac{5 \zt^{3} - 8 \zt^{2} + 1}{2 \left(\zt - 1\right)}\,H_{1,3}(\zt)+\frac{6 \zt^{3} - 15 \zt^{2} + 9 \zt - 2}{2 \left(\zt - 1\right)}\,H_{1,2,0}(\zt)
   \nn\\
   &+\frac{5 \zt^{2} - 11 \zt + 5}{2}\,H_{1,2,1}(\zt)-\frac{\left(\zt - 1\right)^{2}}{2}\,H_{1,1,0,0}(\zt)
   -\frac{8 \zt^{3} - 15 \zt^{2} + \zt + 2}{2 \left(\zt - 1\right)}\,H_{1,1,2}(\zt)
   \nn\\
   &+\frac{\left(\zt - 1\right)^{2}}{2}\,H_{1,1,1,0}(\zt)
   +3 \left(\zt - 1\right)^{2}\,H_{1,1,1,1}(\zt)+\frac{\zt \left(\zt - 1\right)}{2}\,H_{5}(\zt)
   \nn\\
   &
   +\frac{3 \zt \left(\zt^{2} - 2 \zt - 1\right)}{4 \left(\zt - 1\right)}\,H_{4,0}(\zt) +\frac{\zt \left(5 \zt^{2} - 10 \zt + 1\right)}{4 \left(\zt - 1\right)}\,H_{4,1}(\zt)
   -\frac{\zt \left(\zt^{2} - 2 \zt - 1\right)}{4 \left(\zt - 1\right)}\,H_{3,0,0}(\zt)
   \nn\\
   &+\frac{3 \zt \left(\zt - 1\right)}{4}\,H_{3,2}(\zt)
   +\frac{\zt \left(\zt - 1\right)}{2}\,H_{3,1,0}(\zt)+\frac{\zt \left(3 \zt^{2} - 6 \zt - 1\right)}{2 \left(\zt - 1\right)}\,H_{3,1,1}(\zt)
   \nn\\
   &
   +\frac{\zt \left(\zt^{2} - 2 \zt + 2\right)}{\zt - 1}\,H_{2,3}(\zt)
   +\frac{3 \zt \left(\zt - 1\right)}{4}\,H_{2,2,0}(\zt)
   +\frac{\zt \left(7 \zt^{2} - 14 \zt + 9\right)}{4 \left(\zt - 1\right)}\,H_{2,2,1}(\zt)
   \nn\\
   &-\frac{\zt \left(\zt - 1\right)}{4}\,H_{2,1,0,0}(\zt)
   +\frac{\zt \left(5 \zt^{2} - 10 \zt + 7\right)}{4 \left(\zt - 1\right)}\,H_{2,1,2}(\zt)+\frac{\zt \left(\zt - 1\right)}{2}\,H_{2,1,1,0}(\zt)
   \nn\\
   &
   +2 \zt \left(\zt - 1\right)\,H_{2,1,1,1}(\zt)
   +\frac{\zt \left(\zt^{2} - 2 \zt + 5\right)}{2 \left(\zt - 1\right)}\,H_{1,4}(\zt)
   +\frac{\zt \left(3 \zt^{2} - 6 \zt - 7\right)}{4 \left(\zt - 1\right)}\,H_{1,3,0}(\zt)
   \nn\\
   &+\frac{\zt \left(5 \zt^{2} - 10 \zt + 9\right)}{4 \left(\zt - 1\right)}\,H_{1,3,1}(\zt)
   -\frac{\zt \left(\zt^{2} - 2 \zt - 1\right)}{4 \left(\zt - 1\right)}\,H_{1,2,0,0}(\zt)
   \nn\\
   &+\frac{3 \zt \left(\zt^{2} - 2 \zt + 5\right)}{4 \left(\zt - 1\right)}\,H_{1,2,2}(\zt)
   +\frac{\zt \left(\zt - 1\right)}{2}\,H_{1,2,1,0}(\zt)+\frac{3 \zt \left(\zt - 1\right)}{2}\,H_{1,2,1,1}(\zt)
   \nn\\
   &
   +\frac{\zt \left(2 \zt^{2} - 4 \zt + 3\right)}{2 \left(\zt - 1\right)}\,H_{1,1,3}(\zt)
   +\frac{\zt \left(3 \zt^{2} - 6 \zt - 1\right)}{4 \left(\zt - 1\right)}\,H_{1,1,2,0}(\zt)
   +\frac{7 \zt \left(\zt - 1\right)}{4}\,H_{1,1,2,1}(\zt)
   \nn\\
   &-\frac{\zt \left(\zt - 1\right)}{4}\,H_{1,1,1,0,0}(\zt)
   +\frac{\zt \left(5 \zt^{2} - 10 \zt + 11\right)}{4 \left(\zt - 1\right)}\,H_{1,1,1,2}(\zt)+\frac{\zt \left(\zt - 1\right)}{2}\,H_{1,1,1,1,0}(\zt)
   \nn\\
   &
   +2 \zt \left(\zt - 1\right)\,H_{1,1,1,1,1}(\zt)\,.
\label{eq:PfullHPL}
\end{align}

The alphabet for $P(\zeta)$ is $\{\zt,1-\zt\}$, and unlike the NLO result in Ref.~\cite{Belitsky:2013ofa} or the HPL part in Ref.~\cite{Henn:2019gkr}, there is no square-root letter, $(1-\sqrt\zt)/(1+\sqrt\zt)$, appearing in the arguments. The transcendental weight is at most five and is not uniform. 
We verify the analytic expression against numerical integration at randomly chosen $\zt$ values.

\subsection{The elliptic sector}\label{sec:solving}

\subsubsection{Solving the system}

With the further decomposition described in Sec.~\ref{sec:sunrise}, the only remaining integral to compute is $\FK(\zt)$. Recall that we have the following equations from Eq.~\eqref{eq:sysboth},
\begin{equation}
  \cL[\FK]=E+\alpha S_0+\beta\FK\,,\qquad
  \cL[S_0]=\tilde{E}+\tilde\alpha S_0+\tilde\beta \FK\,,
\label{eq:sys2}
\end{equation}
where $E$ and $\tilde{E}$ are both harmonic polylogarithms, given explicitly in Eqs.~\eqref{eq:Efull} and~\eqref{eq:Epfull} of App.~\ref{app:EEp}. 
To solve the system, we can write $\FK$ as the sum of the solution to the Picard-Fuchs operator and the residual terms.
First of all, we write the following ansatz
\begin{equation}
   \FK=c_1 S_0 + c_2 S_0^{'}+\Psi,\quad \text{ with }\quad  S_0^{'}\equiv \partial S_0\,,
   \label{eq:FK1ansatz}
\end{equation}
and $c_{1,2}$ rational. Substituting into the second equation in Eq.~\eqref{eq:sys2}, we find
\begin{equation}
   \big(\cL-\tilde\alpha -\tilde\beta c_1-\tilde\beta c_2\,\partial\big)S_0=\tilde{E}+\tilde\beta \Psi\,.
\end{equation}
Demanding $\cL-\tilde\alpha-\tilde\beta c_1-\tilde\beta c_2\,\partial=L_{\rm PF}$ gives $c_1=\frac{1}{3}\,\zt(\zt-1)(8\zt-5)$ and $c_2=\frac{1}{3}\,\zt(8\zt+1)(\zt-1)^2$. Then the second equation of Eq.~\eqref{eq:sys2} becomes $L_{\rm PF} S_0=\tilde{E}+\tilde\beta \Psi$.
We also substitute the ansatz into the first equation,
\begin{multline}
   \left[\partial^2-\frac{8\zt^2-14\zt-3}{\zt(\zt-1)(8\zt+1)}\,\partial-\frac{64\zt^3-42\zt^2-3\zt-4}{\zt^2(\zt-1)^2(8\zt+1)}\right] \Psi \\= \frac{E}{\zt(\zt-1)(8\zt+1)}-\frac{32\zt^2-3\zt+1}{3\,\zt(8\zt+1)}\,\tilde{E}-\frac{\zt-1}{3}\,\partial \tilde{E}\,.
\end{multline}
The solution $\Psi$ to this equation is purely polylogarithmic. To obtain it, we write an ansatz of HPLs of $\zt$ over $\{0,1\}$ with rational function coefficients and solve it by series expansion in the $\zt\to0$ limit. The final expression reads
\begin{align}
   \Psi(\zt)&=\frac{1}{4}\Big[-8\left(\zt - 1\right) +8\left(\zt - 1\right)\,H_{1}(\zt) +2\left(\zt - 1\right)\,H_{0,0}(\zt)-4\left(2 \zt^2 - 4 \zt + 1\right)\,H_{0,1}(\zt)
   \nn\\
   & +2\left(\zt - 1\right)\,H_{1,0}(\zt) -8\left(\zt - 1\right)^2\,H_{1,1}(\zt)-\left(12 \zt^2 - 8 \zt - 1\right)\,H_{0,0,1}(\zt)
   \nn\\
   & +2\left(\zt - 1\right)\,H_{0,1,0}(\zt) -6\left(\zt - 1\right)\left(2 \zt - 1\right)\,H_{0,1,1}(\zt) -2\left(\zt - 1\right)\,H_{1,0,0}(\zt)
   \nn\\
   & -\left(12 \zt^2 - 14 \zt + 5\right)\,H_{1,0,1}(\zt) -12\left(\zt - 1\right)^2\,H_{1,1,1}(\zt)\Big]-\frac{3\zeta_3}{4}\left(4 \zt^2 - 10 \zt + 7\right)
   \nn\\
   &+\frac{\pi^2}{24}\Big[-4\left(2 \zt^2 + \zt - 2\right) -4\left(\zt - 1\right)\,H_{0}(\zt) +\left(10 \zt - 7\right)\,H_{1}(\zt)\Big]
   \nn\\
   &+\left(\zt - 1\right)\left(\zt - \frac{1}{4}\right)\Bigg\{\,\frac{2}{3}\Big[2\,H_{0,0,0,1}(\zt) -H_{0,0,1,1}(\zt)+H_{0,1,0,1}(\zt) -2\,H_{0,1,1,1}(\zt) +4H_{1,0,0,1}(\zt) 
   \nn\\
   &+H_{1,0,1,1}(\zt) +3H_{1,1,0,1}(\zt)\Big]
   +\frac{\pi^2}{6}\Big[H_{0,0}(\zt) -2H_{0,1}(\zt) +H_{1,0}(\zt) -2\,H_{1,1}(\zt)\Big]
   \nn\\
   &-\frac{2\zeta_3}{3}\Big[7\,H_{0}(\zt) +5\,H_{1}(\zt)\Big] +\frac{2\pi^4}{45}\,\Bigg\}\,.
\label{eq:Psifull}
\end{align}

The second equation, $L_{\rm PF}S_0=\tilde{E}+\tilde\beta \Psi$, now has an explicit right-hand side.
To solve this equation, we pass to the modular frame. Let $\psi_1$ and $\psi_2$ be the two solutions of $L_{\rm PF}$ at $\zt=0$, with $\psi_1$ holomorphic and normalized to $\psi_1(0)=1$ and $\psi_2=\psi_1\log\zt+\cO(\zt)$. Then we also define $\tilde\tau$ to be their ratio, $\tilde\tau=\frac{1}{2}+\frac{1}{2\pi i}\,\psi_2(\zt)/\psi_1(\zt)$, so that $2\pi i\left(\tilde\tau-\frac{1}{2}\right)=\log\zt-3\zt+\frac{21}{2}\zt^2-\dots$. 
The collinear limit $\zt\to0$ is $\tilde\tau\to i\infty$. Since $\zt$ is a Hauptmodul for $\Gamma_1(6)$, this map is invertible, and Eq.~\eqref{eq:frame} below gives the inverse in closed form~\cite{Bloch:2013tra,Adams:2017ejb}. 
The four singular points of the elliptic curve $\zt=0,1,\infty,-\frac{1}{8}$ are then mapped to $\tilde\tau=i\infty,\frac{1}{2},\frac{1}{3},0$.
We further define
\begin{equation}
   q=e^{2\pi i\left(\tilde\tau-\frac{1}{2}\right)}\,,\qquad L=\log q\,,\qquad
   \theta=q\,\frac{d}{dq}=\frac{1}{2\pi i}\frac{d}{d\tilde\tau}\,.
\label{eq:qframe}
\end{equation}
The physical region $0<\zt<1$ is the line $\tilde\tau=\frac{1}{2}+is$, on which $q=e^{-2\pi s}$ is real and positive and $s$ runs from $\infty$ in the collinear limit to $0$ in the back-to-back limit.
Thus we shift the power in $q$ by $\frac{1}{2}$ in
Eq.~\eqref{eq:qframe}, which makes $q$, and with it every series below, real and positive on the physical line, at the price of alternating signs relative to the usual $\tilde q=e^{2\pi i\tilde\tau}=-q$.

The holomorphic solution to $L_{\rm PF}$, $\psi_1$, admits the following Taylor expansion in $\zt$ involving the Franel numbers~\cite{Franel:1894,Zagier:2009apery},
\begin{equation}
   \psi_1(\zt)=\sum_{n\ge0}(-1)^n\Big[\sum_{k=0}^n\binom nk^{3}\Big]\zt^n
   =1-2\zt+10\zt^2-56\zt^3+346\zt^4-\dots\,,
\label{eq:psi1}
\end{equation}
a modular form of weight one for $\Gamma_1(6)$ with a simple zero at the cusp $\zt=\infty$
and no other zero.
Everything in this frame can be built from the Dedekind eta function
$\eta(\tilde\tau)=\tilde q^{1/24}\prod_{n\ge1}\left(1-\tilde q^{\,n}\right)$. With
$\eta_d\equiv\eta(d\tilde\tau)$, the Hauptmodul (which is our angular variable), the period, and the two other factors
that will be needed are eta quotients,
\begin{align}
   &\zt=-\frac{\eta_1^3\eta_6^9}{\eta_2^3\eta_3^9}=q+3q^2+3q^3-5q^4-18q^5+\dots\,,\qquad
   \psi_1=\frac{\eta_2\eta_3^6}{\eta_1^2\eta_6^3}=1-2q+4q^2-2q^3+\dots\,,\nn\\
   &1-\zt=\frac{\eta_2^5\eta_6}{\eta_1\eta_3^5}\,,\qquad
   1+8\zt=\frac{\eta_1^8\eta_6^4}{\eta_2^4\eta_3^8}\,.
\label{eq:frame}
\end{align}
The frame and the operator are tied by the derivative rule
\begin{equation}
   \theta\zt=\psi_1^2\,\zt(1-\zt)(1+8\zt)\,,\qquad\text{equivalently}\qquad
   \log\zt=L+\log\frac{\zt}{q}=L+3q-\frac{3}{2}q^2-5q^3+\dots\,.
\label{eq:thetazeta}
\end{equation}
The second form is the dictionary between the logarithm of the kinematic variable and
the logarithm of $q$. It is also the reason why $L$ carries transcendental weight one. The
Wronskian of the Frobenius basis of $L_{\rm PF}$ at $\zt=0$ is
$\left[\zt(1-\zt)(1+8\zt)\right]^{-1}$, and the mirror map $q(\zt)$ built from that basis inverts
the first eta quotient of Eq.~\eqref{eq:frame}. We verify Eqs.~\eqref{eq:psi1},
\eqref{eq:frame} and \eqref{eq:thetazeta}, together with $L_{\rm PF}\psi_1=0$ and the
mirror-map statement, as exact rational $q$-series through $q^{339}$.

Recall that $\psi_1$ has a single simple zero, and that it sits exactly where $\zt$ has
its pole. The space of modular forms of weight $k$ of the group is therefore
\begin{equation}
   M_k\left(\Gamma_1(6)\right)=\psi_1^{\,k}\cdot\mathrm{span}_{\mathbb Q}
   \{1,\zt,\zt^2,\dots,\zt^{\,k}\}\,,\qquad
   \dim M_k\left(\Gamma_1(6)\right)=k+1\,,
\label{eq:Mk}
\end{equation}
which is the Riemann-Roch count for genus zero with four cusps. Odd weights occur because $-1\notin\Gamma_1(6)$. The first cusp form appears only at $k=4$, so weight-two and weight-three forms are all Eisenstein.
We use Eq.~\eqref{eq:Mk} as an operational test rather than as a dimension count. A function is a modular form of modular weight $k$ if and only if it is $\psi_1^{\,k}$ times
a polynomial in $\zt$ of degree at most $k$.
For the NNLO EEC, we will need the following letters
\begin{align}
   \omega_0&\equiv\theta\log\zt=\psi_1^2\left(1+7\zt-8\zt^2\right)\,,
   &&\deg=2\le2\,,\quad\text{modular weight }2\,,\nn\\
   \omega_1&\equiv-\theta\log(1-\zt)=\psi_1^2\left(\zt+8\zt^2\right)\,,
   &&\deg=2\le2\,,\quad\text{modular weight }2\,,\nn\\
   \psi_1\omega_0&=\psi_1^3(1-\zt)(1+8\zt)\,,
   &&\deg=2\le3\,,\quad\text{modular weight }3\,,\nn\\
   \zt\psi_1\omega_0&=\psi_1\theta\zt=\psi_1^3\zt(1-\zt)(1+8\zt)\,,
   &&\deg=3\le3\,,\quad\text{modular weight }3\,,
\label{eq:letters}
\end{align}
where the first two letters are the HPL letters and the last two are new.

With this notation at hand, we can write for any $g$, 
\begin{equation}
   L_{\rm PF}\left(\psi_1\,g\right)=-\frac{\theta^2g}{\zt\,\psi_1\omega_0}\,,
\end{equation}
where we use $L_{\rm PF}\psi_1=0$ and Eq.~\eqref{eq:thetazeta}. The denominator is the modular-weight-three letter of Eq.~\eqref{eq:letters}, and this is how the elliptic letter enters the answer. Applied to $g=S_0/\psi_1$, the identity turns this equation into
\begin{equation}
   \theta^2\left(\frac{S_0}{\psi_1}\right)=-\zt\,\psi_1\omega_0\left(\tilde{E}+\tilde\beta \Psi\right)\,.
\label{eq:S0theta}
\end{equation}
The second-order operator has become two antiderivatives, with the factor
$\zt\,\psi_1\omega_0$ remaining in the integrand. That factor, like $\omega_0$ and
$\omega_1$, is a power of $\psi_1$ times a polynomial in $\zt$, so the answer will be an
iterated integral whose letters are modular forms. With $\Psi$ of Eq.~\eqref{eq:Psifull}
the bracket on the right of Eq.~\eqref{eq:S0theta} is an explicit polylogarithm. 

\subsubsection{Iterated Eisenstein integrals}

To solve this equation, we need to introduce a set of iterated integrals in $q$~\cite{Adams:2017ejb}. Explicitly, we define the iterated Eisenstein integrals via
\begin{equation}
   I(f_1,\dots,f_n;q)=\theta^{-1}\left(f_1\,I(f_2,\dots,f_n;q)\right)=\theta^{-1}\Big[f_1\,\theta^{-1}\left(f_2\cdots\theta^{-1}(f_n)
   \right)\Big]\,,
\end{equation}
with the convention $I(\,;q)=1$. Here $\theta^{-1}$ plays the role of integration; for example,
\begin{align}
   \theta^{-1}q^{\,m}&=\frac{q^{\,m}}{m}\quad(m\ge1)\,,\qquad
   \theta^{-1}L^{\,j}=\frac{L^{\,j+1}}{j+1}\,,\nn\\[2pt]
   \theta^{-1}\left(q^{\,m}L^{\,j}\right)&=\sum_{i=0}^{j}(-1)^i\,\frac{j!}{(j-i)!}\,
   \frac{q^{\,m}L^{\,j-i}}{m^{\,i+1}}\,.
   \label{eq:Iiter}
\end{align}
We write the integration as $\theta^{-1}$ rather than as $\int dq/q$ for two reasons. First, $\theta^{-1}$ absorbs the measure, so that a letter is exactly what stands to the right-hand side of $\theta^2$ in Eq.~\eqref{eq:S0theta}.
Second, it replaces the regularization at
the base point by the algebraic rule on the first line of Eq.~\eqref{eq:Iiter}. That
regularization is needed because a letter with a constant term, such as
$\psi_1\omega_0$ itself, makes $\int dq/q$ divergent at the base point. The rule is the analog of $\log^kx/k!$ for an ordinary polylogarithm.
The definition of iterated integrals respects the shuffle product,
\begin{equation}
   I(f_1,\dots,f_m;q)\,I(f_{m+1},\dots,f_{m+n};q)=\sum_{\sigma}
   I(f_{\sigma(1)},\dots,f_{\sigma(m+n)};q)\,,
\end{equation}
with the sum running over the $\binom{m+n}{m}$ interleavings of the two sequences that preserve the order within each.
The number of letters in an iterated integral is its transcendental weight. It is raised by one at every $\theta^{-1}$, and $L$ carries weight one by the same rule. 
Note that the modular weight in Eq.~\eqref{eq:Mk} is a different concept, attached to the letters alone. It restricts which kernels may appear, is not raised by integration, and does not enter the weight count of a function. For example, $I(1,\psi_1\omega_0,\omega_0,\omega_1,\omega_1;q)$ is a function of
transcendental weight five, whose letters carry modular weights $0,3,2,2,2$.

As an example, take the transcendental weight-two function that occurs below,
$I(1,\psi_1\omega_0;q)$. By Eqs.~\eqref{eq:frame} and~\eqref{eq:letters} its letter is
$\psi_1\omega_0=\psi_1^3(1-\zt)(1+8\zt)=1+q-5q^2+q^3+11q^4-24q^5+\dots$. Because this letter has a constant term, the inner integration produces a power of $L$ through the second rule of Eq.~\eqref{eq:Iiter}. The outer integration, against the constant letter $1$, produces another:
\begin{align}
   I(\psi_1\omega_0;q)&=L+q-\frac{5}{2}q^2+\frac{1}{3}q^3+\frac{11}{4}q^4
   -\frac{24}{5}q^5+\dots\,,\nn\\
   I(1,\psi_1\omega_0;q)&=\frac{L^2}{2}+q-\frac{5}{4}q^2+\frac{1}{9}q^3+\frac{11}{16}q^4
   -\frac{24}{25}q^5+\dots\,,
\label{eq:qex}
\end{align}
where the $L^2/2$ comes from $\theta^{-1}L$ and every $q^m$ has been divided by $m$ once more.
For iterated Eisenstein integrals whose letters are only $\omega_0$ and $\omega_1$, the integrals reduce to the usual GPLs (HPLs in this case). For example, $I(\omega_1;q)=H_1(\zt)=-\log(1-\zt)$.

Applying $\theta^{-1}$ twice and using iterated Eisenstein integrals, we find that the analytic expression for $S_0$ is 
\begin{align}
   \frac{S_0}{\psi_1}&=\frac{33}{4}\zeta_5-\frac{1}{12}\pi^2\zeta_3-\frac{1}{30}\pi^4L -4\,I(1,\zt\psi_1\omega_0,\omega_0,\omega_0,\omega_1;q) -4\,I(1,\zt\psi_1\omega_0,\omega_0,\omega_1,\omega_1;q)\nn\\
   &-4\,I(1,\zt\psi_1\omega_0,\omega_1,\omega_0,\omega_1;q)
    -4\,I(1,\zt\psi_1\omega_0,\omega_1,\omega_1,\omega_1;q)-I(1,\psi_1\omega_0,\omega_0,\omega_0,\omega_1;q)\nn\\
   &-\frac{5}{2}\,I(1,\psi_1\omega_0,\omega_0,\omega_1,\omega_1;q)-\frac{7}{2}\,I(1,\psi_1\omega_0,\omega_1,\omega_0,\omega_1;q)
    -5\,I(1,\psi_1\omega_0,\omega_1,\omega_1,\omega_1;q)\nn\\
   &-\frac{3}{2}\,I(1,\psi_1\omega_0,\omega_0,\omega_1,\omega_0;q)
    +\frac{3}{2}\,I(1,\psi_1\omega_0,\omega_1,\omega_0,\omega_0;q)-\frac{\pi^2}{4}\,I(1,\psi_1\omega_0,\omega_1;q)\nn\\
   &-4\zeta_3\,I(1,\zt\psi_1\omega_0;q)+\frac{1}{2}\zeta_3\,I(1,\psi_1\omega_0;q)\,,
\label{eq:S0words}
\end{align}
where the first three terms are fixed by the boundary data. The boundary data are the value and the $\log\zt$ coefficient of $S_0$ at $\zt=0$, read off the expansion of the two-fold integral of Eq.~\eqref{eq:S0def} at $\zt=0$. We can also obtain the analytic form for $S_0^{'}$ via
\begin{equation}
   S_0'=\frac{\theta S_0}{\theta\zt}=\frac{\theta S_0}{\psi_1^2\,\zt(1-\zt)(1+8\zt)}\,.
\end{equation}
Finally, the elliptic sector follows from Eq.~\eqref{eq:FK1ansatz}.

\subsubsection{Numerical evaluations}

In this subsection we describe how we evaluate the iterated Eisenstein integrals and $F_{K_1}$ numerically to high precision. The method, truncated $q$-series with the regularization of Eq.~\eqref{eq:Iiter}, is the standard one for iterated integrals of modular forms~\cite{Adams:2017ejb,Bogner:2017vim} and is implemented in \textsc{GiNaC}~\cite{Walden:2020odh,Duhr:2026ell}; we spell out our implementation because the letters, the inversion $\zt\mapsto q$ and the precision control are specific to this problem.
To obtain the value at a point $\zt_0$, we need three things: the point $q_0$ with $\zt(q_0)=\zt_0$, the $q$-expansions of the letters to some order $N$, and the iterated integrals as truncated series. 

We first determine $q_0$. On the physical line $q$ is real with $0<q<1$, and there the map $q\mapsto\zt(q)$ is increasing, because $\theta\zt=\psi_1^2\,\zt(1-\zt)(1+8\zt)>0$ by Eq.~\eqref{eq:thetazeta}. The equation $\zt(q_0)=\zt_0$ therefore has exactly one solution. Near the collinear limit it is given by the inverse of the first series in Eq.~\eqref{eq:frame},
\begin{equation}
   q=\zt-3\zt^2+15\zt^3-85\zt^4+522\zt^5-\dots\,,
\label{eq:qinv}
\end{equation}
which converges for $|\zt|<\frac{1}{8}$, the distance from $\zt=0$ to the cusp $\zt=-\frac{1}{8}$. 
Beyond that radius we solve $\zt(q)=\zt_0$ on the truncated series of the eta quotient, by bisection on $0<q<1$ followed by Newton's method, and four Newton steps suffice. 
For example, at $\zt_0=\frac{3}{11}$ we find $q_0\approx0.1739$ and $L=\log q_0\approx-1.7493$, and at $\zt_0=\frac{1}{2}$ we find $q_0\approx0.2724$.

Next come the letters. In Eq.~\eqref{eq:frame} each $\eta_d$ is $\tilde q^{\,d/24}$ times the product $\prod_{n\ge1}\left(1-\tilde q^{\,dn}\right)$ with $\tilde q=-q$. The fractional powers of $\tilde q$ cancel in $\psi_1$, $1-\zt$ and $1+8\zt$, and they combine into the single factor $-\tilde q=q$ in $\zt$. Each quotient is therefore a product of truncated series and of inverses of truncated series, and $\zt$ and $\psi_1$ come out through order $q^N$ with integer coefficients. Then $\omega_0$, $\omega_1$, $\psi_1\omega_0$ and $\zt\psi_1\omega_0$ follow from Eq.~\eqref{eq:letters} by series multiplication. The identity $\theta\zt=\psi_1^2\,\zt(1-\zt)(1+8\zt)$ must hold order by order, and we use it as a check on the truncated series.

The iterated integrals are built from the inside out. The innermost letter is a truncated series, and $\theta^{-1}$ is applied to it. Each further step multiplies the current result by the next letter, truncates at $q^N$, and applies $\theta^{-1}$ again, term by term with Eq.~\eqref{eq:Iiter}: a term $q^{\,m}L^{\,j}$ with $m\ge1$ returns the terms $q^{\,m}L^{\,j-i}$ with the rational coefficients written there, and a term $L^{\,j}$ without $q$ returns $L^{\,j+1}/(j+1)$. Each integral is therefore a polynomial in $L$ whose coefficients are truncated $q$-series with rational coefficients. A power of $L$ is created only when $\theta^{-1}$ meets a term without $q$, that is, only by the letters with a constant term, which are $1$, $\omega_0$ and $\psi_1\omega_0$. Integrals whose innermost letter is $\omega_1$ or $\zt\psi_1\omega_0$ contain no $L$ at all, and the power of $L$ in an integral is the length of its innermost run of letters with a constant term. In Eq.~\eqref{eq:S0words} the highest power is $L^2$, from $I(1,\psi_1\omega_0,\omega_1,\omega_0,\omega_0;q)$ and $I(1,\psi_1\omega_0;q)$.
Substituting $q_0$ and $L=\log q_0$ gives $S_0$, and its derivative can also be calculated directly. Thus these three steps provide a way to evaluate the elliptic functions. 
The polylogarithmic functions, on the other hand, can be evaluated directly at $\zt_0$ with the standard tools~\cite{Bauer:2000cp,Maitre:2005uu,Duhr:2019tlz}. This completes the evaluation of $\FK$.

The precision is controlled by two components.
Rounding enters only at the end, when $q_0$, $L=\log q_0$, the constants $\pi^2$, $\zeta_3$, $\pi^4$, $\zeta_5$ and $\Psi$ are substituted in floating point, and the working precision of that substitution can be set as high as needed. Truncating every series at $q^N$ drops terms of order $q_0^{N+1}$. The letters are holomorphic modular forms, whose $q$-coefficients grow at most polynomially, and $\theta^{-1}$ only divides by powers of $m$, so the truncation error of the iterated integrals at fixed $q_0$ is a constant times $q_0^{N}$. The number of correct decimal digits after $N$ terms, $N_\text{dig}$, is then
\begin{equation}
   N_\text{dig}\simeq N\log_{10}\frac{1}{q_0}-3\,,
\label{eq:digits}
\end{equation}
and fifty digits need $N\simeq53/\log_{10}(1/q_0)$ terms by this estimate. In practice about $75$ terms are needed at $\zt_0=\frac{3}{11}$ and about $106$ at $\zt_0=\frac{1}{2}$, since Eq.~\eqref{eq:digits} is slightly optimistic at large $N$. We have checked our implementation against the \textsc{GiNaC} one on a sample of the iterated integrals of Eq.~\eqref{eq:S0words}.
Near the back-to-back limit $q_0\to 1$ and the series converges slowly. Therefore, we also apply similar construction and develop a routine for values close to $\zt\to1$.

\subsection{Final result}
\label{sec:final_result}

Putting everything together, we now have the full analytic expression for the EEC at NNLO in $\cN=4$ SYM. We write $F_\text{NNLO}$ as the sum of polylogarithmic and elliptic parts, $F_\text{NNLO}(\zt)=F_\text{P}(\zt)+F_\text{E}(\zt)$, where
\begin{align}
   F_\text{P}(\zt)&=F_\text{HPL}(\zt)+2\,\Psi(\zt)+P(\zt)\,,\nn\\
   F_\text{E}(\zt)&=2\left[c_1S_0+c_2S_0'\right]=2\,\FK(\zt)-2\,\Psi(\zt)\,,
\end{align}
with $F_\text{HPL}(\zt)$ provided by Ref.~\cite{Henn:2019gkr}.
Here we use HPLs in the $\{0,+,-\}$ notation to express $F_\text{P}(\zt)$.
Explicitly, these letters are
\begin{align}
   &H_{a_1,\dots,a_n}(\sqrt{\zt})=\int_0^{\sqrt{\zt}}dt\,f_{a_1}(t)\,H_{a_2,\dots,a_n}(t)\,,\qquad a_i\in\{0,+,-\}\,,\nn\\
   &f_0(t)=\frac{1}{t}\,,\quad f_{+}(t)=\frac{1}{1-t}+\frac{1}{1+t}=\frac{2}{1-t^2}\,,\quad
   f_{-}(t)=\frac{1}{1-t}-\frac{1}{1+t}=\frac{2t}{1-t^2}\,.
\label{eq:pmletters}
\end{align}
We omit the $\sqrt{\zt}$ argument of the HPLs to save space and write $H_{a_1,\dots,a_n}\equiv H_{a_1,\dots,a_n}(\sqrt{\zt})$ below.
Then the full analytic expression for $F_\text{P}(\zt)$ is
\begin{align}
   F_\text{P}(\zt)&= \left(\zt - 1\right)\,\Big[ -4\,H_{-,0,0} - 32\,H_{-,0,-,0} - 2\,H_{+,+,0,0} - 2\,H_{+,0,+,0,0} - 2\,H_{+,+,0,-,-} \Big]
   \nn\\
   &-8 \left(\zt + 1\right)\,H_{-,0,0,-,0} +\sqrt{\zt} \left(\zt - 1\right)\,\Big[ 4\,H_{0,+,0} + 6\,H_{-,+,0} + 4\,H_{+,0,0} + 8\,H_{+,0,-}- 6\,H_{+,-,0}
   \nn\\
   & + 4\,H_{0,0,+,0} + 8\,H_{0,-,+,0} + 8\,H_{0,+,0,-} - 8\,H_{0,+,-,0} + 8\,H_{-,0,+,0}+ 15\,H_{-,-,+,0} + 2\,H_{-,+,0,0}
   \nn\\
   & + 16\,H_{-,+,0,-} - 15\,H_{-,+,-,0} - 32\,H_{+,0,0,-}+ 32\,H_{+,0,-,0} - 2\,H_{+,-,0,0} - 16\,H_{+,-,0,-}
   \nn\\
   & + 15\,H_{+,-,-,0} - 15\,H_{+,+,+,0}+ \frac{1}{3} \pi^2\,H_{+} + \frac{2}{3} \pi^2\,H_{0,+} 
   + \frac{7}{6} \pi^2\,H_{-,+} - \frac{2}{3} \pi^2\,H_{+,0} - \frac{7}{6} \pi^2\,H_{+,-}\nn\\
   & + \zeta_3\,H_{+} \Big]
   -\frac{1}{6} \left(\zt - 1\right) \left(\zt - 5\right) \pi^2\,H_{-,0} +\frac{1}{3} \left(\zt - 4\right) \left(\zt - 1\right) \zeta_3\,H_{-} -\left(\zt^{2} - 4 \zt - 20\right)\,H_{-,-,-,0,-}
   \nn\\
   &+\zt \left(\zt - 3\right)\,\Big[ -8\,H_{0,-,-,0,-} - 2 \zeta_3\,H_{0,-} \Big] +\left(\zt - 1\right) \left(\zt - 2\right)\,\Big[ -4\,H_{0,-,0,0} - 2\,H_{-,-,0,0} \Big]
   \nn\\
   &+\left(\zt - 1\right)^{2}\,\Big[ \frac{47}{3}\,H_{-,0,-,-} + 9\,H_{-,-,-,-} - \frac{1}{2} \zeta_3\,H_{+,+} \Big] +\zt \left(\zt - 1\right)\,\Big[ -16\,H_{0,0,-}+ 16\,H_{0,-,0}
   \nn\\
   & - 8\,H_{-,0,-} - 18\,H_{0,-,-,0} + 24\,H_{0,+,+,0} + 6\,H_{0,0,-,-,0} - \frac{2}{3} \pi^2\,H_{0}- \frac{1}{3} \pi^2\,H_{-}+ \pi^2\,H_{0,0} 
   \nn\\
   & + \frac{1}{3} \pi^2\,H_{+,+,0} + 2 \zeta_3 \Big] +\left(\zt - 1\right) \left(\zt + 1\right)\,\Big[ -4\,H_{+,0,+,0} - 8\,H_{+,+,0,-}- 2\,H_{+,0,0,+,0}
   \nn\\
   & - 4\,H_{+,0,+,0,-} \Big] +\zt^{2}\,\Big[ 24\,H_{0,-,0,-,-} - \pi^2\,H_{0,-,0} \Big]+\zt \left(\zt + 1\right)\,\Big[ -4 \zeta_3\,H_{0,0} + \frac{1}{6} \pi^2\zeta_3 \Big]
   \nn\\
   & +\left(\zt - 1\right) \left(\zt + 3\right)\,\Big[ H_{-,+,+,0,0} - H_{+,-,+,0,0}+ H_{+,+,-,0,0} - \frac{1}{3} \pi^2\,H_{-,0,0} \Big]
   \nn\\
   & +\left(\zt^{2} + 4 \zt - 3\right)\,\Big[ -2\,H_{-,0,-,0,0} - H_{-,-,-,0,0} \Big]+4 \zt \left(\zt + 6\right)\,H_{0,0,-,0,-}
   \nn\\
   & -\left(\zt^{2} + 8 \zt + 1\right)\,H_{-,0,-,-,0} -\left(\zt - 1\right) \left(\zt + 14\right)\,H_{-,-,-,0}-4 \left(2 \zt^{2} - \zt - 7\right)\,H_{-,-,0,0,-}
   \nn\\
   & +\left(\zt - 1\right) \left(2 \zt + 1\right)\,\Big[ -2\,H_{+,+,0} + 8\,H_{+,+,0,0,-}- \frac{1}{6} \pi^2\,H_{+,0,+} \Big]+3 \left(2 \zt^{2} + 3\right)\,H_{-,-,-,-,-}
   \nn\\
   &  -\left(3 \zt^{2} - 2 \zt - 13\right)\,H_{-,0,+,+,0}
   +8 \zt \left(3 \zt - 2\right)\,H_{0,0,0,0,-} +\zt \left(3 \zt - 1\right)\,\Big[ 4\,H_{0,+,0,+,0} + 8\,H_{0,+,+,0,-} \Big]
   \nn\\
   &-\frac{1}{180} \left(\zt - 1\right) \left(3 \zt + 4\right) \pi^4 +2 \zt \left(3 \zt + 1\right)\,H_{0,+,+,0,0} +2 \left(4 \zt^{2} - 6 \zt + 7\right)\,H_{-,0,0,-,-}
   \nn\\
   &+2 \left(\zt - 1\right) \left(4 \zt - 1\right)\,H_{-,-,0} +4 \left(4 \zt^{2} - 4 \zt + 9\right)\,H_{-,0,0,0,-} -8 \left(\zt - 1\right) \left(4 \zt + 1\right)\,H_{0,0,-,0}
   \nn\\
   &+8 \zt \left(4 \zt - 3\right)\,H_{0,0,0,-,0} -\frac{1}{6} \left(4 \zt^{2} - 2 \zt + 3\right) \pi^2\,H_{-,-,0} +\left(\zt - 1\right) \left(4 \zt + 3\right)\,\Big[ H_{-,+,0,+,0}
   \nn\\
   &+ 2\,H_{-,+,+,0,-} - H_{+,0,-,+,0} + H_{+,0,+,-,0} - H_{+,-,0,+,0} - 2\,H_{+,-,+,0,-}- 4\,H_{+,+,0,-,0}
   \nn\\
   & + 2\,H_{+,+,-,0,-} - \frac{1}{6} \pi^2\,H_{+,+} \Big] +\left(4 \zt^{2} - \zt + 15\right)\,H_{-,0,-,-,-}+2 \left(4 \zt^{2} + \zt + 17\right)\,H_{-,0,-,0,-}
   \nn\\
   & +\frac{1}{2} \left(4 \zt^{2} + 6 \zt + 3\right) \zeta_3\,H_{-,-} +\left(4 \zt^{2} + 7 \zt - 12\right) \zeta_3\,H_{-,0}-4 \zt \left(5 \zt - 9\right)\,H_{0,-,0,0,-}
   \nn\\
   & +\frac{1}{6} \left(\zt - 1\right) \left(5 \zt + 2\right) \pi^2\,H_{-,-} +\zt \left(5 \zt - 3\right)\,\Big[ -4\,H_{0,0,-,0,0}- 2\,H_{0,-,-,0,0} \Big]
   \nn\\
   & +\left(\zt - 1\right) \left(5 \zt + 3\right)\,\Big[ -\frac{3}{2}\,H_{-,-,+,+,0} + \frac{3}{2}\,H_{-,+,-,+,0}- \frac{3}{2}\,H_{-,+,+,-,0} - \frac{3}{2}\,H_{+,-,-,+,0}
   \nn\\
   & + \frac{3}{2}\,H_{+,-,+,-,0} - \frac{3}{2}\,H_{+,+,-,-,0}+ \frac{3}{2}\,H_{+,+,+,+,0} \Big] +2 \zt \left(5 \zt + 3\right)\,H_{0,0,-,-,-}
   \nn\\
   & +\frac{1}{3} \left(\zt - 1\right) \left(7 \zt - 1\right) \pi^2\,H_{0,-}+4 \zt \left(7 \zt - 5\right)\,H_{0,0,0,-,-} 
   +\left(\zt - 1\right) \left(7 \zt + 3\right)\,\Big[ \frac{1}{12} \pi^2\,H_{-,+,+} 
   \nn\\
   &- \frac{1}{12} \pi^2\,H_{+,-,+}+ \frac{1}{12} \pi^2\,H_{+,+,-} \Big] +\zt \left(7 \zt - 3\right)\,\Big[ -3\,H_{0,-,+,+,0} + 3\,H_{0,+,-,+,0} - 3\,H_{0,+,+,-,0} \Big]
   \nn\\
   &+\frac{1}{90} \zt \left(7 \zt + 12\right) \pi^4\,H_{0} -\frac{1}{12} \left(2 \zt - 1\right) \left(4 \zt + 1\right) \pi^2\,H_{-,-,-} -\frac{1}{6} \left(\zt + 1\right) \left(8 \zt - 9\right) \pi^2\,H_{-,0,-}
   \nn\\
   &+\left(\zt - 1\right) \left(8 \zt + 7\right)\,\Big[ H_{-,+,+,0} - H_{+,-,+,0} + H_{+,+,-,0} \Big] +2\left(\zt - 1\right) \left(9 \zt - 1\right)\,H_{-,-,0,-}
   \nn\\
   &+\zt \left(9 \zt - 5\right)\,\Big[\frac{1}{6}\pi^2\,H_{0,+,+} -2H_{0,0,+,+,0}\Big] -\frac{1}{3} \zt \left(9 \zt - 4\right) \pi^2\,H_{0,0,-}
   \nn\\
   &+\left(9 \zt^{2} - 2 \zt + 12\right)\,H_{-,-,0,-,-} -\frac{1}{6} \zt \left(9 \zt + 1\right) \pi^2\,H_{0,-,-} -\frac{1}{720} \left(11 \zt^{2} + 27 \zt - 45\right) \pi^4\,H_{-}
   \nn\\
   &-\frac{2}{3} \left(\zt - 1\right) \left(13 \zt - 1\right) \zeta_3\,H_{0} +2 \left(13 \zt^{2} - 4 \zt - 2\right)\,H_{-,-,0,-,0} +\zt \left(13 \zt + 3\right)\,H_{0,-,-,-,-}
   \nn\\
   &+4 \zt \left(17 \zt - 9\right)\,H_{0,-,0,-,0} -\frac{1}{4} \zt \left(17 \zt + 60\right) \zeta_5 +\frac{2}{3} \left(\zt - 1\right) \left(19 \zt - 10\right)\,H_{0,-,-,-}
   \nn\\
   &+\frac{1}{2} \left(25 \zt^{2} - 12 \zt - 1\right)\,H_{-,-,-,-,0} +\zt \left(29 \zt - 15\right)\,H_{0,-,-,-,0}+\frac{2}{3} \left(\zt - 1\right) \left(31 \zt - 10\right)\,H_{0,0,-,-}
   \nn\\
   & +\frac{4}{3} \left(\zt - 1\right) \left(31 \zt - 7\right)\,H_{0,-,0,-}
   +\frac{2}{3} \left(\zt - 1\right) \left(47 \zt + 1\right)\,H_{-,0,0,-} +\frac{4}{3} \left(\zt - 1\right) \left(55 \zt - 4\right)\,H_{0,0,0,-}\,.
\label{eq:FPfull}
\end{align}

For the elliptic part, we write 
\begin{align}
   F_\text{E}(\zt)&=2A(\zt)\,\frac{S_0}{\psi_1}+2B(\zt)\,\theta\left(\frac{S_0}{\psi_1}\right)\,,
\end{align}
with the coefficients
\begin{align}
   A(\zt)&=\frac{\zt\left(\zt-1\right)\left(8\zt-5\right)}{3}\,\psi_1(\zt)+\frac{\zt\left(\zt-1\right)^2\left(8\zt+1\right)}{3}\,\psi_1'(\zt)\,,\nn\\
   B(\zt)&=\frac{c_2}{\psi_1\,\zt(1-\zt)(1+8\zt)}=\frac{1-\zt}{3\,\psi_1(\zt)}\,.
\end{align}
These coefficients contain $\psi_1$ and its derivative,
\begin{align}
   \psi_1'(\zt)&=\frac{d\psi_1}{d\zt}=\sum_{n\ge1}(-1)^n\,n\Big[\sum_{k=0}^n\binom nk^{3}\Big]\zt^{n-1}=-2+20\zt-168\zt^2+1384\zt^3-\dots\,.
\end{align}
The transcendental part $S_0/\psi_1$ is given in Eq.~\eqref{eq:S0words},
and its derivative $\theta\left(S_0/\psi_1\right)$ is
\begin{align}
   \theta\left(\frac{S_0}{\psi_1}\right)&=-\frac{1}{30}\pi^4-4\,I(\zt\psi_1\omega_0,\omega_0,\omega_0,\omega_1;q)-4\,I(\zt\psi_1\omega_0,\omega_0,\omega_1,\omega_1;q)-4\,I(\zt\psi_1\omega_0,\omega_1,\omega_0,\omega_1;q)
   \nn\\
   &-4\,I(\zt\psi_1\omega_0,\omega_1,\omega_1,\omega_1;q)-I(\psi_1\omega_0,\omega_0,\omega_0,\omega_1;q)-\frac{5}{2}\,I(\psi_1\omega_0,\omega_0,\omega_1,\omega_1;q)
   \nn\\
   &-\frac{7}{2}\,I(\psi_1\omega_0,\omega_1,\omega_0,\omega_1;q)-5\,I(\psi_1\omega_0,\omega_1,\omega_1,\omega_1;q)-\frac{3}{2}\,I(\psi_1\omega_0,\omega_0,\omega_1,\omega_0;q)
   \nn\\
   &+\frac{3}{2}\,I(\psi_1\omega_0,\omega_1,\omega_0,\omega_0;q)-\frac{\pi^2}{4}\,I(\psi_1\omega_0,\omega_1;q)
   -4\zeta_3\,I(\zt\psi_1\omega_0;q)+\frac{\zeta_3}{2}\,I(\psi_1\omega_0;q)\,.
\label{eq:thetaS0overpsi}
\end{align}

\begin{figure}[!htbp]
    \centering
    \includegraphics[width=0.8\linewidth]{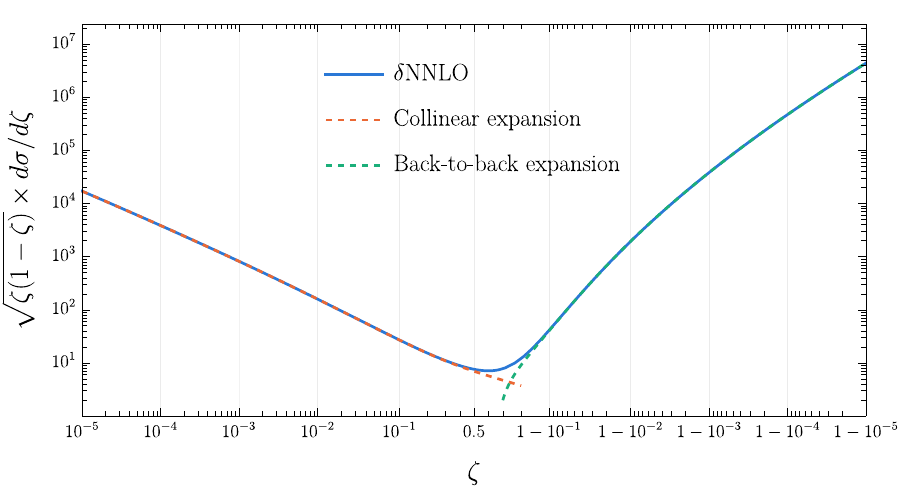}
    \caption{Comparison of the pure NNLO prediction $\delta\text{NNLO}$ and its singular expansions in both the collinear and the back-to-back limits. The distribution is multiplied by $\sqrt{\zt(1-\zt)}$.}
    \label{fig:NNLO_vs_asy}
\end{figure}

Using the analytic form, we can also obtain the singular expansions to very high powers in both the collinear and the back-to-back limits. 
Explicitly, in the collinear limit, we find
\begin{align}
F_{\rm NNLO}(\zt)
&=\zt\Big[\frac{1}{2}\log^{2} \zt+\left(-5+\frac{\pi^{2}}{3}-\zeta_3\right)\log \zt+17-\frac{4\pi^{2}}{3}-\zeta_3+\frac{5\pi^{4}}{144}+\frac{3\zeta_5}{2}\Big]\nn\\
&+\zt^{2}\Big[\left(\frac{349}{144}+\frac{5\pi^{2}}{48}-3\zeta_3\right)\log^{2} \zt+\left(-\frac{6569}{432}-\frac{121\pi^{2}}{144}-\frac{19\zeta_3}{2}+\frac{67\pi^{4}}{180}\right)\log \zt\nn\\
&\qquad +\frac{16361}{864}+\frac{649\pi^{2}}{288}-\frac{7\zeta_3}{12}+\frac{511\pi^{4}}{1440}+\pi^{2}\zeta_3-\frac{347\zeta_5}{4}\Big]\nn\\
&+\zt^{3}\Big[\left(-\frac{560707}{43200}-\frac{\pi^{2}}{24}+11\zeta_3\right)\log^{2} \zt+\left(\frac{7234927}{72000}-\frac{13\pi^{2}}{90}+\frac{149\zeta_3}{4}-\frac{22\pi^{4}}{15}\right)\log \zt\nn\\
&-\frac{2555783}{11250}+\frac{148123\pi^{2}}{32400}+\frac{24107\zeta_3}{360}-\frac{3427\pi^{4}}{1440}-\frac{11\pi^{2}\zeta_3}{3}+363\zeta_5\Big]+\cO\left(\zt^{4}\right)\,.
\label{eq:FNNLOcoll}
\end{align}
In the back-to-back limit, we define $y\equiv 1-\zt$, and the singular expansion reads
\begin{align}
F_{\rm NNLO}(\zt)
&=\Big[-\frac{1}{8}\log^{5} y-\frac{\pi^{2}}{6}\log^{3} y-\frac{11\zeta_3}{4}\log^{2} y-\frac{61\pi^{4}}{720}\log y-\frac{\pi^{2}\zeta_3}{3}-\frac{7\zeta_5}{2}\Big]\nn\\
&+y\Big[\frac{1}{10}\log^{5} y-\frac{1}{3}\log^{4} y+\left(-\frac{1}{6}+\frac{5\pi^{2}}{72}\right)\log^{3} y +\left(\frac{1}{4}-\frac{3\pi^{2}}{8}+2\zeta_3\right)\log^{2} y \nn\\
&+\left(-2-\frac{\pi^{2}}{6}-3\zeta_3-\frac{5\pi^{4}}{72}\right)\log y+14+\frac{13\pi^{2}}{12}-3\pi^{2}\log 2-\zeta_3-\frac{173\pi^{4}}{360}\nn\\
&+2\pi^{2}\log^{2}2+\log^{4}2+24\,\mathrm{Li}_4\left(\frac{1}{2}\right)+\frac{247\pi^{4}\log 2}{240}-\frac{8\pi^{2}\log^{3}2}{3}-\frac{11\pi^{2}\zeta_3}{24}\nn\\
&-\frac{4\log^{5}2}{5}-\frac{2917\zeta_5}{32}+96\,\mathrm{Li}_5\left(\frac{1}{2}\right)-2S_0(1)\Big]\nn\\
&+y^{2}\Big[-\frac{1}{20}\log^{5} y+\frac{13}{16}\log^{4} y+\left(-\frac{97}{24}-\frac{\pi^{2}}{72}\right)\log^{3} y+\left(\frac{431}{32}+\frac{9\pi^{2}}{16}-\frac{3\zeta_3}{4}\right)\log^{2} y\nn\\
&+\left(-\frac{275}{8}-\frac{43\pi^{2}}{24}+9\zeta_3+\frac{5\pi^{4}}{72}\right)\log y+\frac{705}{16}+\frac{113\pi^{2}}{96}+\frac{9\pi^{2}\log 2}{4}-\frac{105\zeta_3}{4}\nn\\
&+\frac{947\pi^{4}}{2880}-\frac{7\pi^{2}\log^{2}2}{3}-\frac{7\log^{4}2}{6}-28\,\mathrm{Li}_4\left(\frac{1}{2}\right)-\frac{2\pi^{4}\log 2}{3}+\frac{5\pi^{2}\log^{3}2}{3}\nn\\
&+\frac{17\pi^{2}\zeta_3}{16}+\frac{\log^{5}2}{2}+\frac{941\zeta_5}{16}-60\,\mathrm{Li}_5\left(\frac{1}{2}\right)+2S_0(1)\Big]+\cO\left(y^{3}\right)\,,
\label{eq:FNNLOb2b}
\end{align}
where
\begin{equation}
    S_0(1)=\frac{611\sqrt{3}\,\pi^{5}}{19440}-3\sqrt{3}\,\mathrm{Im}\,G\left(e^{-i\pi/3},0,0,0,1;1\right)\approx 13.8046414536353167965661617535\,.
\end{equation}
While the collinear expansion only contains zeta values, the back-to-back expansion requires $\log 2$, $\text{Li}_n\left(1/2\right)$ and a special GPL.
In Fig.~\ref{fig:NNLO_vs_asy}, we show the comparisons of both expansions with the pure NNLO (denoted as $\delta\text{NNLO}$). We keep both expansions up to the third order, namely $\mathcal{O}(\zt^3)$ in the collinear limit and $\mathcal{O}((1-\zt)^2)$ in the back-to-back limit. We observe that the collinear expansion provides a reliable prediction even at $\zt\approx 0.5$.

\begin{figure}[!htbp]
    \centering
    \includegraphics[width=0.8\linewidth]{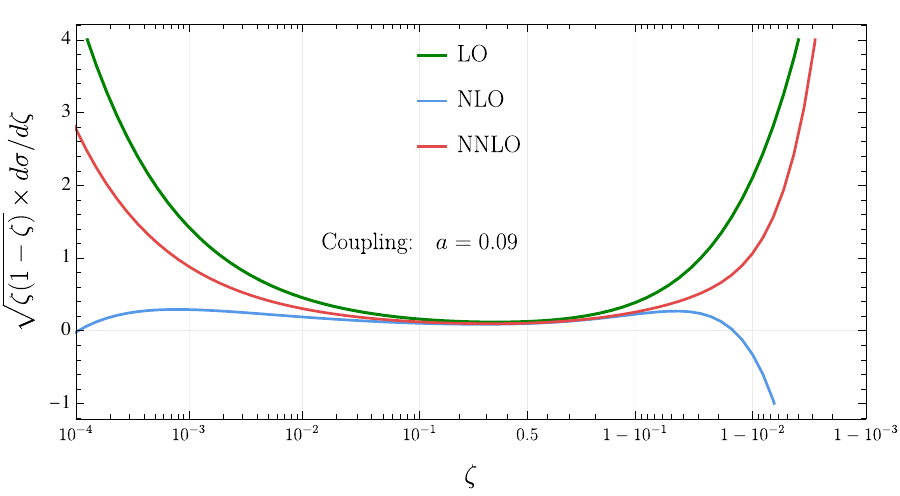}
    \caption{Fixed-order prediction for the EEC up to NNLO. Here we choose the coupling value $a=0.09$. }
    \label{fig:fixed_order}
\end{figure}

In Fig.~\ref{fig:fixed_order}, we show the fixed-order distribution of the EEC up to NNLO, with $a=N_c g_\text{YM}^2/(4\pi^2)=0.09$ as an example. Unlike Fig.~\ref{fig:NNLO_vs_asy}, here NLO includes contributions up to $\mathcal{O}(a^2)$ and NNLO includes contributions up to $\mathcal{O}(a^3)$. 
In the bulk region, we find good order-by-order convergence, and thus fixed-order predictions are reliable.
At both collinear and back-to-back endpoints, the distribution receives large logarithmic enhancements, so we observe instability across orders. In particular, the NLO curve goes negative around $1-\zt\sim 10^{-2}$. Systematic resummation of these large logarithms is required at both endpoints to obtain reliable predictions. 

\section{A first look at the EEC bootstrap}
\label{sec:bootstrap}

We obtained the analytic $F_\text{NNLO}$ from a direct calculation by integrating the polylogarithmic sector directly and by solving the differential system of the elliptic sector. With the answer known, we can ask how much of it a Landau bootstrap in the sense of Ref.~\cite{Hannesdottir:2024hke} would have fixed from the alphabet, the elliptic curve and the known limits alone.
Such an analysis will be useful for the EEC in QCD at this order, where the integrand is known but the integrals are not. 

The idea of perturbative bootstrap programs is to construct the transcendental function space from singularities, write down the ansatz for all rational coefficients, and determine the unknown parameters from either physical constraints or numerical data. 
For scattering amplitudes in planar $\cN=4$ SYM, the coefficients associated with the transcendental basis are rational numbers instead of functions, and there is enough physical information to fix them. 
For QCD observables with less symmetry, one can perform analytic regressions from high-precision numerical data, such as the lattice reduction approach developed in Ref.~\cite{Barrera:2025anr} and its application to the three-point energy correlators~\cite{Gong:2025eec}.
As we saw in Sec.~\ref{sec:final_result}, the EEC up to NNLO requires both the polylogarithmic and the elliptic sectors. In both cases, we need to determine the alphabets, all integrable symbols, and the candidate function space. 
Then we also need to make a reasonable ansatz for the rational coefficients and determine which EEC properties we can make use of.

\subsection{Alphabet and function space}\label{sec:bootspace}

A perturbative bootstrap program starts with the alphabet. For standard Feynman integrals, Landau analysis uses the analytic structures in the integrand, like branch cuts and singularities, and provides a sufficient set of letters. Algorithms have been developed and implemented in \textsc{Pld}~\cite{Fevola:2023fzn} and \textsc{Sofia}~\cite{Correia:2025wtb}.
For observables like differential cross-sections, due to the on-shell constraints from phase space integrals, there is no canonical method to determine the alphabet.
For EEC in $\cN=4$ SYM, we can learn the potential letters from the four-point correlator, whose discontinuity leads to the EEC integrand.
Note that the four-point correlator carries the elementary letters $z$, $\bar{z}$, $1-z$ and $1-\bar{z}$, and its rational prefactor carries the denominator $z-\bar{z}$. Applying Eq.~\eqref{eq:zoft} and the rescaling $x=t/\zb$ turns all of them into polynomials in $x$ of degree at most two over the common denominator $D$: $z$ gives $x(x-1)$, while $1-z$, $z-\zb$, $1-z-\zb$ and $1-z\zb$ give $D_0$, $C$, $D_1$ and $D_2$, with
\begin{equation}
   C=\zt x^{2}+\left(\zb-2\zt\right)x+\zt-1\,,
\label{eq:bootC}
\end{equation}
and other expressions in Eq.~\eqref{eq:ace}. The inner integration over $x\in[0,1]$ then produces logarithms whose arguments are cross ratios of the roots of these polynomials with the two endpoints $x=0,1$, carrying the square root of a discriminant whenever the two roots do not separate rationally.
The outer integration over $\zb$ produces the letters in $\zt$: they are the values the $\zb$-dependent letters take at $\zb=0$ and $\zb=1$.

The map itself gives $\log(-z)=\log\zt+\log\zb+\log x+\log(1-x)-\log D$, whose only $\zt$ letter is $\log\zt$. The linear factor $D$ has the single root $x_D=(\zt-1)/(\zt-\zb)$. After the $x$-integration a root enters only through its cross ratio with the endpoints $x=0,1$, here $x_D/(x_D-1)=(\zt-1)/(\zb-1)$, so $D$ contributes the letters $1-\zt$ and $1-\zb$ and, at the outer endpoint $\zb=0$, only $1-\zt$.
The rational factor of $D_0$ returns the same letter. The first nontrivial object is $C$, and its discriminant is a conic in the outer variable,
\begin{equation}
   R(\zb;\zt)=\operatorname{disc}_xC=\zb^2-4\zt\zb+4\zt\,,\qquad
   \operatorname{disc}_{\zb}R=16\zt\left(\zt-1\right)\,.
\label{eq:bootconic}
\end{equation}
The $x$-integration puts $\sqrt{R(\zb;\zt)}$ inside the $\zb$-letters through the roots $x_\pm=(2\zt-\zb\pm\sqrt R)/(2\zt)$, and the $\zb$-integration evaluates those letters at its endpoints, so the only algebraic quantities that can survive in $\zt$ are $\sqrt{R(1;\zt)}=1$ and $\sqrt{R(0;\zt)}=2\sqrt{\zt}$.
At $\zb=1$ the discriminant is a perfect square, $C$ factors rationally, and that endpoint returns nothing new. 
However, at $\zb=0$, $C$ becomes $\zt(x-1)^{2}-1$, with roots $1\pm1/\sqrt{\zt}$. Their cross ratio with the endpoints of the $x$ integration is the third letter,
\begin{equation}
   \int_0^1\frac{dx}{C(x,0;\zt)}=\frac{1}{2\sqrt{\zt}}\log\frac{1-\sqrt{\zt}}{1+\sqrt{\zt}}\,.
\label{eq:bootcross}
\end{equation}
Putting these together, we conclude the polylogarithmic alphabet is
\begin{equation}
   \mathcal{A}_\text{P}=\left\{\zt,\,1-\zt,\,\frac{1-\sqrt{\zt}}{1+\sqrt{\zt}}\right\}\,.
\label{eq:bootalphabetP}
\end{equation}
We emphasize that at two loops, $z-\bar{z}$ only shows up in the prefactor of the four-point correlator, not in the arguments of transcendental functions. The lesson we learn from this example is that one should always include the rational factors in the potential alphabet for safety.

With the alphabet at hand, we can construct the function space with polylogarithmic functions. Since we have one square root in the alphabet, it is convenient to introduce $\rho\equiv\sqrt{\zt}$, so that the letters become $\{\rho,1-\rho, 1+\rho\}$. 
In the general case, one writes down all possible symbols up to a certain weight, with entries from the alphabet, and applies the integrability condition. The resulting symbol basis can then be lifted to polylogarithmic functions. In this case, since there is only one variable, the integrability condition or any Steinmann-type relations used in the amplitude bootstrap place no constraints. 
With indices $0,\pm 1$, the final functions are simply HPLs.
The discrete symmetry remains in the function space.
Under the transformation $\rho\to-\rho$ the one-forms $f_0\,d\rho$ and $f_-\,d\rho$ of \Eq{eq:pmletters} are even and $f_+\,d\rho$ is odd, so a basis function is even or odd according to the parity of the number of $+$ letters it contains. The observable is a function of $\zt=\rho^{2}$ and must be invariant, so an odd basis function may appear only against an odd power of $\rho$. HPLs of $\zt$ itself lie in the same space, through $H_u(\zt)=2^{\,n_0(u)}H_{u'}(\sqrt{\zt})$ with $0\to0$, $1\to-$ and $n_0(u)$ the number of zeros in $u$.

The elliptic side begins from the same integrand. The two quadratics $D_1$ and $D_2$ of \Eq{eq:ace} have discriminants $Q_1$ and $Q_2$ that are quartic in the outer variable, \Eq{eq:Qexpl}, and a quartic square root cannot be rationalized: the outer integration of a function containing $\sqrt{Q_i(\zb)}$ is a period integral on a curve of genus one.
As discussed above, the two curves are the same by the M\"obius relation of \Eq{eq:mobius}, and the family degenerates where the quartic acquires a double root, which by $\operatorname{disc}_{\zb}Q_1=\operatorname{disc}_{\zb}Q_2=256\,\zt^6(1-\zt)^2(1+8\zt)$ happens at $\zt=0,1,-\frac{1}{8},\infty$. Four maximally unipotent singular fibers identify the family as the equal-mass sunrise family with modular group $\Gamma_1(6)$, as Sec.~\ref{sec:sunrise} shows independently by comparing $j$-invariants, and the cusp widths $1$, $2$, $3$ and $6$ sum to the index $12$.
All these properties follow from the NNLO integrand, and this is also why the elliptic sector only starts at this order.

The elliptic letters then follow from the geometry together with similar physical-singularity rules that fixed \Eq{eq:bootalphabetP}. A logarithmic letter $\ell$ contributes the kernel $\theta\log\ell=(\theta\zt)\,d\log\ell/d\zt$, and by \Eq{eq:thetazeta} the factor $\theta\zt=\psi_1^2\zt(1-\zt)(1+8\zt)$ carries $1+8\zt$. Since $d\log\ell/d\zt$ may be singular only at the physical points $\zt=0$ and $\zt=1$, nothing can cancel that factor and it survives in every admissible letter. At modular weight two this gives
\begin{equation}
   \psi_1^2\left(1+8\zt\right)\cdot\operatorname{span}_{\mathbb{Q}}\left\{1,\zt\right\}
   =\operatorname{span}_{\mathbb{Q}}\left\{\omega_0,\omega_1\right\}\,,
\label{eq:bootw2}
\end{equation}
a two-dimensional subspace of the three-dimensional $M_2$ of \Eq{eq:Mk}.
At modular weight three the letter is the one left behind by variation of parameters. Writing $S_0=\psi_1g$ and using $L_\text{PF}\psi_1=0$, we find that the second-order operator collapses to two antiderivatives, $\theta^2g=-\psi_1^3\zt(1-\zt)(1+8\zt)\,J$ with $J$ the source of the inhomogeneous Picard-Fuchs equation for $S_0$.
We do not have access to the source from the bootstrap perspective; however, there are two properties we can use: 
it is a polylogarithm in $\zt$ over the letters $\zt$ and $1-\zt$ with rational coefficients, and its weight is at most three.
Therefore an admissible kernel is $\psi_1^3\zt(1-\zt)(1+8\zt)\,r(\zt)$ with $r$ rational and singular only at $\zt=0,1$. Requiring it to be a holomorphic modular form of weight three, that is $\psi_1^3$ times a polynomial of degree at most three, bounds the pole orders with no further assumption. This leaves
\begin{equation}
   \mathcal{A}^{(3)}=\psi_1^3\left(1+8\zt\right)\cdot\operatorname{span}_{\mathbb{Q}}\left\{1,\zt,\zt^2\right\}\,,
\label{eq:bootw3}
\end{equation}
three-dimensional inside the four-dimensional $M_3$. 
Together with the constant letter, which the outer integration requires, the elliptic alphabet is
\begin{equation}
   \mathcal{A}_\text{E}=\left\{1\right\}\cup\left\{\omega_0,\omega_1\right\}\cup\mathcal{A}^{(3)}\,.
\label{eq:bootalphabetE}
\end{equation}
In the NNLO bootstrap below, as a robustness test, we will directly build the ansatz on all of $M_3$ and $M_2$, and we will see that the resulting solution confirms this alphabet.

The alphabet fixes the kernels, and the operator $\theta$ fixes the order of iteration.
The iterated integrals are $I(f_1,\dots,f_n;q)=\theta^{-1}[f_1\,I(f_2,\dots,f_n;q)]$ with the leftmost entry outermost.
The elliptic function to be constructed is $g=S_0/\psi_1$, and by the variation of parameters above it is a double antiderivative, $g=\theta^{-2}[f\,J]$ up to solutions of $\theta^2g=0$, of the kernel $f=-\zt\,\psi_1\omega_0\in\mathcal{A}^{(3)}$ times the source $J$, the polylogarithm of weight at most three over the letters $\zt$ and $1-\zt$ characterized above.

Every iterated integral in $g$ therefore has the constant $1$ as its outermost kernel.
Then we have exactly one weight-three kernel as its second entry, and after that only $\omega_0$ and $\omega_1$, which are the one-forms $d\log\zt$ and $-d\log(1-\zt)$ of the polylogarithms inside the source written in the variable $\log q$.
Differentiation respects the order of iteration. 
A weight-5 solution takes the form
\begin{equation}
   g=\lambda_0+\lambda_L L+\sum_{f\in\mathcal{A}^{(3)}}\Big[\sum_{|u|=3}\lambda_{fu}\,I(1,f,\omega_{u_1},\omega_{u_2},\omega_{u_3};q)+\pi^2\sum_{|u|=1}\lambda_{fu}\,I(1,f,\omega_{u_1};q)+\zeta_3\,\lambda_f\,I(1,f;q)\Big]\,,
\label{eq:bootwords}
\end{equation}
with $u=u_1\cdots u_{|u|}$ a string of indices in $\{0,1\}$, $\lambda_{fu}$ the rational coefficients ($\lambda_f$ for the empty string $u$), and the transcendental constants written as $\lambda_0=\lambda_5\zeta_5+\lambda_{23}\pi^2\zeta_3$, $\lambda_L=\lambda_4\pi^4$.

\subsection{Ansatz for rational coefficients}\label{sec:bootcoeff}

What makes a bootstrap program for observables or cross-sections challenging is the rational coefficients. 
For amplitudes in planar $\cN=4$ SYM the coefficient of each transcendental function is a rational number, and thus the whole content of the ansatz is the function list.
Here the coefficients are rational functions of $\zt$, so a second assumption is needed and the number of unknown parameters is dominated by it.
We take the shape proposed for the three-point correlator in Ref.~\cite{Gong:2025eec}, in which the denominator is built from the rational letters of the alphabet with the pole orders left as parameters and the numerator degree tied to them.

Write $\rho=\sqrt{\zt}$, let $T$ run over the basis of total weight at most $2n-1$ for order $\mathcal{O}(a^n)$, and let $p(T)\in\{0,1\}$ be the parity of the type in the sense of Sec.~\ref{sec:bootspace}. Allowing a pole of order $m_0$ at $\zt=0$ and $m_1$ at $\zt=1$, the ansatz is
\begin{equation}
   F^{(n)}(\zt)=\frac{1}{\zt^{\,m_0}\left(1-\zt\right)^{m_1}}\sum_{T}\rho^{\,p(T)}
   \Big[\sum_{j=0}^{\,d-p(T)}c_{T,j}\,\zt^{\,j}\Big]\,T(\rho)\,,\qquad d\ge 2+m_0+m_1\,,
\label{eq:bootansatz}
\end{equation}
with the unknown rational coefficients $c_{T,j}$.
The factor $\rho^{\,p(T)}$ is forced by parity: it makes every term invariant under $\rho\to-\rho$, so that the expansion at $\zt=0$ runs in integer powers of $\zt$ alone. 
Tying the numerator degree to the pole orders by $d=2+m_0+m_1$ fixes the polynomial part of each coefficient, after partial fractioning, to degree two in $\zt$ on the even basis functions and to $\rho$ times degree one on the odd ones. 
Both $m_{0,1}$ and $d$ can be increased in the ansatz, and the final answer should remain the same after imposing the constraints. This also serves as a consistency check.
Note that for the three-point correlator, the leading transcendentality is computed through the spherical contour method and thus one has a sense of how the rational coefficients should look at lower transcendentality. Here we propose the form of \Eq{eq:bootansatz} for every basis function.

The elliptic sector needs the same treatment. Section~\ref{sec:bootspace} constructed $g=S_0/\psi_1$ as a combination $\sum_wa_w\,I(w;q)$ of the iterated integrals of \Eq{eq:bootwords} with rational numbers $a_w$, plus the homogeneous terms. What enters the correlator is not $S_0=\psi_1g$ by itself: the elliptic part of the observable is a combination of $S_0$ and its first derivative with rational-function coefficients, $F_\text{E}=2\left[c_1(\zt)S_0+c_2(\zt)S_0'\right]$ (Sec.~\ref{sec:final_result}), and no higher derivative can occur because the Picard-Fuchs equation is of second order. A bootstrap knows neither $c_1$, $c_2$ nor the $a_w$. Substituting $S_0=\psi_1\sum_wa_wI(w;q)$ shows what it must parametrize: every iterated integral $\psi_1I(w;q)$ and its $\zt$-derivative, each with an unknown rational function in front. The general linear ansatz for the elliptic sector is therefore
\begin{equation}
   F_\text{E}(\zt)=\sum_w\bigg[r_w(\zt)\,\psi_1\,I(w;q)
   +s_w(\zt)\,\frac{d}{d\zt}\Big(\psi_1\,I(w;q)\Big)\bigg]\,,
\label{eq:bootansatzE}
\end{equation}
with $r_w$ and $s_w$ rational functions of the shape~\eqref{eq:bootansatz} with poles at $\zt=0,1$, chosen independently for every iterated integral.

\subsection{Physical and numerical constraints}\label{sec:bootconstraints}

In this subsection, we describe the physical constraints that we will use to determine the unknown parameters in the bootstrap. They are most transparent for the EEC distribution $d\sigma/d\zt$ and can be translated to $F$.
Since $F=2\zt^2(1-\zt)\,d\sigma/d\zt$, a term $f_{k,j}\zt^{\,k}\log^{\,j}\zt$ of $F$ contributes to $d\sigma/d\zt$ at order $\zt^{\,k-2}$, and with $y=1-\zt$ a term $b_{k,j}y^{\,k}\log^{\,j}y$ contributes at order $y^{\,k-1}$. 
Every condition below is imposed channel by channel: the unknown parameters are rational numbers, so we equate the coefficients of $1$, $\pi^2$, $\zeta_3$, $\pi^4$, $\zeta_5$ and $\pi^2\zeta_3$ separately. We list the conditions in order of what they require.

\emph{(C1) Regularity.} The cross-section diverges no faster than $1/\zt$ in the collinear limit and no faster than $1/y$ in the back-to-back limit, which bounds the powers in $F$. The logarithms are bounded by the loop order: the collinear expansion is a single-logarithm series in $a\log\zt$, so at order $a^n$ it carries at most $\log^{\,n-1}\zt$, and the back-to-back expansion is a double-logarithm series, so at order $a^n$ it carries at most $\log^{\,2n-1}y$. These properties follow purely from our understanding of soft and collinear divergences, and we impose them in the endpoint expansion.

A further condition of the same kind is that the local exponents at both endpoints are integers: both expansions of $d\sigma/d\zt$ run in integer powers of the local variable times logarithms. On the polylogarithmic ansatz this is automatic once the parity in $\rho$ is built in. It matters when the elliptic sector is written on a system larger than the Picard-Fuchs operator: the coupled system in Eq.~\eqref{eq:sys2} has four homogeneous solutions, and the two that are not periods behave as $y^{\pm\sqrt{5/3}}$ at $\zt=1$. 
Requiring integer exponents removes these two solutions and thereby selects the physical combination, without using the fact that the system in Eq.~\eqref{eq:sys2} eventually decouples.
It is this requirement that forces the function $\Psi$ of \Eq{eq:Psifull} to be a polylogarithm. The elliptic ansatz in Eq.~\eqref{eq:bootansatzE} satisfies the requirement automatically: $\psi_1$ and the factors $\zt$, $1-\zt$ and $1+8\zt$ of \Eq{eq:frame} are eta quotients, which vanish to integer order at each of the four cusps, so every term of the ansatz has integer local exponents there, up to powers of logarithms.

\emph{(C2) Collinear leading power.} As $\zt\to0$ the light-ray operator product expansion (OPE)~\cite{Hofman:2008ar,Kravchuk:2018htv,Kologlu:2019mfz} organizes $d\sigma/d\zt$ in powers of $\zt$ labeled by the twist of the exchanged light-ray operators. 
The leading twist-two term is $1/\zt$ times a series in $a$ and $\log\zt$ that exponentiates to a power of $\zt$ set by twist-two anomalous dimensions~\cite{Korchemsky:2019nzm,Dixon:2019uzg}. Expanded to order $a^n$, it is $1/\zt$ times a polynomial in $\log\zt$ of degree $n-1$ with predicted coefficients. In $F$ this is the $\zt^1$ bracket, $F^{(n)}=\zt\sum_{j<n}c_{nj}\log^j\zt+\cO(\zt^2)$, at NNLO the first line of \Eq{eq:FNNLOcoll}. 
In QCD the same information comes from the collinear factorization theorem~\cite{Dixon:2019uzg,Chen:2021gdk,Chen:2023zzh}. 

\emph{(C3) Back-to-back leading power.} As $\zt\to1$ the distribution is a Sudakov form factor: $d\sigma/d\zt$ is $1/y$ times a double-logarithmic series whose coefficients follow from effective field theory predictions~\cite{Moult:2018jzp,Korchemsky:2019nzm}, at order $a^n$ a polynomial in $\log y$ of degree $2n-1$. In $F$ this is the $y^0$ bracket, at NNLO the first line of \Eq{eq:FNNLOb2b}. Its constants are multiple zeta values and contain no $\log2$.

Condition (C3) contains a second requirement that is easy to overlook. The HPLs of the ansatz are functions of $\rho=\sqrt\zt$, and their natural local variable at the endpoint is $1-\rho$ rather than $y=(1-\rho)(1+\rho)$. If we re-expand in $y$, their endpoint values contain powers of $\log2$ at every weight, whereas the leading power Sudakov series contains none. The cancellation of all $\log2$ terms at order $y^0$ is therefore a set of equations in its own right, separate from matching the $\log2$-free coefficients. The elliptic iterated integrals obey a stronger version of the same requirement: their values at the cusp $\zt=1$ involve periods of level six outside the multiple zeta values, such as the sixth-root-of-unity constant in $S_0(1)$ of \Eq{eq:FNNLOb2b}, and these must cancel at order $y^0$ as well. Using this needs the expansion of each elliptic iterated integral at the second cusp. 

\emph{(C4) Higher powers.} At next-to-leading power in the back-to-back limit, order $y^0$ in $d\sigma/d\zt$ and order $y^1$ in $F$, the two highest logarithms are known from the subleading-power resummation~\cite{Moult:2019vou,Chen:2023pcr}: at NNLO the coefficients $\tfrac1{10}$ and $-\tfrac13$ of $y\log^5y$ and $y\log^4y$ in \Eq{eq:FNNLOb2b}. 
In principle, further powers in either expansion can be obtained by expanding the integrand and integrating. But they are computations rather than physical constraints. 

\emph{(C5) Sum rules.} The EEC satisfies two sum rules~\cite{Dixon:2019uzg,Korchemsky:2019nzm,Kologlu:2019mfz}:
\begin{equation}
    \int_0^1 d\zt \frac{d\sigma}{d\zt} = \sigma_{\text{tot}},\qquad \int_0^1 d\zt \zt \frac{d\sigma}{d\zt} = \frac{1}{2}\sigma_{\text{tot}}\,.
\end{equation}
Here $\sigma_{\text{tot}}$ is the total cross-section in a given process. Note that in the sum rules, one has to also include the contact terms $\delta(\zt)$ and $\delta(1-\zt)$.
These terms are considered singular terms, similar to the large logarithms $\log(\zt)$ and $\log(1-\zt)$. Therefore, they can be predicted from the light-ray OPE in $\cN=4$ SYM, and from effective field theories in QCD. 

At NLO we will impose the zeroth sum rule, while at NNLO we will impose neither. The moments of the complete elliptic sector are multiple zeta values, as the sum rule requires. However, the integrals of individual elliptic basis elements are not. To properly impose the sum rule, we will need to work out these integrals analytically. We skip this because it does not buy us many constraints.

The above physical constraints cannot determine the unknown parameters of the bootstrap: $130$ of $206$ parameters remain undetermined at NLO and $1963$ of $2336$ at NNLO. This is precisely why bootstrapping cross-section observables or QCD amplitudes is difficult. Unlike for planar $\cN=4$ SYM amplitudes or form factors, we need to determine the rational functions, and each function comes with many unknown parameters. The number of physical constraints is far from enough.
As proposed in Ref.~\cite{Barrera:2025anr}, one can equip a bootstrap program with analytic regression from high-precision numerical data. There are different approaches to performing the analytic regression. 
The most straightforward method is the linear fit, where one obtains floating-point numbers for all unknown parameters. These values can then be reconstructed as rational numbers and transcendental numbers like $\pi$ and $\zeta_n$ through PSLQ~\cite{Ferguson:1992,PSLQref,Bailey:1999nv}.
Alternatively, as introduced in Ref.~\cite{Barrera:2025anr}, one can use lattice reduction~\cite{Lenstra:1982,Nguyen:2001,fplll} to search for integer relations among the function itself and all the transcendental basis functions. Lattice reduction can use either one or multiple numerical points, where the one-point case reduces to PSLQ.

In general, a bootstrap program equipped with analytic regression requires a numerical routine to calculate the observables or amplitudes to high precision. We do not have one for the NNLO EEC independent of the analytic result. However, as a first look, we can generate fake numerical data from the analytic expression and test how robust the bootstrap method can be. The exercises in the next subsection proceed in this spirit.

\subsection{Results}\label{sec:bootresults}
With the function space of Sec.~\ref{sec:bootspace}, the coefficient ansatz of Sec.~\ref{sec:bootcoeff} and the conditions of Sec.~\ref{sec:bootconstraints} in hand, we carry out the bootstrap at the three known orders. Throughout, we use HPLs with argument $\rho=\sqrt{\zt}$ and omit it as in Sec.~\ref{sec:final_result}.
LO and NLO are purely polylogarithmic; NNLO includes the elliptic sector.
We will report the number of unknown parameters at each step of the procedure.

\subsubsection{LO and NLO bootstrap}

The LO EEC is simple enough to bootstrap without any numerical data input.
We first work with the two rational letters and then admit the third letter. 
With $H_0=\frac{1}{2}\log\zt$ and $H_1-H_{-1}=-\log(1-\zt)$ the ansatz is $p_0(\zt)+p_1(\zt)H_0+p_2(\zt)(H_1-H_{-1})$ with quadratic polynomials $p_i$ and the number of unknown parameters is nine.
The logarithm bound of (C1) removes the three coefficients of $p_1$; the collinear condition $F_\text{LO}=\zt+\cO(\zt^2)$ gives two equations; the back-to-back condition $F_\text{LO}\to-\log y$ gives two; the two sum rules give four, of which two are independent of the rest.
The chain is $9\to6\to4\to2\to0$, and the unique solution is
\begin{equation}
   F_\text{LO}(\zt)=-\log(1-\zt)=H_{-}\,,
\label{eq:bootLO}
\end{equation}
in the notation of \Eq{eq:pmletters}. If we include the square-root letter of \Eq{eq:bootalphabetP}, two further terms, odd in $\sqrt\zt$, enter the ansatz and the number of unknown parameters becomes eleven. 
The same conditions give the same answer, with the two new coefficients zero. 

The numerator degree $d$ in \Eq{eq:bootansatz} is unknown beforehand and may differ among coefficients. One picks a $d$, runs the program, and raises $d$ if the conditions are inconsistent; once a $d$ works, raising it further tests the stability of the result. For the LO EEC, $d=2$ already closes the system from the physical constraints alone. We then verify that $d=3$ also gives the same answer.

At NLO the total weight is at most three, so the basis is the $40$ HPLs of weight zero to three carrying the constants $1$, $\pi^2$ and $\zeta_3$, adding up to $45$ basis functions ($26$ even and $19$ odd). 
With $m_0=m_1=1$ and $d=4$, each even basis function carries five numerator coefficients and each odd one four, so \Eq{eq:bootansatz} has $206$ unknown rational parameters. 
Tab.~\ref{tab:bootNLO} summarizes the number of parameters after applying each constraint. 
Regularity alone removes $43$, the leading-power data at the two endpoints remove $22$ more, and the zeroth sum rule removes $11$. In all, $76$ of the $206$ parameters are fixed without sampled values.

\begin{table}[!htbp]
\centering\footnotesize
\setlength{\tabcolsep}{6pt}
\begin{tabular}{@{}p{9.2cm}rr@{}}
\hline\hline
NLO, $206$ parameters & equations & left \\
\hline
ansatz: weights $0$-$3$, $m_0=m_1=1$, $d=4$, parity in $\rho$ & --- & $206$ \\
\hline
no power divergence at $\zt=0$ & $7$ & $199$ \\
\quad nothing at order $\zt^0$ & $7$ & $192$ \\
no power divergence at $\zt=1$ & $14$ & $178$ \\
at most $\log\zt$ at every order in $\zt$ & $15$ & $163$ \\
at most $\log^3y$; integer local exponents & $0$ & $163$ \\
\quad\emph{regularity, subtotal} & $43$ & $163$ \\
\hline
collinear $\zt^1$ bracket, light-ray OPE & $5$ & $158$ \\
back-to-back $y^0$ bracket, Sudakov, $\log2$-free & $7$ & $151$ \\
\quad the $\log2$ structures must cancel & $7$ & $144$ \\
$y^1$ coefficients of $\log^3y$ and $\log^2y$ & $3$ & $141$ \\
\quad\emph{all analytic conditions, subtotal} & $65$ & $141$ \\
\hline
the zeroth sum rule & $11$ & $130$ \\
\hline
linear solve, $126$ points at $180$ digits & --- & $0$ \\
\hline\hline
\end{tabular}
\caption{The NLO EEC bootstrap on the $206$-parameter ansatz. The number of equations and the number of unknown parameters at each step are listed. Finally, the remaining $130$ parameters can be determined by numerical data. }
\label{tab:bootNLO}
\end{table}

The remaining 130 parameters can be fixed from sampled values. As mentioned above, we are doing this study to test the ability of the bootstrap, so we use the numerical evaluation of the true answer to some precision as data. In a realistic bootstrap program, one needs to have an independent numerical evaluator of the observable. 
We sample 126 points to around 180 digits and use a linear solver to determine the unknown parameters. The resulting floating-point numbers are then rationalized to get the analytic form. 
If the sum rule is not applied, then we need 134 points at 200 digits. 
We also perform two further consistency checks: (1) raise the powers of the denominator to $m_0=m_1=2$ or 3, while keeping $d=2+m_0+m_1$; (2) raise the numerator degree to $d=5$ or $6$, with $m_0=m_1=1$ fixed. In both cases, we find the same final answer and thus verify the stability of this procedure.

We also run the lattice reduction for the final steps. We test the performance with different numbers of points and digits of precision. The number of digits required decreases with the number of sample points and saturates at 25. The comparison with linear solve is given in Tab.~\ref{tab:bootfrontier}.

\begin{table}[!htbp]
\centering\footnotesize
\setlength{\tabcolsep}{10pt}
\begin{tabular}{@{}rcc@{}}
\hline\hline
sample points $N$ & lattice reduction & linear solve \\
\hline
$8$   & $50$--$52$ & --- \\
$16$  & $30$--$32$ & --- \\
$32$  & $26$--$28$ & --- \\
$48$  & $25$--$26$ & --- \\
$126$ & $25$ & $174$--$188$ \\
$260$ & $25$ & $160$ \\
\hline\hline
\end{tabular}
\caption{Precision digits needed to determine the remaining parameters in the NLO EEC bootstrap, using both lattice reduction and linear solve. With enough sample points, the required precision saturates at 25 digits.}
\label{tab:bootfrontier}
\end{table}

In the end, we find the following NLO expression $F_{\text{NLO}}=F^{[2]}+F^{[3]}$, with
\begin{align}
   F^{[2]}&=\left(\zt-1\right)\left[4 H_{-,0}-4\sqrt{\zt}H_{+,0}-4\zt H_{0,-}-2\left(\zt-1\right) H_{-,-}-\frac{\pi^2}{3}\zt\right]\,,
\label{eq:bootNLOw2}
\end{align}
and
\begin{align}
   F^{[3]}&=4\zt\left(\zt-2\right)H_{0,0,-}-2\zt\left(4\zt-3\right)H_{0,-,0}-\zt\left(2\zt+1\right)H_{0,-,-}\nn\\
   &+\left(2\zt^2-3\zt-4\right)H_{-,0,-}-\zt\left(4\zt-3\right)H_{-,-,0}-\left(\zt^2+2\right)H_{-,-,-}\nn\\
   &+\left(\zt-1\right)\left(2\zt+1\right)H_{+,+,0}+\frac{\pi^2\zt^2}{3}H_{0}
   +\frac{\pi^2}{12}\left(2\zt^2+\zt-2\right)H_{-}-\frac{\zt\left(4\zt+1\right)}{2}\zeta_3\,,
\label{eq:bootNLOw3}
\end{align}
which agrees term by term with Ref.~\cite{Belitsky:2013ofa}. Exactly one odd harmonic polylogarithm survives, $H_{+,0}$, with its compulsory factor $\sqrt\zt$.

\subsubsection{NNLO bootstrap}

At NNLO both sectors are present and every condition applies to their sum, so the two ans\"atze are added before anything is imposed. In the polylogarithmic sector the basis of total weight at most five has $423$ elements, $220$ even and $203$ odd; with $m_0=m_1=0$ and $d=2$, this gives $1066$ unknown parameters. In the elliptic sector, we do not use the restriction of the kernels to $\mathcal{A}^{(3)}$ or of the tails to $\omega_0,\omega_1$ as input but let the bootstrap decide. With all four kernels $\psi_1^3\{1,\zt,\zt^2,\zt^3\}$ of $M_3$ and all three letters of $M_2$ as tails, \Eq{eq:bootwords} has $124$ iterated integrals plus the three homogeneous terms. Each has the two prefactors of \Eq{eq:bootansatzE} with five coefficients each, which leads to $1270$ unknown parameters. In total, we have $2336$ parameters to be determined. 

\begin{table}[!htbp]
\centering\footnotesize
\setlength{\tabcolsep}{8pt}
\begin{tabular}{@{}p{9.2cm}rr@{}}
\hline\hline
NNLO, joint, $2336$ parameters & equations & left \\
\hline
polylogarithmic ansatz, weights $0$--$5$, $m_0=m_1=0$, $d=2$ & --- & $1066$ \\
elliptic ansatz, all $127$ terms, degree $\le4$ & --- & $1270$ \\
\hline
no power divergence at $\zt=0$, nothing at order $\zt^0$ & $21$ & $2315$ \\
no power divergence at $\zt=1$ & $59$ & $2256$ \\
at most $\log^2\zt$ at every order in $\zt$ & $81$ & $2175$ \\
at most $\log^5y$; integer local exponents & $0$ & $2175$ \\
\hline
collinear $\zt^1$ bracket, light-ray OPE & $13$ & $2162$ \\
back-to-back $y^0$ bracket, Sudakov, $\log2$-free; the $\log2$ structures cancel & $199$ & $1963$ \\
\hline
reorganization & $113$ & $1850$ \\
\hline
lattice reduction:\\
\hspace{1.0cm}  \# of points $= 150$, digits $=350$, time $=119.5$ CPU hours\\
\hspace{1.0cm} \# of points $=300$, digits $=270$, time $=50.4$ CPU hours\\
\hspace{1.0cm} \# of points $=600$, digits $=225$, time $=56.1$ CPU hours\\
\hspace{1.0cm} \# of points $=1200$, digits $=205$, time $=62.1$ CPU hours\\
\hline\hline
\end{tabular}
\caption{The NNLO bootstrap on the joint ansatz. The number of parameters left after each constraint is listed. In the reorganization step, some unknowns appear only in specific linear combinations, which we replace with new variables. The physical constraints leave $1963$ of $2336$ parameters and the reorganization $1850$. We run the lattice reduction in four setups to regress the $1850$ parameters, as shown in the last block.
The CPU hours are recorded on a machine with an Intel Xeon Platinum 8581C (2.30\,GHz, 48 cores, 96 threads), using 32 threads and up to $21$\,GB of memory.
}
\label{tab:bootNNLO}
\end{table}

Tab.~\ref{tab:bootNNLO} lists the number of parameters after each constraint. Regularity removes $161$ and the leading-power data in the two limits $212$, in all $373$ of $2336$. Compared to NLO, only 16\% of the parameters are determined by the physical constraints. The next-to-leading power back-to-back logarithms of (C4) are not needed, since the reduction below closes without them.
After that, it turns out that some of the parameters only appear in certain linear combinations, so we can define new parameters to replace these combinations. This removes a further $113$ parameters, and we are left with $1850$ unknowns.

We again test the lattice reduction on the remaining parameters with fake numerical data generated from the analytic expression. We use \textsc{Flatter}~\cite{Ryan:2023flatter} to perform the reduction. As shown in the last block of Table~\ref{tab:bootNNLO}, with $150$ numerical points and $350$ digits we can determine the remaining $1850$ parameters all at once, and more points buy fewer digits per point: $300$ points at $270$ digits, $600$ at $225$, and $1200$ at $205$ also return the exact coefficient vector. An information count of the sample set (the lattice volume above $10^{-p}$ for $N$ points at $p$ digits) indicates that this gain levels off near $200$ digits per point, and that at $100$ digits or fewer no number of points up to $2400$ suffices.

The lesson we learn from the NNLO bootstrap is that physical constraints are far from sufficient to determine an expression with rational coefficients. In this example, $80\%$ of the parameters are fixed from the numerical data. 
On the one hand, this suggests that we need to explore the analytic structure of energy correlators and come up with more consistency conditions. One direction is to cleanly separate the polylogarithmic sector from the elliptic sector. The NNLO EEC in $\cN=4$ SYM is computed from the four-point correlator, which makes the separation challenging. It will be interesting to see if the integration-by-parts~\cite{Tkachov:1981wb,Chetyrkin:1981qh,Laporta:2000dsw} and differential equations~\cite{Kotikov:1990kg,Remiddi:1997ny,Gehrmann:1999as,Henn:2013pwa} approach provides a handle for identifying elliptic sectors. 
On the other hand, the analytic regression with numerical data is powerful enough to accurately reconstruct the analytic result. This implies that numerical data contain more hidden information and that developing numerical frameworks for precise evaluation is important.

\section{Conclusions}\label{sec:outlook}

In this paper, we have completed the analytic calculation of the energy-energy correlator in $\cN=4$ SYM at NNLO. The result of Ref.~\cite{Henn:2019gkr} consists of HPLs up to weight five, together with one two-fold integral over a polylogarithmic integrand, which had resisted evaluation. A partial fraction decomposition guided by the Galois symmetry of the square roots generated by the inner integration separates it into a piece that integrates to HPLs and two irreducible pieces that live on a single elliptic curve. This curve is the curve of the equal-mass sunrise integral: the two $j$-invariants agree identically under a M\"obius map, so that the elliptic sector of the EEC is built from the modular forms of $\Gamma_1(6)$. Solving the associated Picard-Fuchs system reduces the elliptic part of the correlator to a single function $S_0$ and its first derivative, multiplied by two rational functions. $S_0$ itself is the period $\psi_1$ times a finite combination of iterated Eisenstein integrals whose coefficients are rational multiples of $1$, $\pi^2$, $\zeta_3$, $\pi^4$, $\zeta_5$ and $\pi^2\zeta_3$, as shown in Eq.~\eqref{eq:S0words}. The complete result, $F_\text{NNLO}=F_\text{P}+F_\text{E}$, is given in Sec.~\ref{sec:final_result}. Since the iterated Eisenstein integrals are polynomials in $L=\log q$ with rational $q$-series coefficients, the elliptic sector can be evaluated to high precision as readily as the polylogarithmic one. We also provide the expansions of the NNLO correlator to subleading powers in both the collinear and the back-to-back limits.

The closed form makes possible a first look at a perturbative bootstrap of the EEC. At LO, at NLO and in the polylogarithmic sector at NNLO, the function space is standard, spanned by HPLs in the three letters $\zt$, $1-\zt$ and $(1-\sqrt{\zt})/(1+\sqrt{\zt})$. In the elliptic sector, the curve, the Picard-Fuchs operator and the location of the physical singularities restrict the kernels to six, which allows us to write down an elliptic ansatz. That the resulting function space is finite is far from automatic, and it indicates how bootstrap methods may be extended to observables with elliptic sectors. 
We further formulate a general ansatz for the rational coefficients and collect the physical constraints that the observable supplies. Then we show that the parameters these leave undetermined can be fixed from numerical data by analytic regression, in particular by lattice reduction. 
In this first study, the data for these tests were generated from the analytic result of this paper rather than from an independent numerical evaluator of the EEC.

Both the direct calculation and the bootstrap study carry lessons for QCD. The elliptic function space is fixed by geometry rather than by the details of the theory. The curve and its level follow from the maximal cut of the integrand, and thus it will be interesting to perform the corresponding analysis for the QCD integrand. Once the curve and level are known, the elliptic letters and the function space follow as they did here. 
A complete bootstrap of the NNLO correlator in QCD remains difficult, but knowing the function space in advance can bypass intermediate steps of a direct calculation and simplify its result. 
Finally, the $q$-series evaluation of iterated Eisenstein integrals used here will also allow the QCD result, once obtained, to be evaluated quickly and to high precision, which will be crucial for phenomenological studies at colliders. 
With the elliptic sector of the simplest event shape now under the same control as its polylogarithmic part, both analytically and numerically, the way is open to bring the same tools to bear on precision QCD.

\begin{acknowledgments}
XYZ is supported by the MIT Pappalardo Fellowship.
MDS is supported in part by the U.S. Department of Energy under grant DE-SC0013607. This work was supported in part by the National Science Foundation under Cooperative Agreement PHY-2019786 (The NSF AI Institute for Artificial Intelligence and Fundamental Interactions, \url{https://iaifi.org/}).
Funding for this project was provided by Anthropic. This is not an official Anthropic project and results and views are not endorsed by Anthropic.
\end{acknowledgments}

\appendix

\section{Function classes and conventions}\label{app:defs}

\emph{Generalized polylogarithms.} 
The generalized polylogarithm is defined as~\cite{Goncharov:1998kja}
\begin{equation}
   G(a_1,\dots,a_n;x)=\int_0^x\frac{dt}{t-a_1}\,G(a_2,\dots,a_n;t)\,,\qquad G(\,;x)=1\,,
\label{eq:Gdef}
\end{equation}
with $a_1$ outermost; the length $n$ is the transcendental weight. For $a_n=0$ the innermost integration diverges and we use the standard regularization
\begin{equation}
   G(0^{\,k};x)=\frac{\log^k x}{k!}\,,\qquad
   G(a_1,\dots,a_m;x)\,G(a_{m+1},\dots,a_{m+n};x)=\sum_{\sigma}G\big(a_{\sigma(1)},\dots,a_{\sigma(m+n)};x\big)\,,
\label{eq:Greg}
\end{equation}
the sum running over the interleavings that preserve the order within each factor (the shuffle product), which expresses every $G$ with trailing zeros in terms of convergent ones, e.g.\ $\log x\,G(0,1;x)=2G(0,0,1;x)+G(0,1,0;x)$.

\emph{Harmonic polylogarithms}~\cite{Remiddi:1999ew} are the case $a_i\in\{0,1\}$ with the kernel of the index $1$ taken as $1/(1-t)$, so that
\begin{equation}
   H_u(\zt)=(-1)^{\,n_1(u)}\,G(u;\zt)\,,
\label{eq:HG}
\end{equation}
$n_1(u)$ being the number of $1$'s in $u$; thus $H_0(\zt)=\log\zt$, $H_1(\zt)=-\log(1-\zt)$ and $H_{0^{\,n-1},1}=\mathrm{Li}_n$. Two subscript conventions occur in the text. In the expanded notation (Sec.~\ref{sec:solving}, Eq.~\eqref{eq:Psifull}) the subscript is the full index vector. In the condensed notation of Ref.~\cite{Remiddi:1999ew} (Eq.~\eqref{eq:PfullHPL}) an index $m\ge1$ stands for $m-1$ zeros followed by a $1$, e.g.\ $H_4=H_{0,0,0,1}$, $H_{3,0}=H_{0,0,1,0}$, $H_{1,2}=H_{1,0,1}$; the sign in Eq.~\eqref{eq:HG} is always read off the expanded form. The letters $\pm$ for HPLs of argument $\sqrt\zt$ are defined in Eq.~\eqref{eq:pmletters}. Both conventions are implemented in Ref.~\cite{Maitre:2005uu}.

\emph{Iterated Eisenstein integrals.} In the frame of Eq.~\eqref{eq:qframe}, with $L=\log q$ and $\theta=q\,d/dq$, the iterated integrals of Sec.~\ref{sec:solving} are $I(f_1,\dots,f_n;q)=\theta^{-1}[f_1\,I(f_2,\dots,f_n;q)]$ with $I(\,;q)=1$, $f_1$ outermost, the length again being the transcendental weight; $\theta^{-1}$ acts on $\mathbb{Q}[[q]][L]$ by the rules of Eq.~\eqref{eq:Iiter}, which correspond to the tangential base point at $q=0$, introduce no integration constants, and preserve the shuffle product. ``Weight'' thus has two senses, the transcendental weight of an iterated integral and the modular weight of a letter. Harmonic polylogarithms of $\zt$ are the special case in which every letter has modular weight two: since $\theta\log\zt=\omega_0$ and $\theta[-\log(1-\zt)]=\omega_1$, one has $\log\zt=I(\omega_0;q)$ and $H_u(\zt)=I(\omega_{u_1},\dots,\omega_{u_n};q)$ for $u$ ending in $1$, so the polylogarithmic sector lies in the same function class as the elliptic one. Whether a given $q$-series is an admissible letter of modular weight $k$ is decided by the criterion below Eq.~\eqref{eq:Mk}; the four letters of Eq.~\eqref{eq:letters} were identified in this way.

\section{The sources of the elliptic system}\label{app:EEp}

In this appendix, we provide the analytic expressions for $E$ and $\tilde{E}$, the sources of the elliptic system defined by Eq.~\eqref{eq:sys2}. Both sources are of weight at most four. Explicitly, we have 
\begin{align}
   E(\zt)&=-\frac{64 \zt^{2} + 3 \zt - 1}{2 \zt}-\frac{3 \left(4 \zt - 3\right) \left(4 \zt + 1\right)}{4 \zt}\,G(1;\zt)-\frac{\left(2 \zt + 1\right)^{2}}{\zt}\,G(0;\zt)+\frac{\left(\zt - 1\right) \left(\zt + 2\right)}{\zt}\,G(1,1;\zt)\nn\\
   &-\frac{18 \zt^{3} - 31 \zt^{2} - 16 \zt - 1}{2 \zt \left(\zt - 1\right)}\,G(0,1;\zt)+\frac{6 \zt^{2} - \zt + 1}{2 \zt}\,G(1,0;\zt)+\left(5 \zt + 1\right)\,G(0,0;\zt)\nn\\
   &+\frac{4 \zt^{3} + 66 \zt^{2} - 3 \zt - 7}{24 \zt \left(\zt - 1\right)}\,\pi^2+\frac{2 \zt^{2} + 1}{\zt}\Big[G(0,1,0;\zt)-G(1,0,0;\zt)\Big]\nn\\
   &-\frac{24 \zt^{4} - 288 \zt^{3} + 208 \zt^{2} + 19 \zt - 8}{4 \zt \left(\zt - 1\right)}\,\zeta_3-\frac{2 \left(\zt - 1\right) \left(3 \zt^{2} - 1\right)}{\zt}\,G(0,1,1;\zt)\nn\\
   &-\frac{\left(2 \zt + 1\right) \left(12 \zt^{3} - 54 \zt^{2} + 35 \zt - 8\right)}{4 \zt \left(\zt - 1\right)}\,G(1,0,1;\zt)+\frac{24 \zt^{3} - 88 \zt^{2} + 8 \zt + 11}{4 \left(\zt - 1\right)}\,G(0,0,1;\zt)\nn\\
   &+\frac{2 \left(\zt - 1\right)^{2} \left(3 \zt + 2\right)}{\zt}\,G(1,1,1;\zt)+\frac{40 \zt^{3} + 8 \zt^{2} + 5 \zt - 8}{24 \zt \left(\zt - 1\right)}\,\pi^2\,G(1;\zt)\nn\\
   &-\frac{\left(4 \zt - 1\right) \left(4 \zt + 1\right)}{12 \zt}\,\pi^2\,G(0;\zt)+\frac{2 \zt + 1}{60}\,\mathcal{W}_{1}\,,
\label{eq:Efull}
\end{align}
and
\begin{align}
   \tilde{E}(\zt)&=-\frac{1}{2 \zt}\Big[24+24\,G(1;\zt)-6\,G(0,0;\zt)+6\,G(1,0;\zt)+9\,G(0,1,0;\zt)-9\,G(1,0,0;\zt)\nn\\
   &+2\,\pi^2\,G(0;\zt)\Big] -\frac{12 \left(\zt - 1\right)}{\zt}\,G(1,1;\zt)+\frac{6 \left(2 \zt^{2} - 4 \zt + 1\right)}{\zt \left(\zt - 1\right)}\,G(0,1;\zt)-\frac{2 \zt^{2} + \zt - 2}{\zt \left(\zt - 1\right)}\,\pi^2\nn\\
   &-\frac{28 \zt^{2} - 81 \zt + 62}{2 \zt \left(\zt - 1\right)}\,\zeta_3-\frac{28 \zt - 23}{2 \zt}\,G(0,1,1;\zt)-\frac{28 \zt^{2} - 41 \zt + 22}{2 \zt \left(\zt - 1\right)}\,G(1,0,1;\zt)\nn\\
   &+\frac{28 \zt^{2} - 18 \zt - 1}{2 \zt \left(\zt - 1\right)}\,G(0,0,1;\zt)+\frac{14 \zt - 23}{\zt}\,G(1,1,1;\zt)-\frac{11 \zt - 8}{4 \zt \left(\zt - 1\right)}\,\pi^2\,G(1;\zt)-\frac{4 \zt - 1}{60 \zt}\,\mathcal{W}_{1}\,,
\label{eq:Epfull}
\end{align}
where both $E(\zt)$ and $\tilde{E}(\zt)$ contain
\begin{align}
   \mathcal{W}_{1}\;=\;{}&120\,G(0,0,0,1;\zt)+60\,G(0,0,1,1;\zt)-60\,G(0,1,0,1;\zt)\nn\\
   &-120\,G(0,1,1,1;\zt)-240\,G(1,0,0,1;\zt)+60\,G(1,0,1,1;\zt)+180\,G(1,1,0,1;\zt)\nn\\
   &-15\,\pi^2\,G(0,0;\zt)-30\,\pi^2\,G(0,1;\zt)+15\,\pi^2\,G(1,0;\zt)+30\,\pi^2\,G(1,1;\zt)\nn\\
   &+420\,\zeta_3\,G(0;\zt)-300\,\zeta_3\,G(1;\zt)-4\,\pi^4\,.
\label{eq:Wbrackets}
\end{align}
The weight-four bracket drops out of the combination that drives Eq.~\eqref{eq:S0theta}: by Eq.~\eqref{eq:HG} the last brace of Eq.~\eqref{eq:Psifull} equals $-\mathcal{W}_1/90$, so $\tilde\beta \Psi$ contains $+\frac{4\zt-1}{60\zt}\,\mathcal{W}_{1}$, which cancels the last term of Eq.~\eqref{eq:Epfull}; the source $\tilde{E}+\tilde\beta \Psi$ is therefore of weight at most three, as used in Sec.~\ref{sec:bootspace}.
The sources of the second channel, which enter Eq.~\eqref{eq:sysboth} in the same way, are
\begin{align}
   E_{R_2}(\zt)&=\frac{\zt \left(2 \zt - 1\right) \left(8 \zt + 1\right)}{\zt - 1}+\frac{32 \zt^{3} - 46 \zt^{2} - 12 \zt - 1}{2 \left(\zt - 1\right)}\,G(1;\zt)+\frac{\zt \left(16 \zt^{2} - 23 \zt - 2\right)}{\zt - 1}\,G(0;\zt)\nn\\
   &-\frac{11 \zt^{2} - 2 \zt + 6}{\zt - 1}\,G(1,0;\zt)-\frac{96 \zt^{4} - 134 \zt^{3} + 78 \zt^{2} - 9 \zt - 1}{2 \left(\zt - 1\right)^{2}}\,G(0,1;\zt)\nn\\
   &+\left(24 \zt^{2} - 11 \zt + 2\right)\,G(1,1;\zt)-\frac{\zt \left(24 \zt^{2} - 11 \zt + 2\right)}{\zt - 1}\,G(0,0;\zt)-\frac{\zt \left(24 \zt^{3} + 7 \zt^{2} - 14 \zt - 2\right)}{6 \left(\zt - 1\right)^{2}}\,\pi^2\nn\\
   &+\frac{3 \zt \left(12 \zt^{3} - 36 \zt^{2} + 24 \zt + 13\right)}{2 \left(\zt - 1\right)^{2}}\,\zeta_3+\frac{28 \zt^{2} + 3 \zt - 7}{2 \left(\zt - 1\right)}\,G(1,0,0;\zt)\nn\\
   &+\frac{4 \zt^{3} - 3 \zt^{2} + 62 \zt + 3}{2 \left(\zt - 1\right)^{2}}\,G(0,1,0;\zt)-\frac{36 \zt^{4} - 160 \zt^{3} + 157 \zt^{2} + \zt + 5}{2 \left(\zt - 1\right)^{2}}\,G(0,0,1;\zt)\nn\\
   &+\frac{18 \zt^{3} - 64 \zt^{2} + 30 \zt - 5}{\zt - 1}\,G(0,1,1;\zt)+\frac{36 \zt^{4} - 84 \zt^{3} + 97 \zt^{2} + 27 \zt + 29}{2 \left(\zt - 1\right)^{2}}\,G(1,0,1;\zt)\nn\\
   &-6 \left(3 \zt^{2} - 4 \zt + 2\right)\,G(1,1,1;\zt)-\frac{36 \zt^{3} - 70 \zt^{2} - 69 \zt - 2}{12 \left(\zt - 1\right)^{2}}\,\pi^2\,G(1;\zt)+\frac{\zt \left(32 \zt^{2} + 1\right)}{6 \left(\zt - 1\right)^{2}}\,\pi^2\,G(0;\zt)\nn\\
   &-\frac{36 \zt^{3} - 5 \zt^{2} - 40 \zt + 3}{2 \left(\zt - 1\right)^{2}}\Big[G(1,0,0,1;\zt)-\zeta_3\,G(1;\zt)\Big]\nn\\
   &+\frac{148 \zt^{3} - 168 \zt^{2} - 9 \zt - 1}{12 \left(\zt - 1\right)^{2}}\Big[6\,G(1,1,0,1;\zt)+\pi^2\,G(1,1;\zt)\Big]\nn\\
   &+\frac{\zt \left(160 \zt^{2} - 173 \zt - 45\right)}{80 \left(\zt - 1\right)^{2}}\,\pi^4-\frac{3 \zt \left(10 \zt^{2} - 14 \zt - 3\right)}{\left(\zt - 1\right)^{2}}\,\zeta_3\,G(0;\zt)\nn\\
   &-\frac{32 \zt^{3} - 40 \zt^{2} + 25 \zt + 1}{2 \left(\zt - 1\right)^{2}}\,G(0,1,0,0;\zt)-\frac{\zt \left(90 \zt^{2} - 92 \zt - 25\right)}{\left(\zt - 1\right)^{2}}\,G(0,0,1,0;\zt)\nn\\
   &+\frac{84 \zt^{3} - 112 \zt^{2} + 7 \zt - 3}{2 \left(\zt - 1\right)^{2}}\,G(1,0,1,0;\zt)+\frac{92 \zt^{3} - 158 \zt^{2} + 5 \zt + 1}{2 \left(\zt - 1\right)^{2}}\,G(0,0,0,1;\zt)\nn\\
   &-\frac{182 \zt^{3} - 228 \zt^{2} + 15 \zt + 1}{\left(\zt - 1\right)^{2}}\,G(0,1,0,1;\zt)-\frac{\zt \left(30 \zt^{2} - 71 \zt + 23\right)}{\left(\zt - 1\right)^{2}}\,G(0,0,1,1;\zt)\nn\\
   &-\frac{28 \zt^{2} - 23 \zt + 1}{2 \left(\zt - 1\right)}\,G(1,0,1,1;\zt)+\frac{12 \zt \left(5 \zt - 4\right)}{\zt - 1}\,G(0,1,1,1;\zt)\nn\\
   &-\frac{\zt \left(60 \zt^{2} - 67 \zt - 17\right)}{4 \left(\zt - 1\right)^{2}}\,\pi^2\,G(0,1;\zt)+\frac{\zt \left(41 \zt - 65\right)}{12 \left(\zt - 1\right)^{2}}\,\pi^2\,G(1,0;\zt)-\frac{\zt \left(17 \zt + 1\right)}{12 \left(\zt - 1\right)^{2}}\,\pi^2\,G(0,0;\zt)\nn\\
   &-\frac{32 \zt^{3} - 24 \zt^{2} - 3 \zt + 1}{240 \left(\zt - 1\right)^{2}}\,\mathcal{W}_{2}\,,
\label{eq:ER2full}
\end{align}
and
\begin{align}
   \tilde{E}_{R_2}(\zt)&=\frac{2 \zt^{2} - 2 \zt + 1}{\left(\zt - 1\right)^{2}}\Big[6\,G(0,1;\zt)+\pi^2\Big]-\frac{6 \left(2 \zt^{2} - 4 \zt + 1\right)}{\zt \left(\zt - 1\right)}\,G(1,0;\zt)-\frac{6 \left(2 \zt - 1\right)}{\zt}\,G(1,1;\zt)\nn\\
   &+\frac{6 \left(2 \zt - 1\right)}{\zt - 1}\,G(0,0;\zt)+\frac{1}{40 \left(\zt - 1\right)^{2}}\,\mathcal{W}_{2}-\frac{1}{\left(\zt - 1\right)^{2}}\,\pi^2\,G(0;\zt)\nn\\
   &-\frac{28 \zt^{3} - 49 \zt^{2} + 14 \zt + 1}{2 \zt \left(\zt - 1\right)^{2}}\,\zeta_3\nn\\
   &+\frac{5 \zt^{2} - 6 \zt - 1}{4 \zt \left(\zt - 1\right)^{2}}\,\pi^2\,G(1;\zt)+\frac{3 \left(3 \zt - 5\right)}{2 \zt \left(\zt - 1\right)}\,G(1,0,0;\zt)+\frac{3 \left(5 \zt^{2} - 8 \zt - 1\right)}{2 \zt \left(\zt - 1\right)^{2}}\,G(0,1,0;\zt)\nn\\
   &+\frac{\left(2 \zt - 1\right) \left(7 \zt^{2} - 11 \zt + 1\right)}{\zt \left(\zt - 1\right)^{2}}\,G(0,0,1;\zt)-\frac{\left(4 \zt - 5\right) \left(7 \zt - 1\right)}{2 \zt \left(\zt - 1\right)}\,G(0,1,1;\zt)\nn\\
   &-\frac{28 \zt^{3} - 105 \zt^{2} + 126 \zt - 43}{2 \zt \left(\zt - 1\right)^{2}}\,G(1,0,1;\zt)+\frac{14 \zt - 17}{\zt}\,G(1,1,1;\zt)\nn\\
   &-\frac{4 \zt^{3} - 9 \zt^{2} - 1}{30 \zt \left(\zt - 1\right)^{2}}\Big[90\,G(0,0,1,0;\zt)+15\,\pi^2\,G(0,1;\zt)+30\,\zeta_3\,G(0;\zt)-2\,\pi^4\Big]\nn\\
   & +\frac{4 \zt^{3} - 9 \zt^{2} + 9 \zt + 2}{\zt \left(\zt - 1\right)^{2}}\Big[G(1,0,0,1;\zt)-\zeta_3\,G(1;\zt)\Big]\nn\\
   &+\frac{4 \zt^{2} - 9 \zt + 1}{2 \left(\zt - 1\right)^{2}}\Big[6\,G(1,1,0,1;\zt)+\pi^2\,G(1,1;\zt)\Big]\nn\\
   &-\frac{4 \zt^{2} - 5 \zt - 5}{\zt \left(\zt - 1\right)}\Big[G(0,0,1,1;\zt)-2\,G(0,1,1,1;\zt)\Big]-\frac{1}{\zt \left(\zt - 1\right)^{2}}\,\pi^2\,G(1,0;\zt)\nn\\
   &+\frac{3}{\zt \left(\zt - 1\right)}\,G(0,1,0,0;\zt)+\frac{3 \left(4 \zt^{2} - 9 \zt + 3\right)}{\left(\zt - 1\right)^{2}}\,G(1,0,1,0;\zt)\nn\\
   &+\frac{4 \zt^{3} - 9 \zt^{2} - 3 \zt + 2}{\zt \left(\zt - 1\right)^{2}}\,G(0,0,0,1;\zt)-\frac{20 \zt^{3} - 45 \zt^{2} - 6 \zt + 7}{\zt \left(\zt - 1\right)^{2}}\,G(0,1,0,1;\zt)\nn\\
   &-\frac{4 \zt^{2} - 5 \zt - 2}{\zt \left(\zt - 1\right)}\,G(1,0,1,1;\zt)\,.
\label{eq:EpR2full}
\end{align}
Both $E_{R_2}(\zt)$ and $\tilde{E}_{R_2}(\zt)$ contain the weight-five bracket
\begin{align}
   \mathcal{W}_{2}\;=\;{}&360\,G(0,0,0,1,0;\zt)-240\,G(0,0,0,1,1;\zt)-120\,G(0,0,1,0,0;\zt)+480\,G(0,0,1,1,1;\zt)\nn\\
   &+240\,G(0,1,0,0,1;\zt)-120\,G(0,1,0,1,1;\zt)-120\,G(0,1,1,0,1;\zt)+480\,G(1,0,0,0,1;\zt)\nn\\
   &-600\,G(1,0,0,1,0;\zt)-240\,G(1,0,0,1,1;\zt)+120\,G(1,0,1,0,0;\zt)-720\,G(1,0,1,0,1;\zt)\nn\\
   &-120\,G(1,1,0,0,1;\zt)+240\,G(1,1,0,1,0;\zt)+360\,G(1,1,1,0,1;\zt)\nn\\
   &+60\,\pi^2\,G(0,0,1;\zt)-60\,\pi^2\,G(0,1,0;\zt)-20\,\pi^2\,G(0,1,1;\zt)+20\,\pi^2\,G(1,0,0;\zt)\nn\\
   &-120\,\pi^2\,G(1,0,1;\zt)+40\,\pi^2\,G(1,1,0;\zt)+60\,\pi^2\,G(1,1,1;\zt)\nn\\
   &+120\,\zeta_3\,G(0,0;\zt)-240\,\zeta_3\,G(0,1;\zt)-360\,\zeta_3\,G(1,0;\zt)+120\,\zeta_3\,G(1,1;\zt)\nn\\
   &-8\,\pi^4\,G(0;\zt)+15\,\pi^4\,G(1;\zt)+1980\,\zeta_5-20\,\pi^2\zeta_3\,.
\label{eq:W2bracket}
\end{align}

\bibliography{EEC}

\begin{thebibliography}{100}%
\makeatletter
\providecommand \@ifxundefined [1]{%
 \@ifx{#1\undefined}
}%
\providecommand \@ifnum [1]{%
 \ifnum #1\expandafter \@firstoftwo
 \else \expandafter \@secondoftwo
 \fi
}%
\providecommand \@ifx [1]{%
 \ifx #1\expandafter \@firstoftwo
 \else \expandafter \@secondoftwo
 \fi
}%
\providecommand \natexlab [1]{#1}%
\providecommand \enquote  [1]{``#1''}%
\providecommand \bibnamefont  [1]{#1}%
\providecommand \bibfnamefont [1]{#1}%
\providecommand \citenamefont [1]{#1}%
\providecommand \href@noop [0]{\@secondoftwo}%
\providecommand \href [0]{\begingroup \@sanitize@url \@href}%
\providecommand \@href[1]{\@@startlink{#1}\@@href}%
\providecommand \@@href[1]{\endgroup#1\@@endlink}%
\providecommand \@sanitize@url [0]{\catcode `\\12\catcode `\$12\catcode
  `\&12\catcode `\#12\catcode `\^12\catcode `\_12\catcode `\%12\relax}%
\providecommand \@@startlink[1]{}%
\providecommand \@@endlink[0]{}%
\providecommand \url  [0]{\begingroup\@sanitize@url \@url }%
\providecommand \@url [1]{\endgroup\@href {#1}{\urlprefix }}%
\providecommand \urlprefix  [0]{URL }%
\providecommand \Eprint [0]{\href }%
\providecommand \doibase [0]{http://dx.doi.org/}%
\providecommand \selectlanguage [0]{\@gobble}%
\providecommand \bibinfo  [0]{\@secondoftwo}%
\providecommand \bibfield  [0]{\@secondoftwo}%
\providecommand \translation [1]{[#1]}%
\providecommand \BibitemOpen [0]{}%
\providecommand \bibitemStop [0]{}%
\providecommand \bibitemNoStop [0]{.\EOS\space}%
\providecommand \EOS [0]{\spacefactor3000\relax}%
\providecommand \BibitemShut  [1]{\csname bibitem#1\endcsname}%
\let\auto@bib@innerbib\@empty
\bibitem [{\citenamefont {Henn}\ \emph {et~al.}(2019)\citenamefont {Henn},
  \citenamefont {Sokatchev}, \citenamefont {Yan},\ and\ \citenamefont
  {Zhiboedov}}]{Henn:2019gkr}%
  \BibitemOpen
  \bibfield  {author} {\bibinfo {author} {\bibfnamefont {J.~M.}\ \bibnamefont
  {Henn}}, \bibinfo {author} {\bibfnamefont {E.}~\bibnamefont {Sokatchev}},
  \bibinfo {author} {\bibfnamefont {K.}~\bibnamefont {Yan}}, \ and\ \bibinfo
  {author} {\bibfnamefont {A.}~\bibnamefont {Zhiboedov}},\ }\href {\doibase
  10.1103/PhysRevD.100.036010} {\bibfield  {journal} {\bibinfo  {journal}
  {Phys. Rev. D}\ }\textbf {\bibinfo {volume} {100}},\ \bibinfo {pages}
  {036010} (\bibinfo {year} {2019})},\ \Eprint
  {http://arxiv.org/abs/1903.05314} {arXiv:1903.05314 [hep-th]} \BibitemShut
  {NoStop}%
\bibitem [{\citenamefont {Basham}\ \emph
  {et~al.}(1978{\natexlab{a}})\citenamefont {Basham}, \citenamefont {Brown},
  \citenamefont {Ellis},\ and\ \citenamefont {Love}}]{Basham:1978bw}%
  \BibitemOpen
  \bibfield  {author} {\bibinfo {author} {\bibfnamefont {C.~L.}\ \bibnamefont
  {Basham}}, \bibinfo {author} {\bibfnamefont {L.~S.}\ \bibnamefont {Brown}},
  \bibinfo {author} {\bibfnamefont {S.~D.}\ \bibnamefont {Ellis}}, \ and\
  \bibinfo {author} {\bibfnamefont {S.~T.}\ \bibnamefont {Love}},\ }\href
  {\doibase 10.1103/PhysRevLett.41.1585} {\bibfield  {journal} {\bibinfo
  {journal} {Phys. Rev. Lett.}\ }\textbf {\bibinfo {volume} {41}},\ \bibinfo
  {pages} {1585} (\bibinfo {year} {1978}{\natexlab{a}})}\BibitemShut {NoStop}%
\bibitem [{\citenamefont {Basham}\ \emph
  {et~al.}(1979{\natexlab{a}})\citenamefont {Basham}, \citenamefont {Brown},
  \citenamefont {Ellis},\ and\ \citenamefont {Love}}]{Basham:1978zq}%
  \BibitemOpen
  \bibfield  {author} {\bibinfo {author} {\bibfnamefont {C.~L.}\ \bibnamefont
  {Basham}}, \bibinfo {author} {\bibfnamefont {L.~S.}\ \bibnamefont {Brown}},
  \bibinfo {author} {\bibfnamefont {S.~D.}\ \bibnamefont {Ellis}}, \ and\
  \bibinfo {author} {\bibfnamefont {S.~T.}\ \bibnamefont {Love}},\ }\href
  {\doibase 10.1103/PhysRevD.19.2018} {\bibfield  {journal} {\bibinfo
  {journal} {Phys. Rev. D}\ }\textbf {\bibinfo {volume} {19}},\ \bibinfo
  {pages} {2018} (\bibinfo {year} {1979}{\natexlab{a}})}\BibitemShut {NoStop}%
\bibitem [{\citenamefont {Basham}\ \emph
  {et~al.}(1979{\natexlab{b}})\citenamefont {Basham}, \citenamefont {Brown},
  \citenamefont {Ellis},\ and\ \citenamefont {Love}}]{Basham:1979gh}%
  \BibitemOpen
  \bibfield  {author} {\bibinfo {author} {\bibfnamefont {C.~L.}\ \bibnamefont
  {Basham}}, \bibinfo {author} {\bibfnamefont {L.~S.}\ \bibnamefont {Brown}},
  \bibinfo {author} {\bibfnamefont {S.~D.}\ \bibnamefont {Ellis}}, \ and\
  \bibinfo {author} {\bibfnamefont {S.~T.}\ \bibnamefont {Love}},\ }\href
  {\doibase 10.1016/0370-2693(79)90601-4} {\bibfield  {journal} {\bibinfo
  {journal} {Phys. Lett. B}\ }\textbf {\bibinfo {volume} {85}},\ \bibinfo
  {pages} {297} (\bibinfo {year} {1979}{\natexlab{b}})}\BibitemShut {NoStop}%
\bibitem [{\citenamefont {Basham}\ \emph
  {et~al.}(1978{\natexlab{b}})\citenamefont {Basham}, \citenamefont {Brown},
  \citenamefont {Ellis},\ and\ \citenamefont {Love}}]{Basham:1977iq}%
  \BibitemOpen
  \bibfield  {author} {\bibinfo {author} {\bibfnamefont {C.~L.}\ \bibnamefont
  {Basham}}, \bibinfo {author} {\bibfnamefont {L.~S.}\ \bibnamefont {Brown}},
  \bibinfo {author} {\bibfnamefont {S.~D.}\ \bibnamefont {Ellis}}, \ and\
  \bibinfo {author} {\bibfnamefont {S.~T.}\ \bibnamefont {Love}},\ }\href
  {\doibase 10.1103/PhysRevD.17.2298} {\bibfield  {journal} {\bibinfo
  {journal} {Phys. Rev. D}\ }\textbf {\bibinfo {volume} {17}},\ \bibinfo
  {pages} {2298} (\bibinfo {year} {1978}{\natexlab{b}})}\BibitemShut {NoStop}%
\bibitem [{\citenamefont {Belitsky}\ \emph
  {et~al.}(2014{\natexlab{a}})\citenamefont {Belitsky}, \citenamefont
  {Hohenegger}, \citenamefont {Korchemsky}, \citenamefont {Sokatchev},\ and\
  \citenamefont {Zhiboedov}}]{Belitsky:2013ofa}%
  \BibitemOpen
  \bibfield  {author} {\bibinfo {author} {\bibfnamefont {A.~V.}\ \bibnamefont
  {Belitsky}}, \bibinfo {author} {\bibfnamefont {S.}~\bibnamefont
  {Hohenegger}}, \bibinfo {author} {\bibfnamefont {G.~P.}\ \bibnamefont
  {Korchemsky}}, \bibinfo {author} {\bibfnamefont {E.}~\bibnamefont
  {Sokatchev}}, \ and\ \bibinfo {author} {\bibfnamefont {A.}~\bibnamefont
  {Zhiboedov}},\ }\href {\doibase 10.1103/PhysRevLett.112.071601} {\bibfield
  {journal} {\bibinfo  {journal} {Phys. Rev. Lett.}\ }\textbf {\bibinfo
  {volume} {112}},\ \bibinfo {pages} {071601} (\bibinfo {year}
  {2014}{\natexlab{a}})},\ \Eprint {http://arxiv.org/abs/1311.6800}
  {arXiv:1311.6800 [hep-th]} \BibitemShut {NoStop}%
\bibitem [{\citenamefont {Dixon}\ \emph {et~al.}(2018)\citenamefont {Dixon},
  \citenamefont {Luo}, \citenamefont {Shtabovenko}, \citenamefont {Yang},\ and\
  \citenamefont {Zhu}}]{Dixon:2018qgp}%
  \BibitemOpen
  \bibfield  {author} {\bibinfo {author} {\bibfnamefont {L.~J.}\ \bibnamefont
  {Dixon}}, \bibinfo {author} {\bibfnamefont {M.-X.}\ \bibnamefont {Luo}},
  \bibinfo {author} {\bibfnamefont {V.}~\bibnamefont {Shtabovenko}}, \bibinfo
  {author} {\bibfnamefont {T.-Z.}\ \bibnamefont {Yang}}, \ and\ \bibinfo
  {author} {\bibfnamefont {H.~X.}\ \bibnamefont {Zhu}},\ }\href {\doibase
  10.1103/PhysRevLett.120.102001} {\bibfield  {journal} {\bibinfo  {journal}
  {Phys. Rev. Lett.}\ }\textbf {\bibinfo {volume} {120}},\ \bibinfo {pages}
  {102001} (\bibinfo {year} {2018})},\ \Eprint
  {http://arxiv.org/abs/1801.03219} {arXiv:1801.03219 [hep-ph]} \BibitemShut
  {NoStop}%
\bibitem [{\citenamefont {Luo}\ \emph {et~al.}(2019)\citenamefont {Luo},
  \citenamefont {Shtabovenko}, \citenamefont {Yang},\ and\ \citenamefont
  {Zhu}}]{Luo:2019nig}%
  \BibitemOpen
  \bibfield  {author} {\bibinfo {author} {\bibfnamefont {M.-X.}\ \bibnamefont
  {Luo}}, \bibinfo {author} {\bibfnamefont {V.}~\bibnamefont {Shtabovenko}},
  \bibinfo {author} {\bibfnamefont {T.-Z.}\ \bibnamefont {Yang}}, \ and\
  \bibinfo {author} {\bibfnamefont {H.~X.}\ \bibnamefont {Zhu}},\ }\href
  {\doibase 10.1007/JHEP06(2019)037} {\bibfield  {journal} {\bibinfo  {journal}
  {JHEP}\ }\textbf {\bibinfo {volume} {06}},\ \bibinfo {pages} {037} (\bibinfo
  {year} {2019})},\ \Eprint {http://arxiv.org/abs/1903.07277} {arXiv:1903.07277
  [hep-ph]} \BibitemShut {NoStop}%
\bibitem [{\citenamefont {Gao}\ \emph {et~al.}(2021)\citenamefont {Gao},
  \citenamefont {Shtabovenko},\ and\ \citenamefont {Yang}}]{Gao:2020vyx}%
  \BibitemOpen
  \bibfield  {author} {\bibinfo {author} {\bibfnamefont {J.}~\bibnamefont
  {Gao}}, \bibinfo {author} {\bibfnamefont {V.}~\bibnamefont {Shtabovenko}}, \
  and\ \bibinfo {author} {\bibfnamefont {T.-Z.}\ \bibnamefont {Yang}},\ }\href
  {\doibase 10.1007/JHEP02(2021)210} {\bibfield  {journal} {\bibinfo  {journal}
  {JHEP}\ }\textbf {\bibinfo {volume} {02}},\ \bibinfo {pages} {210} (\bibinfo
  {year} {2021})},\ \Eprint {http://arxiv.org/abs/2012.14188} {arXiv:2012.14188
  [hep-ph]} \BibitemShut {NoStop}%
\bibitem [{\citenamefont {Dixon}\ \emph {et~al.}(2019)\citenamefont {Dixon},
  \citenamefont {Moult},\ and\ \citenamefont {Zhu}}]{Dixon:2019uzg}%
  \BibitemOpen
  \bibfield  {author} {\bibinfo {author} {\bibfnamefont {L.~J.}\ \bibnamefont
  {Dixon}}, \bibinfo {author} {\bibfnamefont {I.}~\bibnamefont {Moult}}, \ and\
  \bibinfo {author} {\bibfnamefont {H.~X.}\ \bibnamefont {Zhu}},\ }\href
  {\doibase 10.1103/PhysRevD.100.014009} {\bibfield  {journal} {\bibinfo
  {journal} {Phys. Rev. D}\ }\textbf {\bibinfo {volume} {100}},\ \bibinfo
  {pages} {014009} (\bibinfo {year} {2019})},\ \Eprint
  {http://arxiv.org/abs/1905.01310} {arXiv:1905.01310 [hep-ph]} \BibitemShut
  {NoStop}%
\bibitem [{\citenamefont {Kologlu}\ \emph {et~al.}(2021)\citenamefont
  {Kologlu}, \citenamefont {Kravchuk}, \citenamefont {Simmons-Duffin},\ and\
  \citenamefont {Zhiboedov}}]{Kologlu:2019mfz}%
  \BibitemOpen
  \bibfield  {author} {\bibinfo {author} {\bibfnamefont {M.}~\bibnamefont
  {Kologlu}}, \bibinfo {author} {\bibfnamefont {P.}~\bibnamefont {Kravchuk}},
  \bibinfo {author} {\bibfnamefont {D.}~\bibnamefont {Simmons-Duffin}}, \ and\
  \bibinfo {author} {\bibfnamefont {A.}~\bibnamefont {Zhiboedov}},\ }\href
  {\doibase 10.1007/JHEP01(2021)128} {\bibfield  {journal} {\bibinfo  {journal}
  {JHEP}\ }\textbf {\bibinfo {volume} {01}},\ \bibinfo {pages} {128} (\bibinfo
  {year} {2021})},\ \Eprint {http://arxiv.org/abs/1905.01311} {arXiv:1905.01311
  [hep-th]} \BibitemShut {NoStop}%
\bibitem [{\citenamefont {Korchemsky}(2020)}]{Korchemsky:2019nzm}%
  \BibitemOpen
  \bibfield  {author} {\bibinfo {author} {\bibfnamefont {G.~P.}\ \bibnamefont
  {Korchemsky}},\ }\href {\doibase 10.1007/JHEP01(2020)008} {\bibfield
  {journal} {\bibinfo  {journal} {JHEP}\ }\textbf {\bibinfo {volume} {01}},\
  \bibinfo {pages} {008} (\bibinfo {year} {2020})},\ \Eprint
  {http://arxiv.org/abs/1905.01444} {arXiv:1905.01444 [hep-th]} \BibitemShut
  {NoStop}%
\bibitem [{\citenamefont {Collins}\ and\ \citenamefont
  {Soper}(1981)}]{COLLINS1981381}%
  \BibitemOpen
  \bibfield  {author} {\bibinfo {author} {\bibfnamefont {J.~C.}\ \bibnamefont
  {Collins}}\ and\ \bibinfo {author} {\bibfnamefont {D.~E.}\ \bibnamefont
  {Soper}},\ }\href {\doibase https://doi.org/10.1016/0550-3213(81)90339-4}
  {\bibfield  {journal} {\bibinfo  {journal} {Nuclear Physics B}\ }\textbf
  {\bibinfo {volume} {193}},\ \bibinfo {pages} {381} (\bibinfo {year}
  {1981})}\BibitemShut {NoStop}%
\bibitem [{\citenamefont {Ellis}\ \emph {et~al.}(1984)\citenamefont {Ellis},
  \citenamefont {Richards},\ and\ \citenamefont {Stirling}}]{ELLIS198499}%
  \BibitemOpen
  \bibfield  {author} {\bibinfo {author} {\bibfnamefont {S.}~\bibnamefont
  {Ellis}}, \bibinfo {author} {\bibfnamefont {D.}~\bibnamefont {Richards}}, \
  and\ \bibinfo {author} {\bibfnamefont {W.}~\bibnamefont {Stirling}},\ }\href
  {\doibase https://doi.org/10.1016/0370-2693(84)92064-1} {\bibfield  {journal}
  {\bibinfo  {journal} {Physics Letters B}\ }\textbf {\bibinfo {volume}
  {136}},\ \bibinfo {pages} {99} (\bibinfo {year} {1984})}\BibitemShut
  {NoStop}%
\bibitem [{\citenamefont {de~Florian}\ and\ \citenamefont
  {Grazzini}(2005)}]{deFlorian:2004mp}%
  \BibitemOpen
  \bibfield  {author} {\bibinfo {author} {\bibfnamefont {D.}~\bibnamefont
  {de~Florian}}\ and\ \bibinfo {author} {\bibfnamefont {M.}~\bibnamefont
  {Grazzini}},\ }\href {\doibase 10.1016/j.nuclphysb.2004.10.051} {\bibfield
  {journal} {\bibinfo  {journal} {Nucl. Phys. B}\ }\textbf {\bibinfo {volume}
  {704}},\ \bibinfo {pages} {387} (\bibinfo {year} {2005})},\ \Eprint
  {http://arxiv.org/abs/hep-ph/0407241} {arXiv:hep-ph/0407241} \BibitemShut
  {NoStop}%
\bibitem [{\citenamefont {Tulip\'ant}\ \emph {et~al.}(2017)\citenamefont
  {Tulip\'ant}, \citenamefont {Kardos},\ and\ \citenamefont
  {Somogyi}}]{Tulipant:2017ybb}%
  \BibitemOpen
  \bibfield  {author} {\bibinfo {author} {\bibfnamefont {Z.}~\bibnamefont
  {Tulip\'ant}}, \bibinfo {author} {\bibfnamefont {A.}~\bibnamefont {Kardos}},
  \ and\ \bibinfo {author} {\bibfnamefont {G.}~\bibnamefont {Somogyi}},\ }\href
  {\doibase 10.1140/epjc/s10052-017-5320-9} {\bibfield  {journal} {\bibinfo
  {journal} {Eur. Phys. J. C}\ }\textbf {\bibinfo {volume} {77}},\ \bibinfo
  {pages} {749} (\bibinfo {year} {2017})},\ \Eprint
  {http://arxiv.org/abs/1708.04093} {arXiv:1708.04093 [hep-ph]} \BibitemShut
  {NoStop}%
\bibitem [{\citenamefont {Moult}\ and\ \citenamefont
  {Zhu}(2018)}]{Moult:2018jzp}%
  \BibitemOpen
  \bibfield  {author} {\bibinfo {author} {\bibfnamefont {I.}~\bibnamefont
  {Moult}}\ and\ \bibinfo {author} {\bibfnamefont {H.~X.}\ \bibnamefont
  {Zhu}},\ }\href {\doibase 10.1007/JHEP08(2018)160} {\bibfield  {journal}
  {\bibinfo  {journal} {JHEP}\ }\textbf {\bibinfo {volume} {08}},\ \bibinfo
  {pages} {160} (\bibinfo {year} {2018})},\ \Eprint
  {http://arxiv.org/abs/1801.02627} {arXiv:1801.02627 [hep-ph]} \BibitemShut
  {NoStop}%
\bibitem [{\citenamefont {Moult}\ \emph {et~al.}(2020)\citenamefont {Moult},
  \citenamefont {Vita},\ and\ \citenamefont {Yan}}]{Moult:2019vou}%
  \BibitemOpen
  \bibfield  {author} {\bibinfo {author} {\bibfnamefont {I.}~\bibnamefont
  {Moult}}, \bibinfo {author} {\bibfnamefont {G.}~\bibnamefont {Vita}}, \ and\
  \bibinfo {author} {\bibfnamefont {K.}~\bibnamefont {Yan}},\ }\href {\doibase
  10.1007/JHEP07(2020)005} {\bibfield  {journal} {\bibinfo  {journal} {JHEP}\
  }\textbf {\bibinfo {volume} {07}},\ \bibinfo {pages} {005} (\bibinfo {year}
  {2020})},\ \Eprint {http://arxiv.org/abs/1912.02188} {arXiv:1912.02188
  [hep-ph]} \BibitemShut {NoStop}%
\bibitem [{\citenamefont {Ebert}\ \emph {et~al.}(2021)\citenamefont {Ebert},
  \citenamefont {Mistlberger},\ and\ \citenamefont {Vita}}]{Ebert:2020sfi}%
  \BibitemOpen
  \bibfield  {author} {\bibinfo {author} {\bibfnamefont {M.~A.}\ \bibnamefont
  {Ebert}}, \bibinfo {author} {\bibfnamefont {B.}~\bibnamefont {Mistlberger}},
  \ and\ \bibinfo {author} {\bibfnamefont {G.}~\bibnamefont {Vita}},\ }\href
  {\doibase 10.1007/JHEP08(2021)022} {\bibfield  {journal} {\bibinfo  {journal}
  {JHEP}\ }\textbf {\bibinfo {volume} {08}},\ \bibinfo {pages} {022} (\bibinfo
  {year} {2021})},\ \Eprint {http://arxiv.org/abs/2012.07859} {arXiv:2012.07859
  [hep-ph]} \BibitemShut {NoStop}%
\bibitem [{\citenamefont {Moult}\ \emph {et~al.}(2022)\citenamefont {Moult},
  \citenamefont {Zhu},\ and\ \citenamefont {Zhu}}]{Moult:2022xzt}%
  \BibitemOpen
  \bibfield  {author} {\bibinfo {author} {\bibfnamefont {I.}~\bibnamefont
  {Moult}}, \bibinfo {author} {\bibfnamefont {H.~X.}\ \bibnamefont {Zhu}}, \
  and\ \bibinfo {author} {\bibfnamefont {Y.~J.}\ \bibnamefont {Zhu}},\
  }\href@noop {} {\  (\bibinfo {year} {2022})},\ \Eprint
  {http://arxiv.org/abs/2205.02249} {arXiv:2205.02249 [hep-ph]} \BibitemShut
  {NoStop}%
\bibitem [{\citenamefont {Duhr}\ \emph {et~al.}(2022)\citenamefont {Duhr},
  \citenamefont {Mistlberger},\ and\ \citenamefont {Vita}}]{Duhr:2022yyp}%
  \BibitemOpen
  \bibfield  {author} {\bibinfo {author} {\bibfnamefont {C.}~\bibnamefont
  {Duhr}}, \bibinfo {author} {\bibfnamefont {B.}~\bibnamefont {Mistlberger}}, \
  and\ \bibinfo {author} {\bibfnamefont {G.}~\bibnamefont {Vita}},\ }\href@noop
  {} {\  (\bibinfo {year} {2022})},\ \Eprint {http://arxiv.org/abs/2205.02242}
  {arXiv:2205.02242 [hep-ph]} \BibitemShut {NoStop}%
\bibitem [{\citenamefont {Chen}\ \emph
  {et~al.}(2020{\natexlab{a}})\citenamefont {Chen}, \citenamefont {Luo},
  \citenamefont {Moult}, \citenamefont {Yang}, \citenamefont {Zhang},\ and\
  \citenamefont {Zhu}}]{Chen:2019bpb}%
  \BibitemOpen
  \bibfield  {author} {\bibinfo {author} {\bibfnamefont {H.}~\bibnamefont
  {Chen}}, \bibinfo {author} {\bibfnamefont {M.-X.}\ \bibnamefont {Luo}},
  \bibinfo {author} {\bibfnamefont {I.}~\bibnamefont {Moult}}, \bibinfo
  {author} {\bibfnamefont {T.-Z.}\ \bibnamefont {Yang}}, \bibinfo {author}
  {\bibfnamefont {X.}~\bibnamefont {Zhang}}, \ and\ \bibinfo {author}
  {\bibfnamefont {H.~X.}\ \bibnamefont {Zhu}},\ }\href {\doibase
  10.1007/JHEP08(2020)028} {\bibfield  {journal} {\bibinfo  {journal} {JHEP}\
  }\textbf {\bibinfo {volume} {08}},\ \bibinfo {pages} {028} (\bibinfo {year}
  {2020}{\natexlab{a}})},\ \Eprint {http://arxiv.org/abs/1912.11050}
  {arXiv:1912.11050 [hep-ph]} \BibitemShut {NoStop}%
\bibitem [{\citenamefont {Chen}\ \emph
  {et~al.}(2020{\natexlab{b}})\citenamefont {Chen}, \citenamefont {Moult},
  \citenamefont {Zhang},\ and\ \citenamefont {Zhu}}]{Chen:2020vvp}%
  \BibitemOpen
  \bibfield  {author} {\bibinfo {author} {\bibfnamefont {H.}~\bibnamefont
  {Chen}}, \bibinfo {author} {\bibfnamefont {I.}~\bibnamefont {Moult}},
  \bibinfo {author} {\bibfnamefont {X.}~\bibnamefont {Zhang}}, \ and\ \bibinfo
  {author} {\bibfnamefont {H.~X.}\ \bibnamefont {Zhu}},\ }\href {\doibase
  10.1103/PhysRevD.102.054012} {\bibfield  {journal} {\bibinfo  {journal}
  {Phys. Rev. D}\ }\textbf {\bibinfo {volume} {102}},\ \bibinfo {pages}
  {054012} (\bibinfo {year} {2020}{\natexlab{b}})},\ \Eprint
  {http://arxiv.org/abs/2004.11381} {arXiv:2004.11381 [hep-ph]} \BibitemShut
  {NoStop}%
\bibitem [{\citenamefont {Yan}\ and\ \citenamefont
  {Zhang}(2022)}]{Yan:2022cye}%
  \BibitemOpen
  \bibfield  {author} {\bibinfo {author} {\bibfnamefont {K.}~\bibnamefont
  {Yan}}\ and\ \bibinfo {author} {\bibfnamefont {X.}~\bibnamefont {Zhang}},\
  }\href {\doibase 10.1103/PhysRevLett.129.021602} {\bibfield  {journal}
  {\bibinfo  {journal} {Phys. Rev. Lett.}\ }\textbf {\bibinfo {volume} {129}},\
  \bibinfo {pages} {021602} (\bibinfo {year} {2022})},\ \Eprint
  {http://arxiv.org/abs/2203.04349} {arXiv:2203.04349 [hep-th]} \BibitemShut
  {NoStop}%
\bibitem [{\citenamefont {Yang}\ and\ \citenamefont
  {Zhang}(2022)}]{Yang:2022tgm}%
  \BibitemOpen
  \bibfield  {author} {\bibinfo {author} {\bibfnamefont {T.-Z.}\ \bibnamefont
  {Yang}}\ and\ \bibinfo {author} {\bibfnamefont {X.}~\bibnamefont {Zhang}},\
  }\href {\doibase 10.1007/JHEP09(2022)006} {\bibfield  {journal} {\bibinfo
  {journal} {JHEP}\ }\textbf {\bibinfo {volume} {09}},\ \bibinfo {pages} {006}
  (\bibinfo {year} {2022})},\ \Eprint {http://arxiv.org/abs/2208.01051}
  {arXiv:2208.01051 [hep-ph]} \BibitemShut {NoStop}%
\bibitem [{\citenamefont {Yang}\ and\ \citenamefont
  {Zhang}(2024)}]{Yang:2024gcn}%
  \BibitemOpen
  \bibfield  {author} {\bibinfo {author} {\bibfnamefont {T.-Z.}\ \bibnamefont
  {Yang}}\ and\ \bibinfo {author} {\bibfnamefont {X.}~\bibnamefont {Zhang}},\
  }\href {\doibase 10.1103/PhysRevD.109.114036} {\bibfield  {journal} {\bibinfo
   {journal} {Phys. Rev. D}\ }\textbf {\bibinfo {volume} {109}},\ \bibinfo
  {pages} {114036} (\bibinfo {year} {2024})},\ \Eprint
  {http://arxiv.org/abs/2402.05174} {arXiv:2402.05174 [hep-ph]} \BibitemShut
  {NoStop}%
\bibitem [{\citenamefont {Chicherin}\ \emph {et~al.}(2024)\citenamefont
  {Chicherin}, \citenamefont {Moult}, \citenamefont {Sokatchev}, \citenamefont
  {Yan},\ and\ \citenamefont {Zhu}}]{Chicherin:2024ifn}%
  \BibitemOpen
  \bibfield  {author} {\bibinfo {author} {\bibfnamefont {D.}~\bibnamefont
  {Chicherin}}, \bibinfo {author} {\bibfnamefont {I.}~\bibnamefont {Moult}},
  \bibinfo {author} {\bibfnamefont {E.}~\bibnamefont {Sokatchev}}, \bibinfo
  {author} {\bibfnamefont {K.}~\bibnamefont {Yan}}, \ and\ \bibinfo {author}
  {\bibfnamefont {Y.}~\bibnamefont {Zhu}},\ }\href {\doibase
  10.1103/PhysRevD.110.L091901} {\bibfield  {journal} {\bibinfo  {journal}
  {Phys. Rev. D}\ }\textbf {\bibinfo {volume} {110}},\ \bibinfo {pages}
  {L091901} (\bibinfo {year} {2024})},\ \Eprint
  {http://arxiv.org/abs/2401.06463} {arXiv:2401.06463 [hep-th]} \BibitemShut
  {NoStop}%
\bibitem [{\citenamefont {Ma}\ \emph {et~al.}(2026)\citenamefont {Ma},
  \citenamefont {Gong}, \citenamefont {Lin}, \citenamefont {Yan}, \citenamefont
  {Yang},\ and\ \citenamefont {Zhang}}]{Ma:2025qtx}%
  \BibitemOpen
  \bibfield  {author} {\bibinfo {author} {\bibfnamefont {R.}~\bibnamefont
  {Ma}}, \bibinfo {author} {\bibfnamefont {J.}~\bibnamefont {Gong}}, \bibinfo
  {author} {\bibfnamefont {J.}~\bibnamefont {Lin}}, \bibinfo {author}
  {\bibfnamefont {K.}~\bibnamefont {Yan}}, \bibinfo {author} {\bibfnamefont
  {G.}~\bibnamefont {Yang}}, \ and\ \bibinfo {author} {\bibfnamefont
  {Y.}~\bibnamefont {Zhang}},\ }\href {\doibase 10.1007/JHEP02(2026)025}
  {\bibfield  {journal} {\bibinfo  {journal} {JHEP}\ }\textbf {\bibinfo
  {volume} {02}},\ \bibinfo {pages} {025} (\bibinfo {year} {2026})},\ \Eprint
  {http://arxiv.org/abs/2506.02061} {arXiv:2506.02061 [hep-ph]} \BibitemShut
  {NoStop}%
\bibitem [{\citenamefont {Sabry}(1962)}]{Sabry:1962}%
  \BibitemOpen
  \bibfield  {author} {\bibinfo {author} {\bibfnamefont {A.}~\bibnamefont
  {Sabry}},\ }\href {\doibase 10.1016/0029-5582(62)90535-7} {\bibfield
  {journal} {\bibinfo  {journal} {Nucl. Phys.}\ }\textbf {\bibinfo {volume}
  {33}},\ \bibinfo {pages} {401} (\bibinfo {year} {1962})}\BibitemShut
  {NoStop}%
\bibitem [{\citenamefont {Broadhurst}\ \emph {et~al.}(1993)\citenamefont
  {Broadhurst}, \citenamefont {Fleischer},\ and\ \citenamefont
  {Tarasov}}]{Broadhurst:1993mw}%
  \BibitemOpen
  \bibfield  {author} {\bibinfo {author} {\bibfnamefont {D.~J.}\ \bibnamefont
  {Broadhurst}}, \bibinfo {author} {\bibfnamefont {J.}~\bibnamefont
  {Fleischer}}, \ and\ \bibinfo {author} {\bibfnamefont {O.~V.}\ \bibnamefont
  {Tarasov}},\ }\href {\doibase 10.1007/BF01474625} {\bibfield  {journal}
  {\bibinfo  {journal} {Z. Phys. C}\ }\textbf {\bibinfo {volume} {60}},\
  \bibinfo {pages} {287} (\bibinfo {year} {1993})},\ \Eprint
  {http://arxiv.org/abs/hep-ph/9304303} {arXiv:hep-ph/9304303} \BibitemShut
  {NoStop}%
\bibitem [{\citenamefont {Laporta}\ and\ \citenamefont
  {Remiddi}(2005)}]{Laporta:2004rb}%
  \BibitemOpen
  \bibfield  {author} {\bibinfo {author} {\bibfnamefont {S.}~\bibnamefont
  {Laporta}}\ and\ \bibinfo {author} {\bibfnamefont {E.}~\bibnamefont
  {Remiddi}},\ }\href {\doibase 10.1016/j.nuclphysb.2004.10.044} {\bibfield
  {journal} {\bibinfo  {journal} {Nucl. Phys. B}\ }\textbf {\bibinfo {volume}
  {704}},\ \bibinfo {pages} {349} (\bibinfo {year} {2005})},\ \Eprint
  {http://arxiv.org/abs/hep-ph/0406160} {arXiv:hep-ph/0406160} \BibitemShut
  {NoStop}%
\bibitem [{\citenamefont {Bloch}\ and\ \citenamefont
  {Vanhove}(2015)}]{Bloch:2013tra}%
  \BibitemOpen
  \bibfield  {author} {\bibinfo {author} {\bibfnamefont {S.}~\bibnamefont
  {Bloch}}\ and\ \bibinfo {author} {\bibfnamefont {P.}~\bibnamefont
  {Vanhove}},\ }\href {\doibase 10.1016/j.jnt.2014.09.032} {\bibfield
  {journal} {\bibinfo  {journal} {J. Number Theor.}\ }\textbf {\bibinfo
  {volume} {148}},\ \bibinfo {pages} {328} (\bibinfo {year} {2015})},\ \Eprint
  {http://arxiv.org/abs/1309.5865} {arXiv:1309.5865 [hep-th]} \BibitemShut
  {NoStop}%
\bibitem [{\citenamefont {Adams}\ \emph {et~al.}(2013)\citenamefont {Adams},
  \citenamefont {Bogner},\ and\ \citenamefont {Weinzierl}}]{Adams:2013nia}%
  \BibitemOpen
  \bibfield  {author} {\bibinfo {author} {\bibfnamefont {L.}~\bibnamefont
  {Adams}}, \bibinfo {author} {\bibfnamefont {C.}~\bibnamefont {Bogner}}, \
  and\ \bibinfo {author} {\bibfnamefont {S.}~\bibnamefont {Weinzierl}},\ }\href
  {\doibase 10.1063/1.4804996} {\bibfield  {journal} {\bibinfo  {journal} {J.
  Math. Phys.}\ }\textbf {\bibinfo {volume} {54}},\ \bibinfo {pages} {052303}
  (\bibinfo {year} {2013})},\ \Eprint {http://arxiv.org/abs/1302.7004}
  {arXiv:1302.7004 [hep-ph]} \BibitemShut {NoStop}%
\bibitem [{\citenamefont {Goncharov}\ \emph {et~al.}(2010)\citenamefont
  {Goncharov}, \citenamefont {Spradlin}, \citenamefont {Vergu},\ and\
  \citenamefont {Volovich}}]{Goncharov:2010jf}%
  \BibitemOpen
  \bibfield  {author} {\bibinfo {author} {\bibfnamefont {A.~B.}\ \bibnamefont
  {Goncharov}}, \bibinfo {author} {\bibfnamefont {M.}~\bibnamefont {Spradlin}},
  \bibinfo {author} {\bibfnamefont {C.}~\bibnamefont {Vergu}}, \ and\ \bibinfo
  {author} {\bibfnamefont {A.}~\bibnamefont {Volovich}},\ }\href {\doibase
  10.1103/PhysRevLett.105.151605} {\bibfield  {journal} {\bibinfo  {journal}
  {Phys. Rev. Lett.}\ }\textbf {\bibinfo {volume} {105}},\ \bibinfo {pages}
  {151605} (\bibinfo {year} {2010})},\ \Eprint {http://arxiv.org/abs/1006.5703}
  {arXiv:1006.5703 [hep-th]} \BibitemShut {NoStop}%
\bibitem [{\citenamefont {Bauer}\ \emph {et~al.}(2002)\citenamefont {Bauer},
  \citenamefont {Frink},\ and\ \citenamefont {Kreckel}}]{Bauer:2000cp}%
  \BibitemOpen
  \bibfield  {author} {\bibinfo {author} {\bibfnamefont {C.~W.}\ \bibnamefont
  {Bauer}}, \bibinfo {author} {\bibfnamefont {A.}~\bibnamefont {Frink}}, \ and\
  \bibinfo {author} {\bibfnamefont {R.}~\bibnamefont {Kreckel}},\ }\href
  {\doibase 10.1006/jsco.2001.0494} {\bibfield  {journal} {\bibinfo  {journal}
  {J. Symb. Comput.}\ }\textbf {\bibinfo {volume} {33}},\ \bibinfo {pages} {1}
  (\bibinfo {year} {2002})},\ \Eprint {http://arxiv.org/abs/cs/0004015}
  {arXiv:cs/0004015 [cs.SC]} \BibitemShut {NoStop}%
\bibitem [{\citenamefont {Naterop}\ \emph {et~al.}(2020)\citenamefont
  {Naterop}, \citenamefont {Signer},\ and\ \citenamefont
  {Ulrich}}]{Naterop:2019xaf}%
  \BibitemOpen
  \bibfield  {author} {\bibinfo {author} {\bibfnamefont {L.}~\bibnamefont
  {Naterop}}, \bibinfo {author} {\bibfnamefont {A.}~\bibnamefont {Signer}}, \
  and\ \bibinfo {author} {\bibfnamefont {Y.}~\bibnamefont {Ulrich}},\ }\href
  {\doibase 10.1016/j.cpc.2020.107165} {\bibfield  {journal} {\bibinfo
  {journal} {Comput. Phys. Commun.}\ }\textbf {\bibinfo {volume} {253}},\
  \bibinfo {pages} {107165} (\bibinfo {year} {2020})},\ \Eprint
  {http://arxiv.org/abs/1909.01656} {arXiv:1909.01656 [hep-ph]} \BibitemShut
  {NoStop}%
\bibitem [{\citenamefont {Wang}\ \emph {et~al.}(2021)\citenamefont {Wang},
  \citenamefont {Yang},\ and\ \citenamefont {Zhou}}]{Wang:2021imw}%
  \BibitemOpen
  \bibfield  {author} {\bibinfo {author} {\bibfnamefont {Y.}~\bibnamefont
  {Wang}}, \bibinfo {author} {\bibfnamefont {L.~L.}\ \bibnamefont {Yang}}, \
  and\ \bibinfo {author} {\bibfnamefont {B.}~\bibnamefont {Zhou}},\ }\href@noop
  {} {\  (\bibinfo {year} {2021})},\ \Eprint {http://arxiv.org/abs/2112.04122}
  {arXiv:2112.04122 [hep-ph]} \BibitemShut {NoStop}%
\bibitem [{\citenamefont {Kotikov}(1991)}]{Kotikov:1990kg}%
  \BibitemOpen
  \bibfield  {author} {\bibinfo {author} {\bibfnamefont {A.~V.}\ \bibnamefont
  {Kotikov}},\ }\href {\doibase 10.1016/0370-2693(91)90413-K} {\bibfield
  {journal} {\bibinfo  {journal} {Phys. Lett. B}\ }\textbf {\bibinfo {volume}
  {254}},\ \bibinfo {pages} {158} (\bibinfo {year} {1991})}\BibitemShut
  {NoStop}%
\bibitem [{\citenamefont {Remiddi}(1997)}]{Remiddi:1997ny}%
  \BibitemOpen
  \bibfield  {author} {\bibinfo {author} {\bibfnamefont {E.}~\bibnamefont
  {Remiddi}},\ }\href {\doibase 10.1007/BF03185566} {\bibfield  {journal}
  {\bibinfo  {journal} {Nuovo Cim. A}\ }\textbf {\bibinfo {volume} {110}},\
  \bibinfo {pages} {1435} (\bibinfo {year} {1997})},\ \Eprint
  {http://arxiv.org/abs/hep-th/9711188} {arXiv:hep-th/9711188} \BibitemShut
  {NoStop}%
\bibitem [{\citenamefont {Gehrmann}\ and\ \citenamefont
  {Remiddi}(2000)}]{Gehrmann:1999as}%
  \BibitemOpen
  \bibfield  {author} {\bibinfo {author} {\bibfnamefont {T.}~\bibnamefont
  {Gehrmann}}\ and\ \bibinfo {author} {\bibfnamefont {E.}~\bibnamefont
  {Remiddi}},\ }\href {\doibase 10.1016/S0550-3213(00)00223-6} {\bibfield
  {journal} {\bibinfo  {journal} {Nucl. Phys. B}\ }\textbf {\bibinfo {volume}
  {580}},\ \bibinfo {pages} {485} (\bibinfo {year} {2000})},\ \Eprint
  {http://arxiv.org/abs/hep-ph/9912329} {arXiv:hep-ph/9912329} \BibitemShut
  {NoStop}%
\bibitem [{\citenamefont {Henn}(2013)}]{Henn:2013pwa}%
  \BibitemOpen
  \bibfield  {author} {\bibinfo {author} {\bibfnamefont {J.~M.}\ \bibnamefont
  {Henn}},\ }\href {\doibase 10.1103/PhysRevLett.110.251601} {\bibfield
  {journal} {\bibinfo  {journal} {Phys. Rev. Lett.}\ }\textbf {\bibinfo
  {volume} {110}},\ \bibinfo {pages} {251601} (\bibinfo {year} {2013})},\
  \Eprint {http://arxiv.org/abs/1304.1806} {arXiv:1304.1806} \BibitemShut
  {NoStop}%
\bibitem [{\citenamefont {Goncharov}(1998)}]{Goncharov:1998kja}%
  \BibitemOpen
  \bibfield  {author} {\bibinfo {author} {\bibfnamefont {A.~B.}\ \bibnamefont
  {Goncharov}},\ }\href {\doibase 10.4310/MRL.1998.v5.n4.a7} {\bibfield
  {journal} {\bibinfo  {journal} {Math. Res. Lett.}\ }\textbf {\bibinfo
  {volume} {5}},\ \bibinfo {pages} {497} (\bibinfo {year} {1998})},\ \Eprint
  {http://arxiv.org/abs/1105.2076} {arXiv:1105.2076 [math.AG]} \BibitemShut
  {NoStop}%
\bibitem [{\citenamefont {Walden}\ and\ \citenamefont
  {Weinzierl}(2021)}]{Walden:2020odh}%
  \BibitemOpen
  \bibfield  {author} {\bibinfo {author} {\bibfnamefont {M.}~\bibnamefont
  {Walden}}\ and\ \bibinfo {author} {\bibfnamefont {S.}~\bibnamefont
  {Weinzierl}},\ }\href {\doibase 10.1016/j.cpc.2021.108020} {\bibfield
  {journal} {\bibinfo  {journal} {Comput. Phys. Commun.}\ }\textbf {\bibinfo
  {volume} {265}},\ \bibinfo {pages} {108020} (\bibinfo {year} {2021})},\
  \Eprint {http://arxiv.org/abs/2010.05271} {arXiv:2010.05271 [hep-ph]}
  \BibitemShut {NoStop}%
\bibitem [{\citenamefont {Duhr}\ \emph {et~al.}(2026)\citenamefont {Duhr},
  \citenamefont {Lorkowski}, \citenamefont {Marzucca}, \citenamefont {Mauc},\
  and\ \citenamefont {Weinzierl}}]{Duhr:2026ell}%
  \BibitemOpen
  \bibfield  {author} {\bibinfo {author} {\bibfnamefont {C.}~\bibnamefont
  {Duhr}}, \bibinfo {author} {\bibfnamefont {F.}~\bibnamefont {Lorkowski}},
  \bibinfo {author} {\bibfnamefont {R.}~\bibnamefont {Marzucca}}, \bibinfo
  {author} {\bibfnamefont {S.}~\bibnamefont {Mauc}}, \ and\ \bibinfo {author}
  {\bibfnamefont {S.}~\bibnamefont {Weinzierl}},\ }\href@noop {} {\  (\bibinfo
  {year} {2026})},\ \Eprint {http://arxiv.org/abs/2602.09956} {arXiv:2602.09956
  [hep-ph]} \BibitemShut {NoStop}%
\bibitem [{\citenamefont {Dixon}\ \emph {et~al.}(2017)\citenamefont {Dixon},
  \citenamefont {Drummond}, \citenamefont {Harrington}, \citenamefont {McLeod},
  \citenamefont {Papathanasiou},\ and\ \citenamefont
  {Spradlin}}]{Dixon:2016nkn}%
  \BibitemOpen
  \bibfield  {author} {\bibinfo {author} {\bibfnamefont {L.~J.}\ \bibnamefont
  {Dixon}}, \bibinfo {author} {\bibfnamefont {J.}~\bibnamefont {Drummond}},
  \bibinfo {author} {\bibfnamefont {T.}~\bibnamefont {Harrington}}, \bibinfo
  {author} {\bibfnamefont {A.~J.}\ \bibnamefont {McLeod}}, \bibinfo {author}
  {\bibfnamefont {G.}~\bibnamefont {Papathanasiou}}, \ and\ \bibinfo {author}
  {\bibfnamefont {M.}~\bibnamefont {Spradlin}},\ }\href {\doibase
  10.1007/JHEP02(2017)137} {\bibfield  {journal} {\bibinfo  {journal} {JHEP}\
  }\textbf {\bibinfo {volume} {02}},\ \bibinfo {pages} {137} (\bibinfo {year}
  {2017})},\ \Eprint {http://arxiv.org/abs/1612.08976} {arXiv:1612.08976
  [hep-th]} \BibitemShut {NoStop}%
\bibitem [{\citenamefont {Caron-Huot}\ \emph {et~al.}(2016)\citenamefont
  {Caron-Huot}, \citenamefont {Dixon}, \citenamefont {McLeod},\ and\
  \citenamefont {von Hippel}}]{Caron-Huot:2016owq}%
  \BibitemOpen
  \bibfield  {author} {\bibinfo {author} {\bibfnamefont {S.}~\bibnamefont
  {Caron-Huot}}, \bibinfo {author} {\bibfnamefont {L.~J.}\ \bibnamefont
  {Dixon}}, \bibinfo {author} {\bibfnamefont {A.}~\bibnamefont {McLeod}}, \
  and\ \bibinfo {author} {\bibfnamefont {M.}~\bibnamefont {von Hippel}},\
  }\href {\doibase 10.1103/PhysRevLett.117.241601} {\bibfield  {journal}
  {\bibinfo  {journal} {Phys. Rev. Lett.}\ }\textbf {\bibinfo {volume} {117}},\
  \bibinfo {pages} {241601} (\bibinfo {year} {2016})},\ \Eprint
  {http://arxiv.org/abs/1609.00669} {arXiv:1609.00669 [hep-th]} \BibitemShut
  {NoStop}%
\bibitem [{\citenamefont {Caron-Huot}\ \emph {et~al.}(2019)\citenamefont
  {Caron-Huot}, \citenamefont {Dixon}, \citenamefont {Dulat}, \citenamefont
  {von Hippel}, \citenamefont {McLeod},\ and\ \citenamefont
  {Papathanasiou}}]{Caron-Huot:2019vjl}%
  \BibitemOpen
  \bibfield  {author} {\bibinfo {author} {\bibfnamefont {S.}~\bibnamefont
  {Caron-Huot}}, \bibinfo {author} {\bibfnamefont {L.~J.}\ \bibnamefont
  {Dixon}}, \bibinfo {author} {\bibfnamefont {F.}~\bibnamefont {Dulat}},
  \bibinfo {author} {\bibfnamefont {M.}~\bibnamefont {von Hippel}}, \bibinfo
  {author} {\bibfnamefont {A.~J.}\ \bibnamefont {McLeod}}, \ and\ \bibinfo
  {author} {\bibfnamefont {G.}~\bibnamefont {Papathanasiou}},\ }\href {\doibase
  10.1007/JHEP08(2019)016} {\bibfield  {journal} {\bibinfo  {journal} {JHEP}\
  }\textbf {\bibinfo {volume} {08}},\ \bibinfo {pages} {016} (\bibinfo {year}
  {2019})},\ \Eprint {http://arxiv.org/abs/1903.10890} {arXiv:1903.10890
  [hep-th]} \BibitemShut {NoStop}%
\bibitem [{\citenamefont {Dixon}\ \emph {et~al.}(2021)\citenamefont {Dixon},
  \citenamefont {McLeod},\ and\ \citenamefont {Wilhelm}}]{Dixon:2020bbt}%
  \BibitemOpen
  \bibfield  {author} {\bibinfo {author} {\bibfnamefont {L.~J.}\ \bibnamefont
  {Dixon}}, \bibinfo {author} {\bibfnamefont {A.~J.}\ \bibnamefont {McLeod}}, \
  and\ \bibinfo {author} {\bibfnamefont {M.}~\bibnamefont {Wilhelm}},\ }\href
  {\doibase 10.1007/JHEP04(2021)147} {\bibfield  {journal} {\bibinfo  {journal}
  {JHEP}\ }\textbf {\bibinfo {volume} {04}},\ \bibinfo {pages} {147} (\bibinfo
  {year} {2021})},\ \Eprint {http://arxiv.org/abs/2012.12286} {arXiv:2012.12286
  [hep-th]} \BibitemShut {NoStop}%
\bibitem [{\citenamefont {Dixon}\ \emph {et~al.}(2022)\citenamefont {Dixon},
  \citenamefont {Gurdogan}, \citenamefont {McLeod},\ and\ \citenamefont
  {Wilhelm}}]{Dixon:2022rse}%
  \BibitemOpen
  \bibfield  {author} {\bibinfo {author} {\bibfnamefont {L.~J.}\ \bibnamefont
  {Dixon}}, \bibinfo {author} {\bibfnamefont {O.}~\bibnamefont {Gurdogan}},
  \bibinfo {author} {\bibfnamefont {A.~J.}\ \bibnamefont {McLeod}}, \ and\
  \bibinfo {author} {\bibfnamefont {M.}~\bibnamefont {Wilhelm}},\ }\href
  {\doibase 10.1007/JHEP07(2022)153} {\bibfield  {journal} {\bibinfo  {journal}
  {JHEP}\ }\textbf {\bibinfo {volume} {07}},\ \bibinfo {pages} {153} (\bibinfo
  {year} {2022})},\ \Eprint {http://arxiv.org/abs/2204.11901} {arXiv:2204.11901
  [hep-th]} \BibitemShut {NoStop}%
\bibitem [{\citenamefont {Basso}\ \emph {et~al.}(2025)\citenamefont {Basso},
  \citenamefont {Dixon},\ and\ \citenamefont {Tumanov}}]{Basso:2024hlx}%
  \BibitemOpen
  \bibfield  {author} {\bibinfo {author} {\bibfnamefont {B.}~\bibnamefont
  {Basso}}, \bibinfo {author} {\bibfnamefont {L.~J.}\ \bibnamefont {Dixon}}, \
  and\ \bibinfo {author} {\bibfnamefont {A.~G.}\ \bibnamefont {Tumanov}},\
  }\href {\doibase 10.1007/JHEP02(2025)034} {\bibfield  {journal} {\bibinfo
  {journal} {JHEP}\ }\textbf {\bibinfo {volume} {02}},\ \bibinfo {pages} {034}
  (\bibinfo {year} {2025})},\ \Eprint {http://arxiv.org/abs/2410.22402}
  {arXiv:2410.22402 [hep-th]} \BibitemShut {NoStop}%
\bibitem [{\citenamefont {Cai}\ \emph {et~al.}(2024)\citenamefont {Cai},
  \citenamefont {Merz}, \citenamefont {Charton}, \citenamefont {Nolte},
  \citenamefont {Wilhelm}, \citenamefont {Cranmer},\ and\ \citenamefont
  {Dixon}}]{Cai:2024znx}%
  \BibitemOpen
  \bibfield  {author} {\bibinfo {author} {\bibfnamefont {T.}~\bibnamefont
  {Cai}}, \bibinfo {author} {\bibfnamefont {G.~W.}\ \bibnamefont {Merz}},
  \bibinfo {author} {\bibfnamefont {F.}~\bibnamefont {Charton}}, \bibinfo
  {author} {\bibfnamefont {N.}~\bibnamefont {Nolte}}, \bibinfo {author}
  {\bibfnamefont {M.}~\bibnamefont {Wilhelm}}, \bibinfo {author} {\bibfnamefont
  {K.}~\bibnamefont {Cranmer}}, \ and\ \bibinfo {author} {\bibfnamefont
  {L.~J.}\ \bibnamefont {Dixon}},\ }\href {\doibase 10.1088/2632-2153/ad743e}
  {\bibfield  {journal} {\bibinfo  {journal} {Mach. Learn. Sci. Tech.}\
  }\textbf {\bibinfo {volume} {5}},\ \bibinfo {pages} {035073} (\bibinfo {year}
  {2024})},\ \Eprint {http://arxiv.org/abs/2405.06107} {arXiv:2405.06107
  [cs.LG]} \BibitemShut {NoStop}%
\bibitem [{\citenamefont {Cai}\ \emph {et~al.}(2025)\citenamefont {Cai},
  \citenamefont {Charton}, \citenamefont {Cranmer}, \citenamefont {Dixon},
  \citenamefont {Merz},\ and\ \citenamefont {Wilhelm}}]{Cai:2025atc}%
  \BibitemOpen
  \bibfield  {author} {\bibinfo {author} {\bibfnamefont {T.}~\bibnamefont
  {Cai}}, \bibinfo {author} {\bibfnamefont {F.}~\bibnamefont {Charton}},
  \bibinfo {author} {\bibfnamefont {K.}~\bibnamefont {Cranmer}}, \bibinfo
  {author} {\bibfnamefont {L.~J.}\ \bibnamefont {Dixon}}, \bibinfo {author}
  {\bibfnamefont {G.~W.}\ \bibnamefont {Merz}}, \ and\ \bibinfo {author}
  {\bibfnamefont {M.}~\bibnamefont {Wilhelm}},\ }\href {\doibase
  10.1007/JHEP04(2025)143} {\bibfield  {journal} {\bibinfo  {journal} {JHEP}\
  }\textbf {\bibinfo {volume} {04}},\ \bibinfo {pages} {143} (\bibinfo {year}
  {2025})},\ \Eprint {http://arxiv.org/abs/2501.05743} {arXiv:2501.05743
  [hep-th]} \BibitemShut {NoStop}%
\bibitem [{\citenamefont {Guo}\ \emph {et~al.}(2021)\citenamefont {Guo},
  \citenamefont {Wang},\ and\ \citenamefont {Yang}}]{Guo:2021bym}%
  \BibitemOpen
  \bibfield  {author} {\bibinfo {author} {\bibfnamefont {Y.}~\bibnamefont
  {Guo}}, \bibinfo {author} {\bibfnamefont {L.}~\bibnamefont {Wang}}, \ and\
  \bibinfo {author} {\bibfnamefont {G.}~\bibnamefont {Yang}},\ }\href {\doibase
  10.1103/PhysRevLett.127.151602} {\bibfield  {journal} {\bibinfo  {journal}
  {Phys. Rev. Lett.}\ }\textbf {\bibinfo {volume} {127}},\ \bibinfo {pages}
  {151602} (\bibinfo {year} {2021})},\ \Eprint
  {http://arxiv.org/abs/2106.01374} {arXiv:2106.01374 [hep-th]} \BibitemShut
  {NoStop}%
\bibitem [{\citenamefont {Dersy}\ \emph {et~al.}(2024)\citenamefont {Dersy},
  \citenamefont {Schwartz},\ and\ \citenamefont {Zhang}}]{Dersy:2022bym}%
  \BibitemOpen
  \bibfield  {author} {\bibinfo {author} {\bibfnamefont {A.}~\bibnamefont
  {Dersy}}, \bibinfo {author} {\bibfnamefont {M.~D.}\ \bibnamefont {Schwartz}},
  \ and\ \bibinfo {author} {\bibfnamefont {X.}~\bibnamefont {Zhang}},\ }\href
  {\doibase 10.1142/S2810939223500028} {\bibfield  {journal} {\bibinfo
  {journal} {Int. J. Data Sci. Math. Sci.}\ }\textbf {\bibinfo {volume} {1}},\
  \bibinfo {pages} {135} (\bibinfo {year} {2024})},\ \Eprint
  {http://arxiv.org/abs/2206.04115} {arXiv:2206.04115 [cs.LG]} \BibitemShut
  {NoStop}%
\bibitem [{\citenamefont {Chicherin}\ \emph {et~al.}(2018)\citenamefont
  {Chicherin}, \citenamefont {Henn},\ and\ \citenamefont
  {Mitev}}]{Chicherin:2017dob}%
  \BibitemOpen
  \bibfield  {author} {\bibinfo {author} {\bibfnamefont {D.}~\bibnamefont
  {Chicherin}}, \bibinfo {author} {\bibfnamefont {J.}~\bibnamefont {Henn}}, \
  and\ \bibinfo {author} {\bibfnamefont {V.}~\bibnamefont {Mitev}},\ }\href
  {\doibase 10.1007/JHEP05(2018)164} {\bibfield  {journal} {\bibinfo  {journal}
  {JHEP}\ }\textbf {\bibinfo {volume} {05}},\ \bibinfo {pages} {164} (\bibinfo
  {year} {2018})},\ \Eprint {http://arxiv.org/abs/1712.09610} {arXiv:1712.09610
  [hep-th]} \BibitemShut {NoStop}%
\bibitem [{\citenamefont {Caron-Huot}\ \emph {et~al.}(2018)\citenamefont
  {Caron-Huot}, \citenamefont {Dixon}, \citenamefont {von Hippel},
  \citenamefont {McLeod},\ and\ \citenamefont
  {Papathanasiou}}]{Caron-Huot:2018dsv}%
  \BibitemOpen
  \bibfield  {author} {\bibinfo {author} {\bibfnamefont {S.}~\bibnamefont
  {Caron-Huot}}, \bibinfo {author} {\bibfnamefont {L.~J.}\ \bibnamefont
  {Dixon}}, \bibinfo {author} {\bibfnamefont {M.}~\bibnamefont {von Hippel}},
  \bibinfo {author} {\bibfnamefont {A.~J.}\ \bibnamefont {McLeod}}, \ and\
  \bibinfo {author} {\bibfnamefont {G.}~\bibnamefont {Papathanasiou}},\ }\href
  {\doibase 10.1007/JHEP07(2018)170} {\bibfield  {journal} {\bibinfo  {journal}
  {JHEP}\ }\textbf {\bibinfo {volume} {07}},\ \bibinfo {pages} {170} (\bibinfo
  {year} {2018})},\ \Eprint {http://arxiv.org/abs/1806.01361} {arXiv:1806.01361
  [hep-th]} \BibitemShut {NoStop}%
\bibitem [{\citenamefont {Henn}\ \emph {et~al.}(2018)\citenamefont {Henn},
  \citenamefont {Herrmann},\ and\ \citenamefont
  {Parra-Martinez}}]{Henn:2018cdp}%
  \BibitemOpen
  \bibfield  {author} {\bibinfo {author} {\bibfnamefont {J.}~\bibnamefont
  {Henn}}, \bibinfo {author} {\bibfnamefont {E.}~\bibnamefont {Herrmann}}, \
  and\ \bibinfo {author} {\bibfnamefont {J.}~\bibnamefont {Parra-Martinez}},\
  }\href {\doibase 10.1007/JHEP10(2018)059} {\bibfield  {journal} {\bibinfo
  {journal} {JHEP}\ }\textbf {\bibinfo {volume} {10}},\ \bibinfo {pages} {059}
  (\bibinfo {year} {2018})},\ \Eprint {http://arxiv.org/abs/1806.06072}
  {arXiv:1806.06072 [hep-th]} \BibitemShut {NoStop}%
\bibitem [{\citenamefont {He}\ \emph {et~al.}(2021)\citenamefont {He},
  \citenamefont {Li},\ and\ \citenamefont {Yang}}]{He:2021non}%
  \BibitemOpen
  \bibfield  {author} {\bibinfo {author} {\bibfnamefont {S.}~\bibnamefont
  {He}}, \bibinfo {author} {\bibfnamefont {Z.}~\bibnamefont {Li}}, \ and\
  \bibinfo {author} {\bibfnamefont {Q.}~\bibnamefont {Yang}},\ }\href {\doibase
  10.1007/JHEP12(2021)110} {\bibfield  {journal} {\bibinfo  {journal} {JHEP}\
  }\textbf {\bibinfo {volume} {12}},\ \bibinfo {pages} {110} (\bibinfo {year}
  {2021})},\ \bibinfo {note} {[Erratum: JHEP 05, 075 (2022)]},\ \Eprint
  {http://arxiv.org/abs/2106.09314} {arXiv:2106.09314 [hep-th]} \BibitemShut
  {NoStop}%
\bibitem [{\citenamefont {He}\ \emph {et~al.}(2026)\citenamefont {He},
  \citenamefont {Li},\ and\ \citenamefont {Yang}}]{He:2021eec}%
  \BibitemOpen
  \bibfield  {author} {\bibinfo {author} {\bibfnamefont {S.}~\bibnamefont
  {He}}, \bibinfo {author} {\bibfnamefont {Z.}~\bibnamefont {Li}}, \ and\
  \bibinfo {author} {\bibfnamefont {Q.}~\bibnamefont {Yang}},\ }\href {\doibase
  10.1007/JHEP06(2026)067} {\bibfield  {journal} {\bibinfo  {journal} {JHEP}\
  }\textbf {\bibinfo {volume} {06}},\ \bibinfo {pages} {067} (\bibinfo {year}
  {2026})},\ \Eprint {http://arxiv.org/abs/2112.11842} {arXiv:2112.11842
  [hep-th]} \BibitemShut {NoStop}%
\bibitem [{\citenamefont {Barrera}\ \emph {et~al.}(2025)\citenamefont
  {Barrera}, \citenamefont {Dersy}, \citenamefont {Husain}, \citenamefont
  {Schwartz},\ and\ \citenamefont {Zhang}}]{Barrera:2025anr}%
  \BibitemOpen
  \bibfield  {author} {\bibinfo {author} {\bibfnamefont {O.}~\bibnamefont
  {Barrera}}, \bibinfo {author} {\bibfnamefont {A.}~\bibnamefont {Dersy}},
  \bibinfo {author} {\bibfnamefont {R.}~\bibnamefont {Husain}}, \bibinfo
  {author} {\bibfnamefont {M.~D.}\ \bibnamefont {Schwartz}}, \ and\ \bibinfo
  {author} {\bibfnamefont {X.}~\bibnamefont {Zhang}},\ }\href@noop {} {\
  (\bibinfo {year} {2025})},\ \Eprint {http://arxiv.org/abs/2507.17815}
  {arXiv:2507.17815 [hep-th]} \BibitemShut {NoStop}%
\bibitem [{\citenamefont {Morales}\ \emph {et~al.}(2023)\citenamefont
  {Morales}, \citenamefont {Spiering}, \citenamefont {Wilhelm}, \citenamefont
  {Yang},\ and\ \citenamefont {Zhang}}]{Morales:2022csr}%
  \BibitemOpen
  \bibfield  {author} {\bibinfo {author} {\bibfnamefont {R.}~\bibnamefont
  {Morales}}, \bibinfo {author} {\bibfnamefont {A.}~\bibnamefont {Spiering}},
  \bibinfo {author} {\bibfnamefont {M.}~\bibnamefont {Wilhelm}}, \bibinfo
  {author} {\bibfnamefont {Q.}~\bibnamefont {Yang}}, \ and\ \bibinfo {author}
  {\bibfnamefont {C.}~\bibnamefont {Zhang}},\ }\href {\doibase
  10.1103/PhysRevLett.131.041601} {\bibfield  {journal} {\bibinfo  {journal}
  {Phys. Rev. Lett.}\ }\textbf {\bibinfo {volume} {131}},\ \bibinfo {pages}
  {041601} (\bibinfo {year} {2023})},\ \Eprint
  {http://arxiv.org/abs/2212.09762} {arXiv:2212.09762 [hep-th]} \BibitemShut
  {NoStop}%
\bibitem [{\citenamefont {Schwartz}(2026)}]{Schwartz:2026afx}%
  \BibitemOpen
  \bibfield  {author} {\bibinfo {author} {\bibfnamefont {M.~D.}\ \bibnamefont
  {Schwartz}},\ }\href@noop {} {\  (\bibinfo {year} {2026})},\ \Eprint
  {http://arxiv.org/abs/2610.02286} {arXiv:2610.02286 [hep-ph]} \BibitemShut
  {NoStop}%
\bibitem [{\citenamefont {Brown}(2017)}]{Brown:2015fyf}%
  \BibitemOpen
  \bibfield  {author} {\bibinfo {author} {\bibfnamefont {F.}~\bibnamefont
  {Brown}},\ }\href {\doibase 10.4310/CNTP.2017.v11.n3.a1} {\bibfield
  {journal} {\bibinfo  {journal} {Commun. Num. Theor. Phys.}\ }\textbf
  {\bibinfo {volume} {11}},\ \bibinfo {pages} {453} (\bibinfo {year} {2017})},\
  \Eprint {http://arxiv.org/abs/1512.06409} {arXiv:1512.06409 [math-ph]}
  \BibitemShut {NoStop}%
\bibitem [{\citenamefont {Remiddi}\ and\ \citenamefont
  {Vermaseren}(2000)}]{Remiddi:1999ew}%
  \BibitemOpen
  \bibfield  {author} {\bibinfo {author} {\bibfnamefont {E.}~\bibnamefont
  {Remiddi}}\ and\ \bibinfo {author} {\bibfnamefont {J.~A.~M.}\ \bibnamefont
  {Vermaseren}},\ }\href {\doibase 10.1142/S0217751X00000367} {\bibfield
  {journal} {\bibinfo  {journal} {Int. J. Mod. Phys. A}\ }\textbf {\bibinfo
  {volume} {15}},\ \bibinfo {pages} {725} (\bibinfo {year} {2000})},\ \Eprint
  {http://arxiv.org/abs/hep-ph/9905237} {arXiv:hep-ph/9905237} \BibitemShut
  {NoStop}%
\bibitem [{\citenamefont {Adams}\ and\ \citenamefont
  {Weinzierl}(2018{\natexlab{a}})}]{Adams:2017ejb}%
  \BibitemOpen
  \bibfield  {author} {\bibinfo {author} {\bibfnamefont {L.}~\bibnamefont
  {Adams}}\ and\ \bibinfo {author} {\bibfnamefont {S.}~\bibnamefont
  {Weinzierl}},\ }\href {\doibase 10.4310/CNTP.2018.v12.n2.a1} {\bibfield
  {journal} {\bibinfo  {journal} {Commun. Num. Theor. Phys.}\ }\textbf
  {\bibinfo {volume} {12}},\ \bibinfo {pages} {193} (\bibinfo {year}
  {2018}{\natexlab{a}})},\ \Eprint {http://arxiv.org/abs/1704.08895}
  {arXiv:1704.08895} \BibitemShut {NoStop}%
\bibitem [{\citenamefont {Adams}\ and\ \citenamefont
  {Weinzierl}(2018{\natexlab{b}})}]{Adams:2018yfj}%
  \BibitemOpen
  \bibfield  {author} {\bibinfo {author} {\bibfnamefont {L.}~\bibnamefont
  {Adams}}\ and\ \bibinfo {author} {\bibfnamefont {S.}~\bibnamefont
  {Weinzierl}},\ }\href {\doibase 10.1016/j.physletb.2018.04.002} {\bibfield
  {journal} {\bibinfo  {journal} {Phys. Lett. B}\ }\textbf {\bibinfo {volume}
  {781}},\ \bibinfo {pages} {270} (\bibinfo {year} {2018}{\natexlab{b}})},\
  \Eprint {http://arxiv.org/abs/1802.05020} {arXiv:1802.05020 [hep-ph]}
  \BibitemShut {NoStop}%
\bibitem [{\citenamefont {Bogner}\ \emph {et~al.}(2017)\citenamefont {Bogner},
  \citenamefont {Schweitzer},\ and\ \citenamefont
  {Weinzierl}}]{Bogner:2017vim}%
  \BibitemOpen
  \bibfield  {author} {\bibinfo {author} {\bibfnamefont {C.}~\bibnamefont
  {Bogner}}, \bibinfo {author} {\bibfnamefont {A.}~\bibnamefont {Schweitzer}},
  \ and\ \bibinfo {author} {\bibfnamefont {S.}~\bibnamefont {Weinzierl}},\
  }\href {\doibase 10.1016/j.nuclphysb.2017.07.008} {\bibfield  {journal}
  {\bibinfo  {journal} {Nucl. Phys. B}\ }\textbf {\bibinfo {volume} {922}},\
  \bibinfo {pages} {528} (\bibinfo {year} {2017})},\ \Eprint
  {http://arxiv.org/abs/1705.08952} {arXiv:1705.08952 [hep-ph]} \BibitemShut
  {NoStop}%
\bibitem [{\citenamefont {Broedel}\ \emph {et~al.}(2018)\citenamefont
  {Broedel}, \citenamefont {Duhr}, \citenamefont {Dulat},\ and\ \citenamefont
  {Tancredi}}]{Broedel:2017kkb}%
  \BibitemOpen
  \bibfield  {author} {\bibinfo {author} {\bibfnamefont {J.}~\bibnamefont
  {Broedel}}, \bibinfo {author} {\bibfnamefont {C.}~\bibnamefont {Duhr}},
  \bibinfo {author} {\bibfnamefont {F.}~\bibnamefont {Dulat}}, \ and\ \bibinfo
  {author} {\bibfnamefont {L.}~\bibnamefont {Tancredi}},\ }\href {\doibase
  10.1007/JHEP05(2018)093} {\bibfield  {journal} {\bibinfo  {journal} {JHEP}\
  }\textbf {\bibinfo {volume} {05}},\ \bibinfo {pages} {093} (\bibinfo {year}
  {2018})},\ \Eprint {http://arxiv.org/abs/1712.07089} {arXiv:1712.07089
  [hep-th]} \BibitemShut {NoStop}%
\bibitem [{\citenamefont {Sterman}(1975)}]{Sterman:1975xv}%
  \BibitemOpen
  \bibfield  {author} {\bibinfo {author} {\bibfnamefont {G.~F.}\ \bibnamefont
  {Sterman}},\ }\href@noop {} {\bibfield  {journal} {\bibinfo  {journal}
  {ILL-TH-75-32}\ } (\bibinfo {year} {1975})}\BibitemShut {NoStop}%
\bibitem [{\citenamefont {Sveshnikov}\ and\ \citenamefont
  {Tkachov}(1996)}]{Sveshnikov:1995vi}%
  \BibitemOpen
  \bibfield  {author} {\bibinfo {author} {\bibfnamefont {N.~A.}\ \bibnamefont
  {Sveshnikov}}\ and\ \bibinfo {author} {\bibfnamefont {F.~V.}\ \bibnamefont
  {Tkachov}},\ }\href {\doibase 10.1016/0370-2693(96)00558-8} {\bibfield
  {journal} {\bibinfo  {journal} {Phys. Lett. B}\ }\textbf {\bibinfo {volume}
  {382}},\ \bibinfo {pages} {403} (\bibinfo {year} {1996})},\ \Eprint
  {http://arxiv.org/abs/hep-ph/9512370} {arXiv:hep-ph/9512370} \BibitemShut
  {NoStop}%
\bibitem [{\citenamefont {Tkachov}(1997)}]{Tkachov:1995kk}%
  \BibitemOpen
  \bibfield  {author} {\bibinfo {author} {\bibfnamefont {F.~V.}\ \bibnamefont
  {Tkachov}},\ }\href {\doibase 10.1142/S0217751X97002899} {\bibfield
  {journal} {\bibinfo  {journal} {Int. J. Mod. Phys. A}\ }\textbf {\bibinfo
  {volume} {12}},\ \bibinfo {pages} {5411} (\bibinfo {year} {1997})},\ \Eprint
  {http://arxiv.org/abs/hep-ph/9601308} {arXiv:hep-ph/9601308} \BibitemShut
  {NoStop}%
\bibitem [{\citenamefont {Korchemsky}\ and\ \citenamefont
  {Sterman}(1999)}]{Korchemsky:1999kt}%
  \BibitemOpen
  \bibfield  {author} {\bibinfo {author} {\bibfnamefont {G.~P.}\ \bibnamefont
  {Korchemsky}}\ and\ \bibinfo {author} {\bibfnamefont {G.~F.}\ \bibnamefont
  {Sterman}},\ }\href {\doibase 10.1016/S0550-3213(99)00308-9} {\bibfield
  {journal} {\bibinfo  {journal} {Nucl. Phys. B}\ }\textbf {\bibinfo {volume}
  {555}},\ \bibinfo {pages} {335} (\bibinfo {year} {1999})},\ \Eprint
  {http://arxiv.org/abs/hep-ph/9902341} {arXiv:hep-ph/9902341} \BibitemShut
  {NoStop}%
\bibitem [{\citenamefont {Hofman}\ and\ \citenamefont
  {Maldacena}(2008)}]{Hofman:2008ar}%
  \BibitemOpen
  \bibfield  {author} {\bibinfo {author} {\bibfnamefont {D.~M.}\ \bibnamefont
  {Hofman}}\ and\ \bibinfo {author} {\bibfnamefont {J.}~\bibnamefont
  {Maldacena}},\ }\href {\doibase 10.1088/1126-6708/2008/05/012} {\bibfield
  {journal} {\bibinfo  {journal} {JHEP}\ }\textbf {\bibinfo {volume} {05}},\
  \bibinfo {pages} {012} (\bibinfo {year} {2008})},\ \Eprint
  {http://arxiv.org/abs/0803.1467} {arXiv:0803.1467 [hep-th]} \BibitemShut
  {NoStop}%
\bibitem [{\citenamefont {Belitsky}\ \emph
  {et~al.}(2014{\natexlab{b}})\citenamefont {Belitsky}, \citenamefont
  {Hohenegger}, \citenamefont {Korchemsky}, \citenamefont {Sokatchev},\ and\
  \citenamefont {Zhiboedov}}]{Belitsky:2013xxa}%
  \BibitemOpen
  \bibfield  {author} {\bibinfo {author} {\bibfnamefont {A.~V.}\ \bibnamefont
  {Belitsky}}, \bibinfo {author} {\bibfnamefont {S.}~\bibnamefont
  {Hohenegger}}, \bibinfo {author} {\bibfnamefont {G.~P.}\ \bibnamefont
  {Korchemsky}}, \bibinfo {author} {\bibfnamefont {E.}~\bibnamefont
  {Sokatchev}}, \ and\ \bibinfo {author} {\bibfnamefont {A.}~\bibnamefont
  {Zhiboedov}},\ }\href {\doibase 10.1016/j.nuclphysb.2014.04.020} {\bibfield
  {journal} {\bibinfo  {journal} {Nucl. Phys. B}\ }\textbf {\bibinfo {volume}
  {884}},\ \bibinfo {pages} {305} (\bibinfo {year} {2014}{\natexlab{b}})},\
  \Eprint {http://arxiv.org/abs/1309.0769} {arXiv:1309.0769 [hep-th]}
  \BibitemShut {NoStop}%
\bibitem [{\citenamefont {Belitsky}\ \emph
  {et~al.}(2014{\natexlab{c}})\citenamefont {Belitsky}, \citenamefont
  {Hohenegger}, \citenamefont {Korchemsky}, \citenamefont {Sokatchev},\ and\
  \citenamefont {Zhiboedov}}]{Belitsky:2013bja}%
  \BibitemOpen
  \bibfield  {author} {\bibinfo {author} {\bibfnamefont {A.~V.}\ \bibnamefont
  {Belitsky}}, \bibinfo {author} {\bibfnamefont {S.}~\bibnamefont
  {Hohenegger}}, \bibinfo {author} {\bibfnamefont {G.~P.}\ \bibnamefont
  {Korchemsky}}, \bibinfo {author} {\bibfnamefont {E.}~\bibnamefont
  {Sokatchev}}, \ and\ \bibinfo {author} {\bibfnamefont {A.}~\bibnamefont
  {Zhiboedov}},\ }\href {\doibase 10.1016/j.nuclphysb.2014.04.019} {\bibfield
  {journal} {\bibinfo  {journal} {Nucl. Phys. B}\ }\textbf {\bibinfo {volume}
  {884}},\ \bibinfo {pages} {206} (\bibinfo {year} {2014}{\natexlab{c}})},\
  \Eprint {http://arxiv.org/abs/1309.1424} {arXiv:1309.1424 [hep-th]}
  \BibitemShut {NoStop}%
\bibitem [{\citenamefont {Cheng}\ and\ \citenamefont
  {Wu}(1987)}]{Cheng:1987ga}%
  \BibitemOpen
  \bibfield  {author} {\bibinfo {author} {\bibfnamefont {H.}~\bibnamefont
  {Cheng}}\ and\ \bibinfo {author} {\bibfnamefont {T.~T.}\ \bibnamefont {Wu}},\
  }\href@noop {} {\emph {\bibinfo {title} {{Expanding Protons: Scattering at
  High Energies}}}}\ (\bibinfo  {publisher} {MIT Press},\ \bibinfo {address}
  {Cambridge, MA},\ \bibinfo {year} {1987})\BibitemShut {NoStop}%
\bibitem [{\citenamefont {M{\"u}ller-Stach}\ \emph {et~al.}(2012)\citenamefont
  {M{\"u}ller-Stach}, \citenamefont {Weinzierl},\ and\ \citenamefont
  {Zayadeh}}]{MullerStach:2011ru}%
  \BibitemOpen
  \bibfield  {author} {\bibinfo {author} {\bibfnamefont {S.}~\bibnamefont
  {M{\"u}ller-Stach}}, \bibinfo {author} {\bibfnamefont {S.}~\bibnamefont
  {Weinzierl}}, \ and\ \bibinfo {author} {\bibfnamefont {R.}~\bibnamefont
  {Zayadeh}},\ }\href {\doibase 10.4310/CNTP.2012.v6.n1.a5} {\bibfield
  {journal} {\bibinfo  {journal} {Commun. Num. Theor. Phys.}\ }\textbf
  {\bibinfo {volume} {6}},\ \bibinfo {pages} {203} (\bibinfo {year} {2012})},\
  \Eprint {http://arxiv.org/abs/1112.4360} {arXiv:1112.4360 [hep-ph]}
  \BibitemShut {NoStop}%
\bibitem [{\citenamefont {Panzer}(2015)}]{Panzer:2014caa}%
  \BibitemOpen
  \bibfield  {author} {\bibinfo {author} {\bibfnamefont {E.}~\bibnamefont
  {Panzer}},\ }\href {\doibase 10.1016/j.cpc.2014.10.019} {\bibfield  {journal}
  {\bibinfo  {journal} {Comput. Phys. Commun.}\ }\textbf {\bibinfo {volume}
  {188}},\ \bibinfo {pages} {148} (\bibinfo {year} {2015})},\ \Eprint
  {http://arxiv.org/abs/1403.3385} {arXiv:1403.3385 [hep-th]} \BibitemShut
  {NoStop}%
\bibitem [{\citenamefont {Franel}(1894)}]{Franel:1894}%
  \BibitemOpen
  \bibfield  {author} {\bibinfo {author} {\bibfnamefont {J.}~\bibnamefont
  {Franel}},\ }\href@noop {} {\bibfield  {journal} {\bibinfo  {journal}
  {L'Interm\'ediaire des Math\'ematiciens}\ }\textbf {\bibinfo {volume} {1}},\
  \bibinfo {pages} {45} (\bibinfo {year} {1894})}\BibitemShut {NoStop}%
\bibitem [{\citenamefont {Zagier}(2009)}]{Zagier:2009apery}%
  \BibitemOpen
  \bibfield  {author} {\bibinfo {author} {\bibfnamefont {D.}~\bibnamefont
  {Zagier}},\ }in\ \href@noop {} {\emph {\bibinfo {booktitle} {{Groups and
  Symmetries: From Neolithic Scots to John McKay}}}},\ \bibinfo {series} {CRM
  Proceedings and Lecture Notes}, Vol.~\bibinfo {volume} {47}\ (\bibinfo
  {publisher} {American Mathematical Society},\ \bibinfo {address} {Providence,
  RI},\ \bibinfo {year} {2009})\ pp.\ \bibinfo {pages} {349--366}\BibitemShut
  {NoStop}%
\bibitem [{\citenamefont {Maitre}(2006)}]{Maitre:2005uu}%
  \BibitemOpen
  \bibfield  {author} {\bibinfo {author} {\bibfnamefont {D.}~\bibnamefont
  {Maitre}},\ }\href {\doibase 10.1016/j.cpc.2005.10.008} {\bibfield  {journal}
  {\bibinfo  {journal} {Comput. Phys. Commun.}\ }\textbf {\bibinfo {volume}
  {174}},\ \bibinfo {pages} {222} (\bibinfo {year} {2006})},\ \Eprint
  {http://arxiv.org/abs/hep-ph/0507152} {arXiv:hep-ph/0507152} \BibitemShut
  {NoStop}%
\bibitem [{\citenamefont {Duhr}\ and\ \citenamefont
  {Dulat}(2019)}]{Duhr:2019tlz}%
  \BibitemOpen
  \bibfield  {author} {\bibinfo {author} {\bibfnamefont {C.}~\bibnamefont
  {Duhr}}\ and\ \bibinfo {author} {\bibfnamefont {F.}~\bibnamefont {Dulat}},\
  }\href {\doibase 10.1007/JHEP08(2019)135} {\bibfield  {journal} {\bibinfo
  {journal} {JHEP}\ }\textbf {\bibinfo {volume} {08}},\ \bibinfo {pages} {135}
  (\bibinfo {year} {2019})},\ \Eprint {http://arxiv.org/abs/1904.07279}
  {arXiv:1904.07279 [hep-th]} \BibitemShut {NoStop}%
\bibitem [{\citenamefont {Hannesdottir}\ \emph {et~al.}(2025)\citenamefont
  {Hannesdottir}, \citenamefont {McLeod}, \citenamefont {Schwartz},\ and\
  \citenamefont {Vergu}}]{Hannesdottir:2024hke}%
  \BibitemOpen
  \bibfield  {author} {\bibinfo {author} {\bibfnamefont {H.~S.}\ \bibnamefont
  {Hannesdottir}}, \bibinfo {author} {\bibfnamefont {A.~J.}\ \bibnamefont
  {McLeod}}, \bibinfo {author} {\bibfnamefont {M.~D.}\ \bibnamefont
  {Schwartz}}, \ and\ \bibinfo {author} {\bibfnamefont {C.}~\bibnamefont
  {Vergu}},\ }\href {\doibase 10.1103/PhysRevD.111.085003} {\bibfield
  {journal} {\bibinfo  {journal} {Phys. Rev. D}\ }\textbf {\bibinfo {volume}
  {111}},\ \bibinfo {pages} {085003} (\bibinfo {year} {2025})},\ \Eprint
  {http://arxiv.org/abs/2410.02424} {arXiv:2410.02424 [hep-ph]} \BibitemShut
  {NoStop}%
\bibitem [{\citenamefont {Gong}\ \emph {et~al.}(2025)\citenamefont {Gong},
  \citenamefont {Pokraka}, \citenamefont {Yan},\ and\ \citenamefont
  {Zhang}}]{Gong:2025eec}%
  \BibitemOpen
  \bibfield  {author} {\bibinfo {author} {\bibfnamefont {J.}~\bibnamefont
  {Gong}}, \bibinfo {author} {\bibfnamefont {A.}~\bibnamefont {Pokraka}},
  \bibinfo {author} {\bibfnamefont {K.}~\bibnamefont {Yan}}, \ and\ \bibinfo
  {author} {\bibfnamefont {X.}~\bibnamefont {Zhang}},\ }\href@noop {} {\
  (\bibinfo {year} {2025})},\ \Eprint {http://arxiv.org/abs/2509.22782}
  {arXiv:2509.22782 [hep-th]} \BibitemShut {NoStop}%
\bibitem [{\citenamefont {Fevola}\ \emph {et~al.}(2024)\citenamefont {Fevola},
  \citenamefont {Mizera},\ and\ \citenamefont {Telen}}]{Fevola:2023fzn}%
  \BibitemOpen
  \bibfield  {author} {\bibinfo {author} {\bibfnamefont {C.}~\bibnamefont
  {Fevola}}, \bibinfo {author} {\bibfnamefont {S.}~\bibnamefont {Mizera}}, \
  and\ \bibinfo {author} {\bibfnamefont {S.}~\bibnamefont {Telen}},\ }\href
  {\doibase 10.1016/j.cpc.2024.109278} {\bibfield  {journal} {\bibinfo
  {journal} {Comput. Phys. Commun.}\ }\textbf {\bibinfo {volume} {303}},\
  \bibinfo {pages} {109278} (\bibinfo {year} {2024})},\ \Eprint
  {http://arxiv.org/abs/2311.16219} {arXiv:2311.16219 [math-ph]} \BibitemShut
  {NoStop}%
\bibitem [{\citenamefont {Correia}\ \emph {et~al.}(2026)\citenamefont
  {Correia}, \citenamefont {Giroux},\ and\ \citenamefont
  {Mizera}}]{Correia:2025wtb}%
  \BibitemOpen
  \bibfield  {author} {\bibinfo {author} {\bibfnamefont {M.}~\bibnamefont
  {Correia}}, \bibinfo {author} {\bibfnamefont {M.}~\bibnamefont {Giroux}}, \
  and\ \bibinfo {author} {\bibfnamefont {S.}~\bibnamefont {Mizera}},\ }\href
  {\doibase 10.1016/j.cpc.2025.109970} {\bibfield  {journal} {\bibinfo
  {journal} {Comput. Phys. Commun.}\ }\textbf {\bibinfo {volume} {320}},\
  \bibinfo {pages} {109970} (\bibinfo {year} {2026})},\ \Eprint
  {http://arxiv.org/abs/2503.16601} {arXiv:2503.16601 [hep-th]} \BibitemShut
  {NoStop}%
\bibitem [{\citenamefont {Kravchuk}\ and\ \citenamefont
  {Simmons-Duffin}(2018)}]{Kravchuk:2018htv}%
  \BibitemOpen
  \bibfield  {author} {\bibinfo {author} {\bibfnamefont {P.}~\bibnamefont
  {Kravchuk}}\ and\ \bibinfo {author} {\bibfnamefont {D.}~\bibnamefont
  {Simmons-Duffin}},\ }\href {\doibase 10.1007/JHEP11(2018)102} {\bibfield
  {journal} {\bibinfo  {journal} {JHEP}\ }\textbf {\bibinfo {volume} {11}},\
  \bibinfo {pages} {102} (\bibinfo {year} {2018})},\ \Eprint
  {http://arxiv.org/abs/1805.00098} {arXiv:1805.00098 [hep-th]} \BibitemShut
  {NoStop}%
\bibitem [{\citenamefont {Chen}\ \emph {et~al.}(2022)\citenamefont {Chen},
  \citenamefont {Moult},\ and\ \citenamefont {Zhu}}]{Chen:2021gdk}%
  \BibitemOpen
  \bibfield  {author} {\bibinfo {author} {\bibfnamefont {H.}~\bibnamefont
  {Chen}}, \bibinfo {author} {\bibfnamefont {I.}~\bibnamefont {Moult}}, \ and\
  \bibinfo {author} {\bibfnamefont {H.~X.}\ \bibnamefont {Zhu}},\ }\href
  {\doibase 10.1007/JHEP08(2022)233} {\bibfield  {journal} {\bibinfo  {journal}
  {JHEP}\ }\textbf {\bibinfo {volume} {08}},\ \bibinfo {pages} {233} (\bibinfo
  {year} {2022})},\ \Eprint {http://arxiv.org/abs/2104.00009} {arXiv:2104.00009
  [hep-ph]} \BibitemShut {NoStop}%
\bibitem [{\citenamefont {Chen}(2024)}]{Chen:2023zzh}%
  \BibitemOpen
  \bibfield  {author} {\bibinfo {author} {\bibfnamefont {H.}~\bibnamefont
  {Chen}},\ }\href {\doibase 10.1007/JHEP01(2024)035} {\bibfield  {journal}
  {\bibinfo  {journal} {JHEP}\ }\textbf {\bibinfo {volume} {01}},\ \bibinfo
  {pages} {035} (\bibinfo {year} {2024})},\ \Eprint
  {http://arxiv.org/abs/2311.00350} {arXiv:2311.00350 [hep-ph]} \BibitemShut
  {NoStop}%
\bibitem [{\citenamefont {Chen}\ \emph {et~al.}(2023)\citenamefont {Chen},
  \citenamefont {Zhou},\ and\ \citenamefont {Zhu}}]{Chen:2023pcr}%
  \BibitemOpen
  \bibfield  {author} {\bibinfo {author} {\bibfnamefont {H.}~\bibnamefont
  {Chen}}, \bibinfo {author} {\bibfnamefont {X.}~\bibnamefont {Zhou}}, \ and\
  \bibinfo {author} {\bibfnamefont {H.~X.}\ \bibnamefont {Zhu}},\ }\href@noop
  {} {\  (\bibinfo {year} {2023})},\ \Eprint {http://arxiv.org/abs/2301.03616}
  {arXiv:2301.03616 [hep-th]} \BibitemShut {NoStop}%
\bibitem [{\citenamefont {Ferguson}\ and\ \citenamefont
  {Bailey}(1992)}]{Ferguson:1992}%
  \BibitemOpen
  \bibfield  {author} {\bibinfo {author} {\bibfnamefont {H.}~\bibnamefont
  {Ferguson}}\ and\ \bibinfo {author} {\bibfnamefont {D.}~\bibnamefont
  {Bailey}},\ }\href@noop {} {\emph {\bibinfo {title} {A Polynomial Time,
  Numerically Stable Integer Relation Algorithm}}},\ \bibinfo {type} {Tech.
  Rep.}\ (\bibinfo  {institution} {NASA Ames Research Center},\ \bibinfo {year}
  {1992})\BibitemShut {NoStop}%
\bibitem [{\citenamefont {Ferguson}\ \emph {et~al.}(1999)\citenamefont
  {Ferguson}, \citenamefont {Bailey},\ and\ \citenamefont {Arno}}]{PSLQref}%
  \BibitemOpen
  \bibfield  {author} {\bibinfo {author} {\bibfnamefont {H.~R.~P.}\
  \bibnamefont {Ferguson}}, \bibinfo {author} {\bibfnamefont {D.~H.}\
  \bibnamefont {Bailey}}, \ and\ \bibinfo {author} {\bibfnamefont
  {S.}~\bibnamefont {Arno}},\ }\href {\doibase 10.1090/S0025-5718-99-00995-3}
  {\bibfield  {journal} {\bibinfo  {journal} {Math. Comput.}\ }\textbf
  {\bibinfo {volume} {68}},\ \bibinfo {pages} {351–369} (\bibinfo {year}
  {1999})}\BibitemShut {NoStop}%
\bibitem [{\citenamefont {Bailey}\ and\ \citenamefont
  {Broadhurst}(2001)}]{Bailey:1999nv}%
  \BibitemOpen
  \bibfield  {author} {\bibinfo {author} {\bibfnamefont {D.~H.}\ \bibnamefont
  {Bailey}}\ and\ \bibinfo {author} {\bibfnamefont {D.~J.}\ \bibnamefont
  {Broadhurst}},\ }\href {\doibase 10.1090/S0025-5718-00-01278-3} {\bibfield
  {journal} {\bibinfo  {journal} {Math. Comput.}\ }\textbf {\bibinfo {volume}
  {70}},\ \bibinfo {pages} {1719} (\bibinfo {year} {2001})},\ \Eprint
  {http://arxiv.org/abs/math/9905048} {arXiv:math/9905048} \BibitemShut
  {NoStop}%
\bibitem [{\citenamefont {Lenstra}\ \emph {et~al.}(1982)\citenamefont
  {Lenstra}, \citenamefont {Lenstra},\ and\ \citenamefont
  {Lov{\'a}sz}}]{Lenstra:1982}%
  \BibitemOpen
  \bibfield  {author} {\bibinfo {author} {\bibfnamefont {A.~K.}\ \bibnamefont
  {Lenstra}}, \bibinfo {author} {\bibfnamefont {H.~W.}\ \bibnamefont
  {Lenstra}}, \ and\ \bibinfo {author} {\bibfnamefont {L.}~\bibnamefont
  {Lov{\'a}sz}},\ }\href {\doibase 10.1007/BF01457454} {\bibfield  {journal}
  {\bibinfo  {journal} {Mathematische Annalen}\ }\textbf {\bibinfo {volume}
  {261}},\ \bibinfo {pages} {515} (\bibinfo {year} {1982})}\BibitemShut
  {NoStop}%
\bibitem [{\citenamefont {Nguyen}\ and\ \citenamefont
  {Stern}(2001)}]{Nguyen:2001}%
  \BibitemOpen
  \bibfield  {author} {\bibinfo {author} {\bibfnamefont {P.~Q.}\ \bibnamefont
  {Nguyen}}\ and\ \bibinfo {author} {\bibfnamefont {J.}~\bibnamefont {Stern}},\
  }in\ \href@noop {} {\emph {\bibinfo {booktitle} {Cryptography and
  Lattices}}},\ \bibinfo {editor} {edited by\ \bibinfo {editor} {\bibfnamefont
  {J.~H.}\ \bibnamefont {Silverman}}}\ (\bibinfo  {publisher} {Springer Berlin
  Heidelberg},\ \bibinfo {address} {Berlin, Heidelberg},\ \bibinfo {year}
  {2001})\ pp.\ \bibinfo {pages} {146--180}\BibitemShut {NoStop}%
\bibitem [{\citenamefont {development team}(2023)}]{fplll}%
  \BibitemOpen
  \bibfield  {author} {\bibinfo {author} {\bibfnamefont {T.~F.}\ \bibnamefont
  {development team}},\ }\href {https://github.com/fplll/fplll} {\enquote
  {\bibinfo {title} {{fplll}, a lattice reduction library, {Version}: 5.5.0},}\
  } (\bibinfo {year} {2023}),\ \bibinfo {note} {available at
  \url{https://github.com/fplll/fplll}}\BibitemShut {NoStop}%
\bibitem [{\citenamefont {Ryan}\ and\ \citenamefont
  {Heninger}(2023)}]{Ryan:2023flatter}%
  \BibitemOpen
  \bibfield  {author} {\bibinfo {author} {\bibfnamefont {K.}~\bibnamefont
  {Ryan}}\ and\ \bibinfo {author} {\bibfnamefont {N.}~\bibnamefont
  {Heninger}},\ }in\ \href {\doibase 10.1007/978-3-031-38548-3_1} {\emph
  {\bibinfo {booktitle} {{Advances in Cryptology -- CRYPTO 2023}}}},\ \bibinfo
  {series and number} {Lecture Notes in Computer Science}\ (\bibinfo
  {publisher} {Springer},\ \bibinfo {year} {2023})\ pp.\ \bibinfo {pages}
  {3--36},\ \bibinfo {note} {cryptology ePrint Archive 2023/237}\BibitemShut
  {NoStop}%
\bibitem [{\citenamefont {Tkachov}(1981)}]{Tkachov:1981wb}%
  \BibitemOpen
  \bibfield  {author} {\bibinfo {author} {\bibfnamefont {F.~V.}\ \bibnamefont
  {Tkachov}},\ }\href {\doibase 10.1016/0370-2693(81)90288-4} {\bibfield
  {journal} {\bibinfo  {journal} {Phys. Lett. B}\ }\textbf {\bibinfo {volume}
  {100}},\ \bibinfo {pages} {65} (\bibinfo {year} {1981})}\BibitemShut
  {NoStop}%
\bibitem [{\citenamefont {Chetyrkin}\ and\ \citenamefont
  {Tkachov}(1981)}]{Chetyrkin:1981qh}%
  \BibitemOpen
  \bibfield  {author} {\bibinfo {author} {\bibfnamefont {K.~G.}\ \bibnamefont
  {Chetyrkin}}\ and\ \bibinfo {author} {\bibfnamefont {F.~V.}\ \bibnamefont
  {Tkachov}},\ }\href {\doibase 10.1016/0550-3213(81)90199-1} {\bibfield
  {journal} {\bibinfo  {journal} {Nucl. Phys. B}\ }\textbf {\bibinfo {volume}
  {192}},\ \bibinfo {pages} {159} (\bibinfo {year} {1981})}\BibitemShut
  {NoStop}%
\bibitem [{\citenamefont {Laporta}(2000)}]{Laporta:2000dsw}%
  \BibitemOpen
  \bibfield  {author} {\bibinfo {author} {\bibfnamefont {S.}~\bibnamefont
  {Laporta}},\ }\href {\doibase 10.1142/S0217751X00002159} {\bibfield
  {journal} {\bibinfo  {journal} {Int. J. Mod. Phys. A}\ }\textbf {\bibinfo
  {volume} {15}},\ \bibinfo {pages} {5087} (\bibinfo {year} {2000})},\ \Eprint
  {http://arxiv.org/abs/hep-ph/0102033} {arXiv:hep-ph/0102033} \BibitemShut
  {NoStop}%
\end{thebibliography}%

\end{document}